\documentclass[english]{article}

\usepackage[T1]{fontenc}
\usepackage[latin9]{inputenc}
\usepackage{color}
\usepackage{mathtools}
\usepackage{amsmath}
\usepackage{amssymb}
\usepackage{cancel}
\usepackage{mathdots}
\usepackage{stmaryrd}
\usepackage{stackrel}
\usepackage{undertilde}
\PassOptionsToPackage{version=3}{mhchem}
\usepackage{mhchem}
\usepackage{esint}
\usepackage{jheppub}
\usepackage{lmodern}
\usepackage{slashed} 
\usepackage{bbm}
\usepackage{babel}
\usepackage{xparse}
\usepackage{physics}
\usepackage{tensor}
\usepackage{slashed}
\usepackage{flowchart}
\usetikzlibrary{arrows}

\usepackage{epsfig}
\usepackage{amsfonts}
\usepackage{graphicx}
\usepackage{fancybox}
\usepackage{enumerate}
\usepackage{dsfont}
\usepackage{verbatim}
\usepackage{wrapfig}
\usepackage{shuffle}
\usepackage{braket}

\makeatletter
\numberwithin{equation}{section}

\renewcommand\[{\begin{equation}}
\renewcommand\]{\end{equation}}
\renewenvironment{gather*}{\gather}{\endgather}
\renewenvironment{align*}{\align}{\endalign}

\newcommand{\btheta}{\bar{\theta}}
\newcommand{\bepsilon}{\bar{\epsilon}}
\newcommand{\e}{\mathrm{e}}
\newcommand{\cq}{\mathcal{Q}^+}
\newcommand{\fq}{\mathcal{Q}^-}

\author[a]{Luca Griguolo,} 
\author[a]{Luigi Guerrini,} 
\author[b]{and Itamar Yaakov} 
\affiliation[a]{Dipartimento di Scienze Matematiche Fisiche e Informatiche, Universit\`a di Parma \& INFN Gruppo Collegato di Parma, Parco Area delle Scienze 7/A, 43124 Parma, Italy} 
\affiliation[b]{INFN - Sezione di Milano Bicocca, Dipartimento di Fisica, Edificio U2, Piazza della Scienza 3, I-20126 Milano, MI, Italy} 
\emailAdd{luca.griguolo@unipr.it} 
\emailAdd{luigi.guerrini@unipr.it} 
\emailAdd{itamar.yaakov@mib.infn.it}

\abstract{We investigate several aspects of BPS latitude Wilson loops in gauge theories in three dimensions with $\mathcal{N}\ge 4$ supersymmetry. We derive a matrix model for the bosonic latitude Wilson loop in ABJM using supersymmetric localization, and show how to extend the computation to more general Chern-Simons-matter theories. We then define latitude type Wilson and vortex loop operators in theories without Chern-Simons terms, and explore a connection to the recently derived superalgebra defining local Higgs and Coulomb branch operators in these theories. Finally, we identify a BPS loop operator dual to the bosonic latitude Wilson loop which is a novel bound state of Wilson and vortex loops, defined using a worldvolume supersymmetric quantum mechanics.}

\makeatother

\usepackage{babel}
\begin{document}
\title{Localization and Duality for ABJM Latitude Wilson Loops}
\maketitle

\section{Introduction}

Extended operators or defects, those supported on sub-manifolds of
positive dimension, play a central role in the study of gauge theory.
The prototypical example is the Wilson loop, defined as the trace of the holonomy of a gauge field in a chosen representation. Insertions
of the Wilson loop operator capture the effects of a non-dynamical
electrically charged probe particle. The behavior of large Wilson loop operators
can famously be used to characterize the low energy phase of gauge
theories in four spacetime dimensions \cite{tHooft:1977nqb}, and it has recently been argued that operators supported on higher dimensional submanifolds can aid in this process \cite{Gukov:2013zka}. 

Depending on the spacetime dimension, gauge theories may admit additional
loop operators which are not of Wilson type. Vortex operators are an example of such defect operators in gauge theory in three dimensions \cite{Witten:1988hf,Moore:1989yh}. It is interesting, in this situation,
to study the relationships within the collection of all loop operators vis-\`{a}-vis
the operator product expansion and duality. In the presence of supersymmetry,
it is also natural to consider BPS line operators, those which are invariant
under a subset of the supercharges. Under favorable circumstances, such operators can be studied
using supersymmetric localization which yields finite dimensional
expressions capturing exact non-perturbative expectation values. 

In superconformal field theories, BPS Wilson loops are directly related to other physical observables like the cusp anomalous dimension and the Bremsstrahlung function \cite{Correa:2012at,Correa:2012hh,Drukker:2012de,Gromov:2012eu}. The quantum expectation value of these observables can be exactly determined, in principle, by applying supersymmetric localization to the loop operator. Alternatively, these quantities can often be computed by using integrability techniques. The study of BPS 
Wilson loops therefore represents a potential testing ground for integrability.

In this paper, we investigate a one parameter family of BPS line operators,
the latitude Wilson loops \cite{Cardinali:2012ru, Bianchi:2014laa}, which first appeared in the ABJM model \cite{Aharony:2008ug}. The ABJM
model is a Chern-Simons-Matter theory in three spacetime dimensions
with gauge group $U\left(N_1\right)_{k}\times U\left(N_2\right)_{-k}$,
Chern-Simons levels $k$ and $-k$, and $\mathcal{N}\ge6$ supersymmetry.
When $N_1=N_2$ and $k$ is equal to $1$ or $2$, the supersymmetry of the ABJM
model is enhanced to $\mathcal{N}=8$. Specifically, ABJM at level
$1$ with $N_{1,2}=N$ is a dual description of the infra-red limit of $\mathcal{N}=8$ super-Yang-Mills
(SYM) with gauge group $U\left(N\right)$. An exact relationship between the latitude loops and ABJM 
Bremsstrahlung functions was proposed in \cite{Bianchi:2014laa}, and subsequently proven in 
\cite{Bianchi:2017ozk, Bianchi:2018scb}. 

In \cite{Bianchi:2018bke}, the authors presented a matrix model which
was conjectured to capture the exact expectation value of the latitude
Wilson loop in ABJM. The matrix model is a deformation of the one derived in \cite{Kapustin:2009kz} for the $1/6$ BPS Wilson loop. We will attempt to give a derivation of this
matrix model using supersymmetric localization. Somewhat surprisingly, the main obstruction
we encounter to preforming these localization computations in ABJM is the
difficulty of realizing a single generic supersymmetry transformation
off-shell. This technical difficulty is evident in a larger class
of 3d $\mathcal{N}\ge4$ Chern-Simons-matter theories constructed
by Gaiotto and Witten (GW) in \cite{Gaiotto:2008sd}, of which ABJM is
a generalization \cite{Hosomichi:2008jd}. We will show that theories in this class also support BPS latitude Wilson loops. We will then attempt to overcome
the problems with off-shell closure using a cohomological version of Chern-Simons theory introduced by K\"{a}ll\'{e}n in \cite{Kallen:2011ny}. With some mild assumptions, we will reproduce the matrix model for ABJM put forward in \cite{Bianchi:2018bke}, and provide the tools to derive the generalizations for matrix models of GW theories recently considered in \cite{Drukker:2020dvr}.

Next, we show that latitude type BPS loop operators can also be defined in 3d $\mathcal{N}\ge 4$ theories without Chern-Simons terms. In order to avoid confusion, we will call such theories \textit{standard}. Localization in these theories has a slightly different flavor because of the availability of unconstrained off-shell vector multiplets. The BPS loop operators in question are of both Wilson and vortex type. We also find an interesting
relationship between the superalgebra preserved by latitude loops, in an appropriate limit, and that which governs Higgs and Coulomb branch operators, and the associated topological quantum mechanics,
of standard $\mathcal{N}\ge4$ theories. These local operators were introduced in \cite{Beem:2013sza} and subsequently studied in \cite{Chester:2014mea,Beem:2016cbd,Dedushenko:2016jxl}.

Finally, we perform localization with a latitude type loop operator in a standard non-conformal
$\mathcal{N}=4$ SYM theory in the same universality class as ABJM
at level $1$. We first identify an appropriate dual loop operator to the ABJM latitude Wilson loop. Surprisingly, our results imply that this operator
is a mixture of a BPS Wilson loop and a
BPS vortex loop operator. We define such a mixed type operator using an explicit worldvolume supersymmetric quantum mechanics, which is a deformation of the models for $1/2$-BPS Wilson and vortex loops discussed by Assel and Gomis in \cite{Assel:2015oxa}. We then show that the matrix model obtained from considering this operator, although ostensibly quite different, computes the same expectation value as the one which we derive for the latitude loop in ABJM.

The paper is organized as follows. In Section \ref{sec:The-ABJM-Model},
we review the ABJM model and the most general version of the BPS latitude Wilson loops and their supersymmetry algebra in flat space. We discuss the original and deformed matrix models, and exhibit latitude type BPS Wilson loops for general GW theories. In Section \ref{sec:The-latitude-matrix-model},
we map the latitude loops to the three sphere and use supersymmetric localization to derive the deformed ABJM matrix model. In Section \ref{sec:Latitude-loop-in-standard-N4-theories}, we describe the action of the superalgebra preserved by the latitude loops in standard $\mathcal{N}=4$ gauge theories, and show the relationship to Higgs and Coulomb branch operators. We also classify
two primary types of loop operators, Wilson operators and vortex operators, compatible with the supercharges preserved
by the latitude. We then identify the loop operator dual to the ABJM latitude and use supersymmetric localization to derive a matrix model for its expectation value. Section \ref{sec:Conclusion} contains a summary of the results and a discussion of open problems and interesting future work.

\section{\label{sec:The-ABJM-Model}The ABJM model and the latitude loop}

We describe in this section a class of Wilson loops in $U(N_1)_k \times U(N_2)_{-k}$ ABJM theory that preserve a certain fraction of the original ${\mathcal N} = 6$ supersymmetry. These operators can be constructed by generalizing the ordinary gauge holonomy with the addition of
either scalar matter bilinears (obtaining the so-called ``bosonic'' Wilson loops) or scalar bilinears and fermions (resulting in ``fermionic'' Wilson loops). For the case of straight line and maximal circular contours, a general classification of such BPS operators based on the amount
of preserved supercharges can be found in \cite{Ouyang:2015iza, Ouyang:2015bmy, Mauri:2017whf}. The bosonic case turns out to preserve 1/6 of the supersymmetries of the theory, and the two gauge groups of ABJM are decoupled in the construction \cite{Berenstein:2008dc, Drukker:2008zx, Chen:2008bp, Rey:2008bh}. A more supersymmetric operator, preserving half of the supersymmetries, mixes instead the two gauge groups and includes coupling to the fermionic fields as well \cite{Drukker:2009hy}. Unexpectedly, the supersymmetry variation of this loop does not vanish locally, but it is rather a total derivative along the loop. It is also possible to deform continuously the fermionic Wilson loop into a linear combination of two bosonic ones, preserving 1/6 of the supersymmetries along the flow \cite{Ouyang:2015iza, Ouyang:2015bmy, Drukker:2019bev} (see also \cite{Drukker:2020dvr,Drukker:2020opf} for a discussion of this topic in the case of theories with lower supersymmetry). In this paper, we are instead interested in studying another one-parameter family of deformations of the bosonic and fermionic maximal circles, the latitudes \cite{Cardinali:2012ru, Bianchi:2014laa}.

We start by recalling the general class of Wilson operators introduced in \cite{Cardinali:2012ru}. They feature a parametric dependence on a $\alpha$--angle\footnote{The $\alpha$--angle can be freely chosen in the interval $[0, \tfrac{\pi}{2}]$, see \cite{Cardinali:2012ru}.} that governs the couplings to matter in the internal $R$--symmetry space and a geometric angle $\theta_0 \in [-\tfrac{\pi}{2}, \tfrac{\pi}{2}]$ that fixes the contour to be a latitude circle on the unit sphere   
\begin{equation}
\Gamma: \quad x^\mu = (\cos{\theta_0} \cos{\tau}, \cos{\theta_0}\sin{\tau}, \sin{\theta_0})\,, \qquad \tau \in [0, 2\pi) \,.
\end{equation}
As discussed in \cite{Bianchi:2014laa}, these operators can be constructed in such a way that they depend only on the effective ``latitude parameter'' 
\begin{equation}
\nu \equiv \sin{2\alpha} \cos{\theta_0}\,, \qquad \qquad 0 \leq \nu \leq 1\,.
\end{equation}
The bosonic latitude Wilson loops corresponding to the two gauge groups are explicitly given by 
\begin{align}\label{eq:bosonic}
&& W_B(\nu,\mathfrak{R}) = \frac{1}{ {\rm dim}(\mathfrak{R})}\, \Tr_\mathfrak{R}\, \mathrm{P}\, \exp \left\{-\mathbbm{i}\oint_{\Gamma} d\tau \left(A_{\mu} \dot x^{\mu}-\frac{2 \pi \mathbbm{i}}{k} |\dot x| M_{J}^{\ \ I} C_{I}\bar C^{J}\right) \right\}\,,
\nonumber 
\\ 
&& \hat{W}_{B}(\nu,\hat{\mathfrak{R}}) = \frac{1}{{\rm dim}(\hat{\mathfrak{R}})}\, \Tr_{\hat{\mathfrak{R}}}\, \mathrm{P}\, \exp \left\{-\mathbbm{i}\oint_{\Gamma} d\tau \left( \hat  A_{\mu} \dot x^{\mu}-\frac{2 \pi \mathbbm{i}}{k} |\dot x| M_{J}^{\ \ I} \bar C^{J} C_I \right)  \right\} \,.
\end{align}
where the matrix describing the coupling to the $(C_I, \bar{C}^I)$ scalars reads
\begin{eqnarray} 
 \mbox{\small $\! M_{J}^{\ I}=\left(\!\!
\begin{array}{cccc}
 - \nu  & e^{-\mathbbm{i} \tau } \sqrt{1-\nu ^2} & 0 & 0 \\
e^{\mathbbm{i} \tau }  \sqrt{1-\nu ^2}  & \nu  & 0 & 0 \\
 0 & 0 & -1 & 0 \\
 0 & 0 & 0 & 1 \\
\end{array}
\right)$ }\,.
\end{eqnarray}
The traces in (\ref{eq:bosonic}) are taken over generic representations $\mathfrak{R}, \hat{\mathfrak{R}}$ of $U(N_1)$ and $U(N_2)$, respectively. The overall constants have been chosen in order to normalize the tree level expectation values $\langle W_{B} \rangle^{(0)}$ and $\langle \hat{W}_{B}\rangle^{(0)}$ to one.  

The fermionic latitude Wilson loop is instead defined for a representation\footnote{Special combinations of representations of $U(N_1)\times U(N_2)$ are also representations of $U(N_1|N_2)$: we will consider only the fundamental representation, that actually belongs to these particular cases} $\mathbf{R}$ of the superalgebra $U(N_1|N_2)$
\begin{equation}\label{eq:fermionic}
W_F(\nu,\mathbf{R}) =  {\mathcal{ R}} \; \mathrm{STr}_{\mathbf{R}} \left[ \mathrm{P}\,\exp\left(-\mathbbm{i}\oint_{\Gamma}\mathcal{L}(\tau)d\tau\right) \begin{pmatrix} e^{-\frac{\mathbbm{i}\pi\nu }{2}}  \mathbf{1}_{N_1} & 0\\ 0 &  e^{\frac{\mathbbm{i}\pi\nu}{2}}  \mathbf{1}_{N_2} \end{pmatrix} \right] \,,
\end{equation}
where $\mathcal{L}$ is the $U(N_1|N_2)$ superconnection 
\begin{eqnarray}
\label{supermatrix}
\mathcal{L} &=& \begin{pmatrix}
\mathcal{A}
&\mathbbm{i} \sqrt{\frac{2\pi}{k}}  |\dot x | \eta_{I}\bar\psi^{I}\\
-\mathbbm{i} \sqrt{\frac{2\pi}{k}}   |\dot x | \psi_{I}\bar{\eta}^{I} &
\hat{\mathcal{A}}
\end{pmatrix}\,, \ \  \  \mathrm{with}\ \ \  \left\{\begin{matrix} \mathcal{A}\equiv A_{\mu} \dot x^{\mu}-\frac{2 \pi \mathbbm{i}}{k} |\dot x| {\mathcal M}_{J}^{\ \ I} C_{I}\bar C^{J}\\
\\
\hat{\mathcal{A}}\equiv\hat  A_{\mu} \dot x^{\mu}-\frac{2 \pi \mathbbm{i}}{k} |\dot x| {\mathcal {M}}_{J}^{\ \ I} \bar C^{J} C_{I}
\end{matrix}\ \right.\,, \nonumber \\
\end{eqnarray}
and
\begin{eqnarray}
\label{eq:matrixfermionic}
\mbox{\small $\!{\mathcal{ M}}_{I}^{ \  J}\!=\!\!\left(\!\!
\begin{array}{cccc}
 - \nu  & e^{-\mathbbm{i} \tau } 
   \sqrt{1-\nu ^2} & 0 & 0 \\
e^{\mathbbm{i} \tau }  \sqrt{1-\nu ^2}
   & \nu  & 0 & 0 \\
 0 & 0 & 1 & 0 \\
 0 & 0 & 0 & 1 \\
\end{array}\!
\right)$}\,, \quad & & \quad  \mbox{\small $\begin{array}{l}\eta_I^\alpha \equiv n_I \eta^\alpha = \frac{e^{\frac{\mathbbm{i}\nu \tau}{2}}}{\sqrt{2}}\left(\!\!\!\begin{array}{c}\!\sqrt{1+\nu}\\ -\sqrt{1-\nu} e^{\mathbbm{i}\tau}\\0\\0 \!\end{array}\!\!\right)_{\!I} \! \! \!\!(1, -\mathbbm{i} e^{-\mathbbm{i} \tau})^\alpha
\end{array}$}\!\!\,,
\nonumber \\
&~& \quad \bar\eta_\alpha^I \equiv \bar{n}^I \bar{\eta}_\alpha = \mathbbm{i} (\eta^{\alpha}_{I})^{\dagger}\,.
\end{eqnarray}

The generalized prescription (\ref{eq:fermionic}) that requires taking the supertrace of the superholonomy times a constant ``twist'' matrix was originally proposed to assure invariance under super gauge transformations \cite{Cardinali:2012ru}. Equivalently one can remove the dependence on the twist matrix, the price being the introduction of a suitable background term in the bosonic part of \ref{supermatrix} (see \cite{Drukker:2019bev} for a complete discussion of this topic). The overall constant in (\ref{eq:fermionic}) can be chosen again to normalize the expectation value to 1. In the following discussion, the Wilson loops are understood to be in the fundamental representation. For generic values of the parameters, the latitude bosonic operators in (\ref{eq:bosonic}) preserve 1/12 of the original ${\mathcal {N}}=6$ supercharges, whereas the fermionic one in (\ref{eq:fermionic}) is 1/6--BPS. The explicit expressions of the preserved supercharges will be given in the next section.

A remarkable, and maybe unexpected, feature of these constructions is the following: at the classical level the fermionic latitude Wilson loop (\ref{eq:fermionic}) is cohomologically equivalent to the following linear combination of bosonic latitudes
\begin{equation}\label{cohom}
 W_F(\nu) =  {\mathcal {R}} \left[ N_1 \, e^{-\frac{\mathbbm{i}\pi  \nu}{2}} \, W_B(\nu) + N_2 \, e^{\frac{\mathbbm{i} \pi \nu}{2}} \, \hat W_B(\nu) \right]  + {\mathcal {Q}}(\nu) ({\rm something})\,.
\end{equation}
In the above formula, ${\mathcal{Q}}(\nu)$ is the linear combination of superPoincar\'e and superconformal charges preserved by both bosonic and fermionic Wilson loops \cite{Bianchi:2014laa}. Assuming that this relation holds at the quantum level, the vacuum expectation value $\langle W_{F}  (\nu) \rangle$  of the fermionic operator is equal to the one for a bosonic operator.  However, the problem of understanding how the classical cohomological equivalence gets implemented at the quantum level is strictly interconnected with the problem of understanding framing \cite{Bianchi:2016yzj}. In the circular case it holds at framing one, as originally proposed in \cite{Kapustin:2009kz}, while for latitudes a generic non-integer framing $\nu$ should be assumed (see \cite{Bianchi:2014laa} for an extensive discussion of this point).

The exact computation of the 1/6 BPS circular Wilson loop can be performed by supersymmetric localization \cite{Pestun:2007rz}. Using this procedure, the ABJM partition function on the three--sphere can be reduced to the matrix model integral \cite{Kapustin:2009kz}
\begin{align}
\label{eq:matrixABJM}
Z = &\frac{1}{N_1!N_2!}\int \prod_{a=1}^{N_1}d\lambda _{a} \ e^{\mathbbm{i}\pi k\lambda_{a}^{2}}\prod_{b=1}^{N_2}d\mu_{b} \ e^{-\mathbbm{i}\pi k\mu_{b}^{2}}\,  \frac{\displaystyle\prod_{a<b}^{N_1}\left(2\sinh \pi (\lambda_{a}-\lambda _{b})\right)^2\prod_{a<b}^{N_2}\left(2\sinh \pi (\mu_{a}-\mu_{b})\right)^2}{\displaystyle\prod_{a=1}^{N_1}\prod_{b=1}^{N_2}\left(2\cosh \pi (\lambda _{a}-\mu_{b})\right)^2}\,.
\end{align}
disregarding the precise normalization, which is unimportant for the computation of Wilson loops. 
At $\nu=1$, the expectation values of the 1/6--BPS bosonic Wilson loop and, by assuming the cohomological equivalence, of the 1/2--BPS fermionic Wilson loop can be expressed as matrix model averages \cite{Kapustin:2009kz,Drukker:2009hy,Drukker:2010nc}. In particular, the bosonic 1/6--BPS Wilson loops in the fundamental representation are given by
\begin{align}\label{eq:matrixmodelaverage}
\langle  W_B (1) \rangle = \frac{1}{N_1}\left\langle \sum_{i=1}^{N_1} e^{2\pi \, \lambda_{i} } \right\rangle\,,  \qquad \qquad 
\langle  \hat W_B (1) \rangle = \frac{1}{N_2}\left\langle \sum_{i=1}^{N_2} e^{2\pi \, \mu_{i} } \right\rangle \,.
\end{align}
where the right--hand--side brackets stand for the integration using the matrix model measure defined in \eqref{eq:matrixABJM}, normalized by the partition function. In this language, the 1/2--BPS Wilson loop, originally defined as the average of a supermatrix operator, can be obtained using \eqref{cohom} as a linear combination of the bosonic ones \cite{Drukker:2009hy, Drukker:2010nc}. 

For generic $\nu$ the situation is a little bit different: even though the latitude bosonic Wilson loops are BPS operators, the standard localization arguments of \cite{Kapustin:2009kz} cannot be directly applied \cite{Bianchi:2014laa} to their computation. On the other hand, one could expect that a matrix model representation for their vacuum expectation value exists. A conjectured a matrix model for the latitude Wilson loops has been therefore proposed in \cite{Bianchi:2018bke}, that turned out to be compatible with all the available data points at weak and strong coupling. We briefly describe its genesis. The idea is to start from the matrix model average \eqref{eq:matrixABJM} computing the expectation value of 1/6--BPS Wilson loops and try to deform it by introducing a suitable dependence on the $\nu$  parameter. The structure of the $\theta$--Bremsstrahlung function \cite{Griguolo:2012iq, Bianchi:2017svd} suggests that the latitude Wilson loop should be computed by inserting the operator $\Tr\, e^{2\pi \sqrt{\nu} \lambda}$ into a matrix model which is symmetric under the inversion $\nu\leftrightarrow 1/\nu$. Such an argument is reminiscent of the one proposed in \cite{Lewkowycz:2013laa}, but somehow with a reverse logic. In that case, for the ABJM theory a supersymmetric Wilson loop on a squashed sphere was considered, whose matrix model is invariant under the inversion of the squashing parameter \cite{Hama:2011ea}. This was used to argue that the derivative of this Wilson loop expectation value with respect to the squashing parameter $b$, evaluated at $b= 1$, could be traded with the derivative of the multiply wound 1/6--BPS Wilson loop with respect to the winding number $m$ (see \cite{Bianchi:2014laa, Bianchi:2018scb} for a detailed discussion of these points).  Using also some perturbative inputs, the following matrix model average for the expectation value of a bosonic latitude Wilson loop has been proposed
\begin{align} \label{eq:matrixlat}
\langle W_B(\nu) \rangle  &= \left\langle \frac{1}{N_1} \sum_{1\leq i\leq N_1} e^{2\pi\,  \sqrt{\nu}\, \lambda_{i}} \right\rangle\,,
\end{align}
where the average is evaluated and normalized using the matrix model partition function
\begin{align} \label{eq:partitionf}
&  Z =\frac{1}{N_1!N_2!} \int \prod_{a=1}^{N_1}d\lambda _{a} \ e^{\mathbbm{i}\pi k\lambda_{a}^{2}}\prod_{b=1}^{N_2}d\mu_{b} \ e^{-\mathbbm{i}\pi k\mu_{b}^{2}}  \\
&~~ \times 
 \frac{\displaystyle\prod_{a<b}^{N_1}2\sinh \sqrt{\nu} \pi (\lambda_{a}-\lambda _{b})2\sinh\frac{ \pi (\lambda_{a}-\lambda _{b})}{\sqrt{\nu}} \prod_{a<b}^{N_2}2\sinh\sqrt{\nu} \pi (\mu_{a}-\mu_{b})2\sinh\frac{\pi (\mu_{a}-\mu_{b})}{\sqrt{\nu}} }{\displaystyle\prod_{a=1}^{N_1}\prod_{b=1}^{N_2}2\cosh\sqrt{\nu} \pi (\lambda _{a}-\mu_{b})2 \cosh\frac{\pi (\lambda _{a}-\mu_{b})}{\sqrt{\nu}} } \,.\nonumber
\end{align}
Similarly, $\langle \hat W_B(\nu) \rangle$ corresponds to the insertion of $\frac{1}{N_2} \sum_{1\leq i\leq N_2} e^{2\pi\, \, \sqrt{\nu}\, \mu_{i}} $. According to the discussion above, such a matrix model should arise from a suitable localization of the ABJ(M) theory. We stress that this is the simplest non--trivial deformation of the matrix model \eqref{eq:matrixABJM} reducing to the usual expression at $\nu=1$, and whose measure is symmetric under $\nu\leftrightarrow 1/\nu$, the specific dependence on $\nu$ in the hyperbolic functions and in the operator insertion has been fixed via comparison with the perturbative results \cite{Bianchi:2018bke}. Three-loop weak-coupling computations are consistent with this conjecture: the proposed matrix model is amenable of a Fermi gas reformulation \cite{Marino:2011eh} and strong-coupling results can be obtained that coincide with previous and subsequent string-theory calculation at leading and subleading order \cite{Correa:2014aga, Aguilera-Damia:2014bqa, Aguilera-Damia:2018bam, Medina-Rincon:2019bcc, David:2019lhr}.

Analyzing this matrix model, we immediately realize that the $\nu$-dependence cannot be reabsorbed by a simple
redefinition of the coupling constant: we expect obviously that the deformation could affect the observable that we average to
evaluate the Wilson loop. However to be consistent with a localization result, when we replace the observable with the identity the  matrix integral over the deformed measure must still give the partition function of ABJ(M) on $S^3$. This implies that the
dependence of the partition function on $\nu$ must become trivial, a peculiar feature that should rely on the intimate structure of the matrix model itself. The crucial property that streamlines this miraculous independence is the Cauchy determinant identity
\begin{equation}\label{eq:cauchy}
\frac{\displaystyle\prod_{a<b}^{N}\sinh r\, \pi\, (\lambda_{a}-\lambda _{b})\, \sinh r\, \pi\, (\mu_{a}-\mu_{b})}{\displaystyle\prod_{a=1}^{N}\prod_{b=1}^{N}\cosh\, r\, \pi\, (\lambda _{a}-\mu_{b})} = \sum_{\sigma\in S_N} (-1)^{\sigma} \prod_{a=1}^{N} \frac{1}{\cosh \,r\, \pi\, (\lambda _{a}-\mu_{\sigma(a)})}\,.
\end{equation}
Here the final sum is over all the permutations of the $N$ eigenvalues and  $r$ is an arbitrary parameter. Let us consider the case of equal rank: splitting the integrand of \eqref{eq:partitionf} into two combinations of hyperbolic functions with arguments containing the factors $\sqrt{\nu}$ and $1/\sqrt{\nu}$ respectively, and applying the Cauchy identity separately with an appropriate choice of $r$, after few algebraic steps described in \cite{Kapustin:2010xq, Bianchi:2018bke}, we end up with
\begin{equation}\label{eq:partfuncfermi}
Z = \frac{1}{N!}\, \sum_{\sigma\in S_N} (-1)^{\sigma}\, \int \frac{dy^N}{(2\pi k)^N} \prod_{a=1}^N \left(2 \cosh  \frac{y_a}{2}\right)^{-1}\left( 2 \cosh \frac{y_a-y_{\sigma(a)}}{2k}\right)^{-1}\,,
\end{equation}
where $N=N_1=N_2$.
We thus see that the dependence on $\nu$ drops completely and the partition function has the same expression as for ABJM theory.
For different ranks of the gauge groups  the starting point \eqref{eq:cauchy} must be replaced by  a generalization of the Cauchy determinant identity  discussed in \cite{Awata:2012jb, Matsumoto:2013nya, Honda:2013pea}. Then  if we repeat the steps leading from \eqref{eq:cauchy} to \eqref{eq:partfuncfermi}, we can isolate and evaluate the $\nu-$dependent part of the partition function
\begin{equation}
\label{phaseABJ}
\exp\left(\frac{\pi \mathbbm{i}}{12 k}\left(\nu+\frac{1}{\nu}\right) ((N_1-N_2)^3-(N_1-N_2))\right)\,.
\end{equation}
The only effect of the deformed measure consists in altering  the original phase of the ABJ partition function obtained in \cite{Marino:2009jd, Drukker:2010nc} by a trivial  $\nu-$dependent multiplicative factor.

When we insert the Wilson loop in the $\nu-$deformed matrix model we get a $\nu-$dependent result. This can be shown by repeating the steps leading to \ref{eq:partfuncfermi}, as shown in \cite{Bianchi:2018bke}. For future reference, we write the result for the case $N_1=N_2=N$ and $k=1$:
\begin{align}\label{latitudenu}   
 \langle W_{B}(\nu) \rangle= \frac{1}{Z \,N(N!)}\sum_{\rho\in S_N}(-1)^\rho\int d\sigma^N
 \sum_{c=1}^N\,\,\,\frac{e^{\mathbbm{i}\pi\nu} e^{2\pi\sigma_c} }{\prod_{j=1}^N 2\cosh\pi\sigma_j\prod_{k=1}^N 2\cosh\pi(\sigma_k-\sigma_{\rho(k)}+\mathbbm{i}\nu\delta_k^c)}\,.
 \end{align}
This equation makes manifest the fact that the $\nu-$dependence does not disappear.

\subsection{\label{subsec:the_latitude_algebra}The latitude algebra}

The presence of a BPS defect in a superconformal theory breaks the superconformal algebra to a subalgebra. Below, we detail the symmetry algebra left unbroken by the ABJM latitude Wilson loop in its fermionic and bosonic versions. Our supersymmetry conventions for ABJM are collected in Appendix \ref{app1}. 

The supercharges preserved by the fermionic latitude are given by \cite{Bianchi:2014laa}
\begin{align}\label{fund_superch}
\btheta^{13}_1&=\e^{-\mathbbm{i}\frac{\theta_0}{2}}\sqrt{1-\nu}\,\omega_1+ \e^{\mathbbm{i}\frac{\theta_0}{2}}\sqrt{1+\nu}\,\omega_2\, ,\quad &\btheta^{14}_1&=\e^{-\mathbbm{i}\frac{\theta_0}{2}}\sqrt{1-\nu}\,\omega_3 +\e^{\mathbbm{i}\frac{\theta_0}{2}}\sqrt{1+\nu}\,\omega_4 \, ,\notag\\
\btheta^{23}_2&=-\mathbbm{i}\e^{-\mathbbm{i}\frac{\theta_0}{2}}\sqrt{1+\nu}\, \omega_1-\mathbbm{i}\e^{\mathbbm{i}\frac{\theta_0}{2}}\sqrt{1-\nu}\,\omega_2 \, ,\quad &\btheta^{24}_2&= -\mathbbm{i}\e^{-\mathbbm{i}\frac{\theta_0}{2}}\sqrt{1+\nu}\,\omega_3 -\mathbbm{i}\e^{\mathbbm{i}\frac{\theta_0}{2}}\sqrt{1-\nu}\,\omega_4   \, ,\notag\\
\bepsilon^{13}_1&=\mathbbm{i}\e^{\mathbbm{i}\frac{\theta_0}{2}}\sqrt{1-\nu}\,\omega_1 -\mathbbm{i}\e^{-\mathbbm{i}\frac{\theta_0}{2}}\sqrt{1+\nu}\,\omega_2\, , \quad &\bepsilon^{14}_1&=\mathbbm{i}\e^{\mathbbm{i}\frac{\theta_0}{2}}\sqrt{1-\nu}\,\omega_3 -\mathbbm{i}\e^{-\mathbbm{i}\frac{\theta_0}{2}}\sqrt{1+\nu}\,\omega_4   \, ,\notag\\
\bepsilon^{23}_2&=\e^{-\mathbbm{i}\frac{\theta_0}{2}}\sqrt{1-\nu}\,\omega_2- \e^{\mathbbm{i}\frac{\theta_0}{2}}\sqrt{1+\nu}\,\omega_1\, , \quad &\bepsilon^{24}_2&=\e^{-\mathbbm{i}\frac{\theta_0}{2}}\sqrt{1-\nu}\,\omega_4- \e^{\mathbbm{i}\frac{\theta_0}{2}}\sqrt{1+\nu}\,\omega_3 \, ,
\end{align}
where $\omega_i$, $i=1,\dots,4$ are bosonic parameters and we have taken the $\mathbb{R}^3$ conformal Killing spinor to be 
$\bar{\Theta}^{IJ}_\alpha=\bar{\theta}^{IJ}_\alpha-x^\mu(\gamma_\mu)\indices{_\alpha^\beta}\bar{\epsilon}^{IJ}_\beta$.
We define the corresponding operators $Q_i$, $i=1,\dots,4$,  turning on the corresponding $\omega_i$ and setting the others to zero. We also perform a rescaling on the supercharges by a factor $2\sqrt{\nu\cos\theta_0}$. In the following sections, we will define the action of the supercharges \ref{fund_superch} in several slightly different contexts, where we will use different convenient normalizations. The $Q_i$ action on a generic operator $\mathcal{O}$ is defined as:
\begin{equation}
\delta\mathcal{O}\equiv\comm*{\btheta^{IJ}_\alpha\bar{Q}^\alpha_{IJ}+\bepsilon^{IJ}_\alpha\bar{S}^\alpha_{IJ}}{\mathcal{O}}\,,
\end{equation}
where $\bar{Q}^\alpha_{IJ}$ and $\bar{S}^\alpha_{IJ}$ are the operators generating the $\mathfrak{osp}(6|4)$ superconformal algebra. 

The couplings of the Wilson loop $\mathcal{M}\indices{_I^J}$ and $\eta_I^\alpha$ preserve an $SU(2)$ R-symmetry, and it turns out to be convenient to recast the conserved supercharges into two $SU(2)$ doublets
\begin{equation}
\mathcal{Q}_a^-=\begin{pmatrix} Q_1 \\ Q_3   \end{pmatrix}\,,\qquad
\mathcal{Q}_a^+=\begin{pmatrix} Q_2 \\ Q_4     \end{pmatrix}\,.
\end{equation}
From the ABJM superconformal algebra, we get the following anti-commutation relations
\begin{align}
\acomm*{\mathcal{Q}_a^-}{\mathcal{Q}_b^-}&=0\,,\qquad \acomm*{\mathcal{Q}_a^+}{\mathcal{Q}_b^+}=0\,,\\
\acomm*{\mathcal{Q}_a^-}{\mathcal{Q}_b^+}&=\epsilon_{ab}\left( \mathcal{T}-Z\right)-2L_{ab}\,,
\end{align}
where we have introduced the conserved bosonic generators
\begin{align}
\mathcal{T}&=\frac{\mathbbm{i}}{\cos\theta_0}\left( K_3-P_3+2\sin\theta_0D\right)\, ,\\
Z&=\frac{1}{\nu}\left(2\mathbbm{i}M^{12}+J\indices{_2^2}-J\indices{_1^1}\right)\, ,\\
L\indices{_a^b}&=\begin{pmatrix}
J\indices{_3^3}+\frac{1}{2}J\indices{_1^1}+\frac{1}{2}J\indices{_2^2} & J\indices{_3^4}\\
J\indices{_4^3} &J\indices{_4^4}+\frac{1}{2}J\indices{_1^1}+\frac{1}{2}J\indices{_2^2} \\
 \end{pmatrix}=\begin{pmatrix}
L_z & L_-\\
L_+ & -L_z
 \end{pmatrix}\,.
\end{align}

$L\indices{_a^b}$ are the generators of the $\mathfrak{su}(2)$ R-symmetries preserved by the loop. Their action on the conserved supercharges is given by
\begin{align}
\comm*{L\indices{_a^b}}{\fq_c}&=\frac{1}{2}\delta_a^b\fq_c-\delta_c^b\fq_a \, ,\\
\comm*{L\indices{_a^b}}{\cq_c}&=\frac{1}{2}\delta_a^b\cq_c-\delta_c^b\cq_a\, .
\end{align}
$\mathcal{T}$ generates a $U(1)$ symmetry followed by a dilatation. Its action on the supercharges is
\begin{equation}
\comm*{\mathcal{T}}{\fq_a}=-\fq_a \,, \qquad  \comm*{\mathcal{T}}{\cq_a}=\cq_a\, .
\end{equation}
The presence of the dilatation reflects the possibility of moving the loops at different heights of the $S^2$. Finally, $Z$ is a central element of the subalgebra. It generates a $U(1)$ symmetry which mixes a Lorentz rotation with a specific R-symmetry. The precise combination of the two is fixed by the invariance of the off-diagonal terms in ${\mathcal{M}_I}^J$ (or equivalently ${M_I}^J$). The role of $Z$ will be crucial in the following, especially when we will describe loop operators as mixed 3d/1d systems.
The complete superalgebra is a central extension of $\mathfrak{su}(1|2)$, which is isomorphic to $\mathfrak{osp}(2|2)$. 
 
In the rest of the paper, we will be interested mostly in the bosonic latitude. The bosonic latitude preserves only the subset of the supercharges: those generated by $Q_2$ and $Q_3$.
The resulting $\mathfrak{su}(1|1)$ subalgebra of the full latitude algebra is given by
\begin{align}
\acomm*{Q_2}{Q_3}&=\mathcal{T}+2L_z- Z\, ,\\
\comm*{\mathcal{T}}{Q_2}&=Q_2 \, ,  &\comm*{\mathcal{T}}{Q_3}&=-Q_3\, ,\\
\comm*{L_z}{Q_2}&=-\frac{1}{2}Q_2\, ,  &\comm*{L_z}{Q_3}&=\frac{1}{2}Q_3\, .
\end{align}
From now on, we will refer to the bosonic latitude as ``the latitude''. We will also restrict our analysis to $Q_2$ and $Q_3$. Moreover, we will take the limit $\theta_0\to 0$. In this way we avoid any issues related to the presence of the non compact symmetry $D$ in the latitude algebra. Latitude loops at other values of $\theta_0$ are related to the one at $\theta_0=0$ by the action of the conformal group. We can therefore restrict ourselved to examining the expectation value of the $\theta_0=0$ loop without loss of generality.

\subsubsection{The limit $\nu\to 1$}\label{sub_lat_alg_nu_1}

In the limit $\nu\to1$, the latitude coincides with the well-known 1/6-BPS circular Wilson loops introduced by Gaiotto and Yin in \cite{Gaiotto:2007qi}, and further discussed in \cite{Drukker:2008zx, Berenstein:2008dc, Chen:2008bp}. In this limit there are also two additional supercharges preserved by the latitude. All of the supercharges preserved by the latitude in this limit can be parameterized as follows
\begin{subequations}\label{eqsup}
\begin{align}
\bar{\theta}^{13}_1&=\omega_2\,, &\bar{\theta}^{24}_2&=-\mathbbm{i}\omega_3\,,\\
\bar{\theta}^{13}_2&=\mathbbm{i}\omega_5\,, &\bar{\theta}^{24}_1&=\omega_6\,,
\end{align}
\end{subequations}
with the relations
\begin{equation}  
\bar{\epsilon}^{IJ}_\alpha=\mathbbm{i}M\indices{_K^I}(\tau_3)\indices{_\alpha^\beta}\bar{\theta}^{KJ}_\beta\,.
\end{equation}
where now $M\indices{_K^I}=\mathrm{diag}(-1,1,-1,1)$.

It is interesting to show how the supersymmetry enhancement in the limit $\nu\to1$ modifies the latitude superalgebra. Before providing the explicit derivation, we can make an educated guess based on some known results. First, we note that the $1/6$-BPS circular Wilson loop is related to the $1/6$-BPS Wilson line by a conformal transformation \cite{Griguolo:2012iq}. On the other hand, the superalgebra preserved by the Wilson line is $\mathfrak{su}(1,1|1)$, which is a superconformal algebra on the line \cite{Bianchi:2018scb}. Therefore, we expect to find the same superalgebra for the circular Wilson loop, realized in terms of different generators of the ABJM supersymmetry algebra.\footnote{This is the same situation as for the maximally supersymmetric Wilson loop in $\mathcal{N}=4$ SYM in 4d \cite{Drukker:2007qr}. The maximally supersymmetric circular Wilson loop and the maximally supersymmetric Wilson line are related by a conformal map, therefore they preserve the same algebra. However, the algebra is realized with different bulk generators.}.

Let us verify that this is indeed the case. We introduce operators associated to the supersymmetries \ref{eqsup}. They are denoted 
$Q_2$, $Q_3$, $Q_5$ and $Q_6$, corresponding to the parameters $\omega_2$, $\omega_3$, $\omega_5$, and $\omega_6$. We choose the notation so as to make $Q_2$ and $Q_3$ the limit as $\nu\to1$ of $Q_2$ and $Q_3$ of the previous section, up to an overall factor of $1/2$. The explicit form of the supercharges is
\begin{subequations}\label{eq1}
\begin{align}
Q_2&=\frac{1}{\sqrt{2}}\left(\bar{Q}_{2,13}-\mathbbm{i}\bar{S}_{2,13}\right)\,, &Q_3&=\frac{1}{\sqrt{2}}\left(\bar{S}_{1,24}+\mathbbm{i}\bar{Q}_{1,24}\right)\,,\\
Q_5&=\frac{1}{\sqrt{2}}\left(\bar{S}_{1,13}-\mathbbm{i}\bar{Q}_{1,13}\right)\,, &Q_6&=\frac{1}{\sqrt{2}}\left(\bar{Q}_{2,24}+\mathbbm{i}\bar{S}_{2,24}\right)\,.
\end{align}
\end{subequations}
It is also convenient to introduce the following space time generators
\begin{align}\label{1d_conf_bos}
M=-\mathbbm{i}M_{12}\,, \qquad P=\frac{1}{2}\left(P_++K_+\right)\,, \qquad K=\frac{1}{2}\left(P_-+K_-\right)\,,
\end{align}
where $P_\pm=P_1\pm \mathbbm{i}P_2$ and $K_\pm=K_1\pm \mathbbm{i}K_2$. There is also a mixed $\mathfrak{u}(1)$ charge
\begin{equation}
J=\frac{1}{2}\left( \mathbbm{i}P_3-\mathbbm{i}K_3 -2\left(J\indices{_1^1}+J\indices{_3^3}\right)\right)\,.
\end{equation}

The odd-odd part of the enhanced superalgebra is given by
\begin{subequations}\label{1d_conf_ferm}
\begin{align}
\acomm{Q_2}{Q_3}&=2\left(M-J  \right)\,,      &\acomm{Q_5}{Q_6}&=2\left( M+J\right)\,,\\
\acomm{Q_2}{Q_6}&= -2P \,,        &\acomm{Q_3}{Q_5}&=2K\,,
\end{align}
\end{subequations}
The bosonic subalgebra is
\begin{align}
\comm*{P}{K}=2M\qquad \comm*{M}{P}=P\,, \qquad  \comm*{M}{K}=-K\,,
\end{align}
which is indeed the expected $\mathfrak{su}(1,1)$ 1d conformal algebra.
Finally, the mixed commutators are given by:
\begin{subequations}
\begin{align}
\comm*{M}{Q_2}&=\frac{1}{2}Q_2\,,     & \comm*{M}{Q_3}&=-\frac{1}{2}Q_3 \,, &\comm*{M}{Q_5}&=-\frac{1}{2}Q_5 \,,   & \comm*{M}{Q_6}&=\frac{1}{2}Q_6\,, \\
\comm{K}{Q_2}&=Q_5 \,, &\comm{K}{Q_6}&=Q_3 \,, &\comm{P}{Q_5}&=Q_2 \,,  &\comm{P}{Q_3}&=Q_6\,,  \\ 
\comm{J}{Q_2}&=\frac{1}{2}Q_2  \,,  &\comm{J}{Q_3}&=-\frac{1}{2}Q_3  \,,  &\comm{J}{Q_5}&=\frac{1}{2}Q_5 \,,  &\comm{J}{Q_6}&=-\frac{1}{2}Q_6\, .
\end{align}
\end{subequations}
Note that $Q_2$ and $Q_6$ behaves as Poincar\'e supercharges on the loop, while $Q_3$ and $Q_5$ behave like superconformal ones.

In summary, our analysis shows that the latitude preserves a specific 1d Poincar\'e type superalgebra. In the limit $\nu \to1$, this superalgebra enhances to the conformal superalgebra $\mathfrak{su}(1,1|1)$ preserved by the standard $1/6$-BPS Gaiotto-Yin Wilson loop.

\subsubsection{\label{subsec:Embedding-in-N4}Embedding in $\mathcal{N}=4$}

The 3d $\mathcal{N}$-extended superconformal algebra is isomorphic to $\mathfrak{osp}\left(\mathcal{N}\left|4\right.\right)$. The algebra is generated by $\mathcal{N}$ quadruplets of spinors in the vector
representation of $SO\left(\mathcal{N}\right)_{R}$. For the ABJM
model, the manifest supersymmetry is $\mathcal{N}=6$. It is useful
in this case to break up the $SO\left(6\right)_{R}$ index , or rather
$\text{Spin}\left(6\right)_{R}\simeq SU\left(4\right)_{R}$, into
a doublet of anti-symmetric $SU\left(4\right)_{R}$ indices denoted
by $\left\{ I,J,K,\ldots\right\} $. The transformation between the
two notations is performed using the 6d Euclidean Clifford algebra
and its matrices $\Gamma_{aIJ}$

\begin{gather*}
\Gamma_{1}\equiv\mathbbm{i}\tau_{2}\otimes\mathbbm{1}\, ,\quad\Gamma_{2}\equiv\tau_{2}\otimes\tau_{3}\, ,\quad\Gamma_{3}\equiv\tau_{2}\otimes\tau_{1}\, ,\\
\Gamma_{4}\equiv\mathbbm{i}\tau_{1}\otimes\tau_{2}\, ,\quad\Gamma_{5}\equiv\mathbbm{1}\otimes\tau_{2}\, ,\quad\Gamma_{6}\equiv\mathbbm{i}\tau_{3}\otimes\tau_{2}\, ,\\
\tilde{\Gamma}_{a}^{IJ}\equiv\left(\Gamma_{a}^{\dagger}\right)^{IJ}\, ,
\end{gather*}
satisfying
$$
\Gamma_{aIJ}=-\Gamma_{aJI},\quad
\Gamma_{a}\tilde{\Gamma}_{b}+\Gamma_{b}\tilde{\Gamma}_{a}=2{\delta_{I}}^{J}{\delta_{a b}},\quad
\tilde{\Gamma}_{a}\Gamma_{b}+\tilde{\Gamma}_{b}\Gamma_{a}=2{\delta^{I}}_{J}{\delta_{ab}}\, .
$$
Note that $\left(\Gamma^{a}\right)_{IJ}\left(\Gamma_{a}\right)_{KL}=-2\varepsilon_{IJKL}$,
so that the usual $SO\left(6\right)$ invariant inner product is replaced
by contraction of pairs of indices using the $\varepsilon$ symbol.
A generic $\mathcal{N}=6$ superconformal transformation is then specified
by a tensor 
\begin{gather*}
\bar{\Theta}_{i}^{IJ},\quad I,J\in\left\{ 1\ldots4\right\} ,i\in\left\{ 1\ldots4\right\} ,\\
\bar{\Theta}_{i}^{IJ}=-\bar{\Theta}_{i}^{JI}.
\end{gather*}
The index $i$ can be contracted with a basis for the conformal Killing spinors, thus yielding a Killing spinor in our previous notation $\bar{\Theta}^{IJ}_\alpha$.

One may show that the fermionic latitude supercharges, parameterized by $\omega_{i}$, can be embedded into
an $\mathcal{N}=4$ subalgebra of the $\mathcal{N}=6$ supersymmetry
algebra of ABJM. Contracting the latitude supercharges with the 6d $\Gamma$ matrices, we get
\[
\bar{\Theta}_{\alpha}^{IJ}\left(\omega\right)\Gamma_{a IJ}=0,\quad a\in\left\{ 5,6\right\} .
\]
If we define the $U\left(1\right)$ generator 
\begin{gather*}
{U^{I}}_{J}\equiv\exp\left(t\left(\tilde{\Gamma}_{5}\Gamma_{6}-\tilde{\Gamma}_{6}\Gamma_{5}\right)\right),\quad U^{\dagger}=U^{-1},
\end{gather*}
then 
\[
U\bar{\Theta}_{\alpha}\left(\omega\right)U^{t}=\bar{\Theta}_{\alpha}\left(\omega\right),\quad U^{*}\mathcal{M}U^{t}=\mathcal{M}.
\]
$U$ therefore generates a residual $SO\left(2\right)$ R-symmetry
commuting with all of the latitude supercharges.

In the limit $\nu\rightarrow1$, the supercharges $Q_{2,3}$ satisfy
the stronger condition
\[
\lim_{\nu\rightarrow1}\bar{\Theta}_{\alpha}^{IJ}\left(\omega_{2,3}\right)\Gamma_{a IJ}=0,\quad a\in\left\{ 3\ldots6\right\} ,
\]
and so form a part of an $\mathcal{N}=2$ subalgebra. In this limit,
the latitude Wilson loop is invariant under an $SO\left(4\right)$
subgroup of the R symmetry. There are also $2$ additional supercharges
preserved by the bosonic latitude loop only in the limit $\nu\rightarrow1$,
which were denoted $Q_{5,6}$. These lie in the same $\mathcal{N}=2$ subalgebra as $Q_{2,3}$. 
\subsection{\label{subsec:Gaiotto-Witten-Theories}Gaiotto-Witten Theories}

We would like to demonstrate that bosonic latitude type Wilson loops
exist in Chern-Simons-matter theories of Gaiotto-Witten (GW) type
\cite{Gaiotto:2008sd}. GW theories are generic Chern-Simons-matter
theories preserving $\mathcal{N}=4$ supersymmetry. They can be formulated
starting from $\mathcal{N}=1$ superfields, such that $\mathcal{N}=4$
supersymmetry is realized on-shell after all auxiliary fields have been
integrated out. The ABJM family of models are GW
theories of a generalized type introduced in \cite{Hosomichi:2008jd},
with special properties allowing an enhancement to $\mathcal{N}=6$
\cite{Hosomichi:2008jb}. We define a latitude type loop as a bosonic
Wilson loop which preserves $Q_{2,3}$ inside the $\mathcal{N}=4$
algebra. The notation in this section is that of \cite{Hosomichi:2008jd},
which differs from the rest of the paper.

The data defining the GW theories are a gauge group $G$, which is
a subgroup of $Sp\left(2n\right)$ for some $n$, an invariant quadratic
form $k^{mn}$ on the Lie algebra $\mathfrak{g}$, and a $2n$-dimensional
representation of $G$ constructed using the subset of $Sp\left(2n\right)$
generators ${\left(t^{m}\right)^{A}}_{B}$, such that $t_{[AC]}\equiv\omega_{[AB}{t^{B}}_{C]}=0$,
where $\omega_{AB}$ is the invariant symplectic form of $Sp\left(2n\right)$
and square brackets denote anti-symmetrization. In order to yield
an $\mathcal{N}=4$ supersymmetric theory, the representation matrices
must satisfy the following \textit{fundamental identity} \cite{Gaiotto:2008sd}
\begin{equation}
k_{mn}t_{(AB}^{m}t_{C)D}^{n}=0.\label{eq:fundamental_identity}
\end{equation}

We will consider a generalization of GW theories described in \cite{Hosomichi:2008jd}.
The generalization introduces an additional set of representation
matrices, ${\left(\tilde{t}^{m}\right)^{A}}_{B}$, satisfying the
same conditions as ${\left(t^{m}\right)^{A}}_{B}$. A generalized GW theory contains the following
fields: a $G$ connection $A_{\mu}^{m}$, and scalar/spinor pairs
$q_{\alpha}^{A},\psi_{\dot{\alpha}}^{A}$ and $\tilde{q}_{\dot{\alpha}}^{A},\tilde{\psi}_{\alpha}^{A}$
valued in the first and second representation, respectively. $\alpha,\dot{\alpha}$
denote $SU\left(2\right)_{l}\times SU\left(2\right)_{r}$ R-symmetry
indices. The pair $q_{\alpha}^{A},\psi_{\dot{\alpha}}^{A}$ form an
$\mathcal{N}=4$ on-shell hypermultiplet, while $\tilde{q}_{\dot{\alpha}}^{A},\tilde{\psi}_{\alpha}^{A}$
form an on-shell twisted hypermultiplet. The hypermultiplets and twisted hypermultiplets can be in different
representations. Following \cite{Hosomichi:2008jd}, we nevertheless
denote all representation indices as $A,B,\ldots$. The Killing spinors are denoted $\eta_{\alpha\dot{\alpha}}$. 

The relevant supersymmetry transformations are \cite{Hosomichi:2008jd}
\begin{gather*}
\delta q_{\alpha}^{A}=\mathbbm{i}{\eta_{\alpha}}^{\dot{\alpha}}\psi_{\dot{\alpha}}^{A},\qquad\delta\tilde{q}_{\dot{\alpha}}^{A}=-\mathbbm{i}{\eta^{\alpha}}_{\dot{\alpha}}\psi_{\alpha}^{A},\\
\delta A_{\mu}^{m}=2\pi\mathbbm{i}\eta^{\alpha\dot{\alpha}}\gamma_{\mu}\left(j_{\alpha\dot{\alpha}}^{m}-\tilde{j}_{\dot{\alpha}\alpha}^{m}\right).
\end{gather*}
The composite quantities $j_{\alpha\dot{\alpha}}^{m},\tilde{j}_{\dot{\alpha}\alpha}^{m}$
are defined as 
\[
j_{\alpha\dot{\alpha}}^{m}\equiv q_{\alpha}^{A}{t_{AB}^{m}}\psi_{\dot{\alpha}}^{B},\qquad\tilde{j}_{\dot{\alpha}\alpha}^{m}\equiv\tilde{q}_{\dot{\alpha}}^{A}\tilde{t}_{AB}^{m}\tilde{\psi}_{\alpha}^{A}.
\]
We also define the moment maps
\[
\mu_{\alpha\beta}^{m}\equiv q_{\alpha}^{A}{t_{AB}^{m}}q_{\beta}^{B},\qquad\tilde{\mu}_{\dot{\alpha}\dot{\beta}}^{m}\equiv\tilde{q}_{\dot{\alpha}}^{A}\tilde{t}_{AB}^{m}\tilde{q}_{\dot{\beta}}^{B}.
\]
Note that\footnote{Symmetrized indices are defined as follows $X_{(a}Y_{b)}\equiv\frac{1}{2}\left(X_{a}Y_{b}+X_{b}Y_{a}\right)$.}
\[
\delta\mu_{\alpha\beta}^{m}=2\mathbbm{i}{\eta_{(\alpha}}^{\dot{\alpha}}j_{\beta)\dot{\alpha}}^{m},\qquad\delta\tilde{\mu}_{\dot{\alpha}\dot{\beta}}^{m}=-2\mathbbm{i}{\eta^{\alpha}}_{(\dot{\alpha}}\tilde{j}_{\dot{\beta})\alpha}^{m}.
\]

Define a generic bosonic Wilson loop
\[
\mathcal{W}\equiv\mathcal{P}\text{tr}_{\mathfrak{R}}\exp\oint_{\ell}\left(A_{\mu}\dot{\ell}^{\mu}+2\pi\left|\dot{\ell}\right|\left(C^{\alpha\beta}\mu_{\alpha\beta}+\tilde{C}^{\dot{\alpha}\dot{\beta}}\tilde{\mu}_{\dot{\alpha}\dot{\beta}}\right)\right),
\]
where $C^{\alpha\beta},\tilde{C}^{\dot{\alpha}\dot{\beta}}$ are symmetric
c-number matrices. The supersymmetry variation of $\mathcal{W}$ is
proportional to 
\[
2\pi\mathbbm{i}\eta^{\alpha\dot{\alpha}}\gamma_{\mu}\dot{\ell}^{\mu}\left(j_{\alpha\dot{\alpha}}^{m}-\tilde{j}_{\dot{\alpha}\alpha}^{m}\right)+2\pi\mathbbm{i}\left|\dot{\ell}\right|\left(C^{\alpha\beta}{\eta_{\alpha}}^{\dot{\alpha}}j_{\beta\dot{\alpha}}-\tilde{C}^{\dot{\alpha}\dot{\beta}}{\eta^{\alpha}}_{\dot{\alpha}}\tilde{j}_{\dot{\beta}\alpha}\right).
\]
The conditions for unbroken supersymmetry generated by a spinor $\eta_{\alpha\dot{\alpha}}$
are
\[
\eta_{\alpha\dot{\alpha}}\gamma_{\mu}\dot{\ell}^{\mu}=-\left|\dot{\ell}\right|{C_{\alpha}}^{\beta}{\eta_{\beta\dot{\alpha}}}=-\left|\dot{\ell}\right|{\tilde{C}_{\dot{\alpha}}}^{\dot{\beta}}\eta_{\alpha\dot{\beta}}.
\]

Let $\ell^{\mu}$ be the unit circle in the $x_{1},x_{2}$ plane with
angular coordinate $\tau$, and let $\eta_{\alpha\dot{\alpha}}$ represent
either of the supercharges $Q_{2,3}$ at $\theta_{0}=0$. One can show that the following
matrices solve the supersymmetry equations
\[
C^{\alpha\beta}=\begin{pmatrix}-\mathbbm{i}\sqrt{1-\nu^{2}}e^{\mathbbm{i}\tau} & \nu\\
\nu & -\mathbbm{i}\sqrt{1-\nu^{2}}e^{-\mathbbm{i}\tau}
\end{pmatrix},\qquad\tilde{C}^{\dot{\alpha}\dot{\beta}}=\begin{pmatrix}0 & -1\\
-1 & 0
\end{pmatrix}\, ,
\]
making the associated loop operator a BPS latitude loop. Latitude loops at generic $\theta_{0}$ can be obtained by a conformal transformation. 

\subsubsection{\label{subsubsection:offshell_closure_for_GW_theories}Off-shell closure for GW theories}
 Off-shell closure for GW theories can be achieved in a straightforwards fashion for any given $\mathcal{N}=2$ subalgebra.
It can also be achieved for an $\mathcal{N}=3$ subalgebra by using off-shell $\mathcal{N}=3$
multiplets and the corresponding Chern-Simons and kinetic terms \cite{Kao:1994zj,Kao:1993gs}. In the case of ABJM, the relevant couplings can also be derived in
harmonic superspace \cite{Buchbinder:2008vi}. Additionally, supersymmetries
preserved by the various topological twists of the $\mathcal{N}=4$
superalgebra can be closed off-shell by introducing appropriate auxiliary
fields \cite{Koh:2009um}. 
 
An $\mathcal{N}=4$ superfield construction
of GW theories was described in \cite{Kuzenko:2015lfa}. The construction requires a consistency condition, involving both the
vector multiplets and the hypermultiplets, which appears to prevent the
construction of an $\mathcal{N}=4$ off-shell supersymmetric action
\cite{Kuzenko:2015lfa}. The consistency condition is equivalent to
the fundamental identity in Eq \ref{eq:fundamental_identity}. There is then no way to close the entire $\mathcal{N}=4$ supersymmetry algebra of a GW type theory offshell using superfields. Unfortunately, we do not even know of a construction
for closing a \textit{single} generic $\mathcal{N}=4$ supersymmetry off-shell in
the GW class of theories. Specifically, the latitude supercharges do not fall into any of the categories admitting off-shell closure described above.

Let us describe the situation for the latitude supercharges in more detail. Supersymmetry transformations of off-shell hypermultiplets
are linear in the hypermultiplet fields. Fields belonging to vector
multiplets, as well as those coming from supergravity, may also appear
in the hypermultiplet transformations following partial gauge fixing.
We will show that the square of the latitude supercharge contains
a gauge transformation with a parameter which is quadratic in the
hypermultiplet scalars. In order for the latitude supersymmetry to
close off-shell, this parameter must presumably be the vev of a scalar
in an appropriate vector multiplet. Such a scalar appears in vector
multiplets starting from $\mathcal{N}=2$,
with multiple scalars appearing starting at $\mathcal{N}=3$. We will
also show that the square of the latitude supercharge on the three
sphere contains a translation by a Killing vector $v$. It is easy
to show, however, that $v$ cannot be generated by the square of an
$\mathcal{N}=2$ supercharge unless $\nu=1$. We are left, therefore, with the option of trying to close the latitude supercharges off-shell in an ad hoc manor. We explore this option in Section \ref{subsec:Localization-in-the-ABJM-model}. 
\section{\label{sec:The-latitude-matrix-model}The latitude matrix model}

In this section, we attempt to derive the matrix model for the bosonic
latitude Wilson loop by supersymmetric localization. In supersymmetric localization,
one first deforms a supersymmetric Euclidean action $S$ by adding a term
$t\delta V$, where $\delta$ is a supersymmetry transformation, $V$
a positive semi-definite fermionic functional, and $t$ a positive number. The deformed partition
function, or the expectation value of a $\delta$-closed observable
$\mathcal{O}$, can be shown to be $t$-independent using the following
standard argument
\begin{align*}
\frac{d}{dt}\left\langle \mathcal{O}\right\rangle  & _{t}\equiv\frac{d}{dt}\int\mathcal{D}\Phi\mathcal{O}\exp\left(-S-t\delta V\right)\\
 & =-\int\mathcal{D}\Phi\left(\delta V\right)\mathcal{O}\exp\left(-S-t\delta V\right)\\
 & =-\int\mathcal{D}\Phi\delta\left[V\mathcal{O}\exp\left(-S-t\delta V\right)\right]\\
 & =0.
\end{align*}
Localization onto the moduli space, defined as the space of field configurations where $\delta V$ vanishes, occurs in the limit $t\rightarrow\infty$.

In verifying $t$-independence, we have assumed $\delta S=\delta\mathcal{O}=\delta^{2}V=0$.
We have also assumed that the convergence properties of the path integral
do not depend on $t$, and that the measure $\mathcal{D}\Phi$, where
$\Phi$ stands for all dynamical fields, is $\delta$-invariant, i.e.
the supersymmetry is not anomalous. If the supersymmetry generated
by $\delta$ is closed off-shell, without using the equations of motion
coming from $S$, then $\delta^{2}$ is a bosonic symmetry transformation. In this case, 
one can usually arrange for $V$ to be $\delta^{2}$ invariant. Unfortunately, off-shell closure of even a single supersymmetry transformation can
be tricky. We will deal with this difficulty in the context of the ABJM model in an ad hoc manner.

In Section \ref{subsec:ABJM_sphere_algebra}, we exhibit the superalgebra generated by
the latitude supercharges on the three sphere using the ABJM notation. In Section
\ref{subsec:Localization-in-the-ABJM-model}, we perform localization in the ABJM model and derive a matrix model for the expectation value of the bosonic latitude Wilson loop. This calculation almost proves the conjectured form of the matrix model put forth in \cite{Bianchi:2018bke}. The only caveat is our assumption that a certain procedure can be used to close the latitude supersymmetry transformation off-shell. 
\subsection{\label{subsec:ABJM_sphere_algebra}The ABJM latitude loop on the three sphere}

We exhibit the superalgebra generated by the bosonic latitude supercharges on the three sphere. The ABJM model, and the bosonic latitude Wilson loop, can be mapped to the three sphere using the
change of coordinates in Appendix \ref{sec:Change-of-coordinates}. In order
to avoid producing a conformal or Weyl transformation on $\mathbb{S}^{3}$
in the square of the latitude supercharges, we restrict ourselves to $\theta_{0}=0$.
In the coordinate system introduced in Appendix \ref{sec:Change-of-coordinates}, the latitude loop then sits on the great circle at $\theta=0$ parameterized
by the value of $\tau$ which itself coincides with the affine
parameter used in the flat space notation. The latitude loop operator is defined by
the same expression as in Eq \ref{eq:bosonic} but with the opposite orientation.

We define a latitude supercharge $Q_{\text{L}}\equiv Q_{2}+Q_{3}$. The corresponding
$\mathcal{N}=6$ tensor, in the flat space conformal Killing spinor
basis specified in Appendix \ref{sec:Flat-space-spinor}, is 
\begin{gather*}
\bar{\Theta}_{1}^{\mathbb{R}^{3}}=\left(\begin{array}{cccc}
0 & 0 & \sqrt{\nu+1} & \sqrt{1-\nu}\\
0 & 0 & 0 & 0\\
-\sqrt{\nu+1} & 0 & 0 & 0\\
-\sqrt{1-\nu} & 0 & 0 & 0
\end{array}\right),\qquad\bar{\Theta}_{2}^{\mathbb{R}^{3}}=\left(\begin{array}{cccc}
0 & 0 & \sqrt{\nu+1} & \sqrt{1-\nu}\\
0 & 0 & 0 & 0\\
-\sqrt{\nu+1} & 0 & 0 & 0\\
-\sqrt{1-\nu} & 0 & 0 & 0
\end{array}\right),\\
\bar{\Theta}_{3}^{\mathbb{R}^{3}}=\left(\begin{array}{cccc}
0 & 0 & -\mathbbm{i}\sqrt{\nu+1} & \mathbbm{i}\sqrt{1-\nu}\\
0 & 0 & 0 & 0\\
\mathbbm{i}\sqrt{\nu+1} & 0 & 0 & 0\\
-\mathbbm{i}\sqrt{1-\nu} & 0 & 0 & 0
\end{array}\right),\qquad\bar{\Theta}_{4}^{\mathbb{R}^{3}}=\left(\begin{array}{cccc}
0 & 0 & 0 & 0\\
0 & 0 & \sqrt{1-\nu} & -\sqrt{\nu+1}\\
0 & -\sqrt{1-\nu} & 0 & 0\\
0 & \sqrt{\nu+1} & 0 & 0
\end{array}\right).
\end{gather*}
The change of coordinates described in Appendix \ref{sec:Change-of-coordinates}
transforms this tensor such that the relevant supercharge
on $\mathbb{S}^{3}$ is given by the $\mathbb{S}^{3}$ tensor
\[
\bar{\Theta}_{i}^{IJ}\equiv{R_{i}}^{j}\bar{\Theta}_{j}^{\mathbb{R}^{3},IJ}\, .
\]
Let $\epsilon^{\left(i\right)}$ be the basis for the conformal Killing
spinors defined in Appendix \ref{sec:Geometry-of-S3}. The corresponding
spinor used in the transformation of a field is $\bar{\Theta}_{i}^{IJ}\epsilon^{\left(i\right)}$.
We define the latitude supercharge to act using
the rescaled spinor 
\[\label{eq:ABJM_rescaled_spinor}
\bar{\Theta}_{\text{lat}}^{IJ}\equiv\frac{1}{4\nu^{1/4}}\bar{\Theta}_{i}^{IJ}\epsilon^{\left(i\right)}.
\]

The action of the latitude supercharges generate a subalgebra of the
$\mathbb{S}^{3}$ superconformal algebra. Specifically, the square
of the latitude supercharge generates the following transformations
\begin{enumerate}
\item A diffeomorphism by a Killing vector 
\begin{align*}
\frac{\mathbbm{i}}{2}\bar{\Theta}_{\text{lat}}^{IJ}\gamma^{\mu}\bar{\Theta}_{\text{lat}}^{KL}\varepsilon_{IJKL} & \partial_{\mu}=-\sqrt{\nu}\partial_{\varphi}+\frac{1}{\sqrt{\nu}}\partial_{\tau}.
\end{align*}
\item An R-symmetry transformation acting on $SU\left(4\right)_{R}$ indices
by the matrix 
\[
{\mathcal{R}_{I}}^{J}=2\mathbbm{i}\bar{\Theta}_{\text{lat}}^{KL}\bar{\Theta}_{\text{lat}}^{MJ}\varepsilon_{KLIM}=\frac{\mathbbm{i}}{2}\begin{pmatrix}-\nu^{-1/2} & 0 & 0 & 0\\
0 & \nu^{-1/2} & 0 & 0\\
0 & 0 & -\nu^{1/2} & 0\\
0 & 0 & 0 & \nu^{1/2}
\end{pmatrix}.
\]
\item No Weyl transformation. Note that this is true only when the parameter
$\theta_{0}$ is set to $0$.
\item A gauge transformation given by the gauge parameters

\[
\label{eq:gauge_transformation}
\Phi_1=i_v A-\frac{2\pi\mathbbm{i}}{k\sqrt{\nu}}{\tilde{M}_J}^{\ I}  C_I \bar{C}^J,\quad \Phi_2=-i_v \hat{A}+\frac{2\pi\mathbbm{i}}{k\sqrt{\nu}}{\tilde{M}_J}^{\ I} \bar{C}^J C_I\, ,
\]
where
\begin{eqnarray} 
 \mbox{\small $\! \tilde{M}_{J}^{\ I}=\left(\!\!
\begin{array}{cccc}
 - \nu  & e^{-\mathbbm{i} \tau }\cos\theta \sqrt{1-\nu ^2} & 0 & 0 \\
e^{\mathbbm{i} \tau }\cos\theta  \sqrt{1-\nu ^2}  & \nu  & 0 & 0 \\
 0 & 0 & -1 & e^{-\mathbbm{i} \varphi }\sin\theta \sqrt{1-\nu ^2} \\
 0 & 0 &  -e^{\mathbbm{i} \varphi }\sin\theta \sqrt{1-\nu ^2} & 1 \\
\end{array}
\right)$\, . }
\end{eqnarray}

Note that $\tilde{M}=M$ at $\theta=0$, i.e. on the latitude loop worldvolume. 

\end{enumerate}

One may check that a linear combination of $Q_2$ and $Q_3$ is the supercharge responsible for the cohomological equivalence of the bosonic and the fermionic versions of the latitude Wilson loop, discovered in \cite{Bianchi:2014laa}. The existence of these supercharges on the three sphere, and the absence of a Weyl or conformal transformation in their anti-commutator, implies that the cohomological equivalence holds at the quantum level in this context. This is because, as we will see, there exists a $Q_{2,3}$ exact non-conformal localizing term which can be used to reduce the computation of both operators to a computation in a free theory. 

\subsection{\label{subsec:Localization-in-the-ABJM-model}Localization in the
ABJM model}

In this section, we perform localization of the bosonic latitude Wilson loop in ABJM. We first describe a procedure which we believe may be utilized
in order to obtain off-shell closure and to localize theories of Gaiotto-Witten
type, including the ABJM model. This procedure is only necessary in
the presence of operators like the latitude loop, which do not preserve
supercharges belonging to an $\mathcal{N}\le3$ subalgebra. We will
show that, assuming the procedure works, we are lead to the matrix model for the latitude Wilson loop conjectured in \cite{Bianchi:2018bke}.

\subsubsection{Off-shell closure}

Given the constraints on off-shell closure in GW type theories reviewed in \ref{subsubsection:offshell_closure_for_GW_theories}, we are left with the task of closing
the latitude supersymmetry off-shell in an ad hoc manner. By this we mean utilizing auxiliary fields which do not extend to a full spacetime supersymmetry algebra. Our first task is to close off-shell the transformation of the connection using a vector multiplet. An appropriate multiplet is discussed by K\"{a}ll\'{e}n in \cite{Kallen:2011ny}. This is a sort of cohomological multiplet,
of the type often employed in topological field theory. 

The cohomological vector multiplet contains a connection $A_{\mu}$, a one form $\Psi_{\mu}$,
and a scalar $\Phi$. The cohomological supersymmetry transformation
is 
\[
\delta A=\Psi,\quad\delta\Psi=\mathcal{L}_{v}A+d_{A}\Phi,\quad\delta\Phi=0\, .
\]
Other multiplets are needed in order to construct actions. Specifically, we introduce a projection multiplet
with a fermion $\alpha$ and a scalar $\tilde{D}$
\[
\delta\alpha=\tilde{D},\quad\delta\tilde{D}=\mathcal{L}_{v}\alpha+G_{\Phi}\alpha\, .
\]
A field $X$ in a cohomological multiplet satisfies
\[
\delta^{2}X=\mathcal{L}_{v}X+G_{\Phi}X,
\]
where $G_{\Phi}$ is a gauge transformation with
parameter $\Phi$, and $\mathcal{L}_{v}$ is the Lie derivative with parameter $v$. 

One can write down a supersymmetric Chern-Simons term for the cohomological multiplet \cite{Kallen:2011ny}
 
\[\label{eq:cohomological_CS_term}
S_{\text{CS}}=\frac{\mathbbm{i}k}{4\pi}\int_{\mathbb{S}^{3}}\left(\text{CS}\left(A\right)-\kappa\wedge\Psi\wedge\Psi-2d\kappa\wedge\Psi\alpha+\left(\Phi+i_{v}A\right)\left(\kappa\wedge d\kappa\left(\Phi+i_{v}A-2\tilde{D}\right)-2\kappa\wedge F\right)\right),
\]
where $CS\left(A\right)$ is the usual Chern-Simons density for $A$. The field $\Phi$ transforms inhomogeneously under gauge transformations,
thus ensuring the gauge invariance of $\exp\left(-S_{\text{CS}}\right)$.

K\"{a}ll\'{e}n has shown that these multiplets can be derived by
twisting the fields in an $\mathcal{N}=2$ vector multiplet $\left\{ A_{\mu},\sigma,D\left|\lambda_{\alpha},\tilde{\lambda}_{\alpha}\right)\right\} $,
such that, for instance,
\[
\Phi=\mathbbm{i}\sigma+v^{\mu}A_{\mu}.
\]
However, the validity of the multiplets, and the construction of the Chern-Simons action, is
more general. In fact, as argued in \cite{Kallen:2011ny}, all that
is required is for $v$ to be a Killing vector, and that there exist
a contact form $\kappa$, i.e. a one form such that $\kappa\wedge d\kappa$
is a volume form, for which $v$ is the Reeb vector
\[
i_{v}\kappa=1,\qquad i_{v}d\kappa=0.
\]
In our case, we identify $v$ with the parameter for the diffeomorphism in the square of the latitude supercharge. An appropriate $\kappa$ can be obtained by lowering
the index on $v$ and replacing $\nu\rightarrow\nu^{-1}$. This requires
$\nu\ne0$, and agrees with K\"{a}ll\'{e}n's $\mathcal{N}=2$ twist
when $\nu$ is equal to $1$. 

K\"{a}ll\'{e}n's construction has been extended to $\mathcal{N}=2$
Chern-Simons-matter theories in \cite{Ohta:2012ev}. The authors of
\cite{Ohta:2012ev} also exhibit an invariant action based on cohomological matter
and vector multiplets which is on-shell equivalent to the ABJM model. Unfortunately, both the action and the map between the ABJM fields
and the twisted fields in \cite{Ohta:2012ev} depend on details of
the particular contact form being utilized. This contact form is the one associated to moving along the fiber
of $\mathbb{S}^{3}$, viewed as a circle bundle over $\mathbb{S}^{2}$.
More generally, the authors of \cite{Ohta:2012ev} discuss the contact form associated with
the fiber of a general Seifert manifold, which is a fibration over
a Riemann surface. Our $v$ is not of this type since, for $\nu\ne1$,
it mixes the vector field along the fiber with one associated to an
action on the base. This is possible only because the base, in this
case $\mathbb{S}^{2}$, admits a continuous isometry.\footnote{For $\nu\in\mathbb{Q}_{+}$, $v$ still generates a compact isometry acting freely on $\mathbb{S}^3$.}  

In \cite{Ohta:2012ev}, the square of the cohomological transformation includes an additional
central generator whose eigenvalue is denote by $\Delta$
\[
\delta^{2}X=\mathcal{L}_{v}X+G_{\Phi}X+\Delta X\, .
\] 
In principle then, there is enough freedom in the construction of \cite{Ohta:2012ev} to accommodate the latitude superalgebra given in Section \ref{subsec:ABJM_sphere_algebra}. However, one would have to verify that an action exists, of the type given in \cite{Ohta:2012ev}, which is supersymmetric for the values of $v$ and $\Delta$ implied by the latitude superalgebra, and which is still on-shell
equivalent to ABJM. While we expect that this is possible, we have not shown it explicitly.

\subsubsection{The matrix model}

We will proceed under the assumption that the latitude supersymmetry
can be closed off-shell using the cohomological multiplets of \cite{Kallen:2011ny,Ohta:2012ev} described in the previous section.
The connection appearing in the cohomological multiplet will be identified
with the one appearing in the ABJM action. There are therefore two
independent adjoint valued scalars appearing in the square of the
supersymmetry transformation, one for each gauge group factor. According
to our off-shell closure conjecture, the corresponding cohomological
fields are identified with $\Phi_{1,2}$ in Eq \ref{eq:gauge_transformation}.

The cohomological supersymmetry transformations in \cite{Kallen:2011ny}
make it clear that the moduli space associated with the vector multiplet
is given by solutions to the following equations
\[
F_{\mu\nu}=0,\quad D_{\mu}\Phi=0,\quad\tilde{D}=0.
\]
On $\mathbb{S}^{3}$, the solutions are gauge equivalent to $A_{\mu}=0$
and $\Phi=\mathbbm{i}\sigma$ where $\sigma$ is an arbitrary spacetime
independent adjoint valued parameter \cite{Kapustin:2009kz}. In this
gauge multiplet background, the fermion variations in \cite{Ohta:2012ev}
imply that a chiral multiplet has no moduli whatsoever, as long as
the parameter $\Delta$ is non-zero.\footnote{In the $\mathcal{N}=2$ superalgebra, $\Delta$ corresponds to the
$U\left(1\right)_{R}$ charge of the chiral multiplet. Equivalently,
for a superconformal theory $\Delta$ is the conformal dimension of
the dynamical scalar in the multiplet.} While this conclusion regarding the chiral moduli depends, in principle,
on the contact form being used, we expect it to hold in general. Therefore,
the moduli space is given by constant profiles for some scalars $\sigma^{\left(1,2\right)}$
and the result is a matrix model. The classical contributions to the matrix models, coming from the Chern-Simons terms, can now be read off from Eq \ref{eq:cohomological_CS_term} and coincide with the ones in \ref{eq:matrixlat}. 

Localization also yields an effective action for the moduli coming
from one loop determinants. These determinants are straightforward to calculate in
the cohomological formalism using the equivariant index theorem for
transversely elliptic operators \cite{Pestun:2007rz}.\footnote{Mathematical background for the equivariant index theorem can be found
in \cite{atiyah2006elliptic,Pestun:2016qko}, while reviews of the
application to supersymmetric localization appear in e.g. \cite{Pestun:2016jze,Gomis:2011pf}.} We will follow the application of this theorem to the squashed three sphere partition
function appearing in \cite{Drukker:2012sr}. This is convenient because both the moduli space
and the bosonic symmetry obtained from the square of the latitude
supercharge coincide with those of the squashed sphere, provided we
identify $b=\sqrt{\nu}$. Since these are the only ingredients appearing
in the equivariant index theorem, apart from the implications coming from
the topology of the manifold which are also the same, the one loop
determinants coincide.  The cohomological vector and projection multiplets therefore yield the following determinants \cite{Hama:2011ea,Drukker:2012sr}
\[
\label{eq:one_loop_vector_GW}Z_{\text{vector}}\left(\sigma\right)=\prod_{\alpha>0}\sinh\left(\pi\sqrt{\nu}\alpha\left(\sigma\right)\right)\sinh\left(\pi\frac{\alpha\left(\sigma\right)}{\sqrt{\nu}}\right).
\]

We must now evaluate the effective action for the moduli coming from matter fields. The values for the the $SU\left(4\right)_{R}$
symmetry transformations appearing the the square of the latitude
supercharge for the four scalars $C_{I}$
are $\pm\mathbbm{i}\nu^{\pm1/2}/2$. These should correspond to the parameters $\Delta$ for chiral multiplets in \cite{Ohta:2012ev}. However, as is clear from \cite{Benna:2008zy},
two of the $C_{I}$ are the lowest components of chiral multiplets,
while the other two are the lowest components of anti-chiral multiplets. In order to use directly the results in \cite{Drukker:2012sr} for the one loop determinant of a chiral multiplet, we
must therefore consider the quantum numbers of two $C_{I}$ and two $\bar{C}^{I}$. Of course, the $\bar{C}^I$ fields are also in the complex conjugate
gauge representation. The $\mathcal{N}=4$ version of the R-symmetry transformations \ref{eq:latitude-R-symmetries} makes it clear that $C_{1,2}$ and $C_{3,4}$ are scalar doublets inside an ordinary and a twisted hypermultiplet respectively. Moreover, we can identify $C_1$ and $C_3$ as having the correct R-symmetry charges to be the lowest components of chiral superfields in the limit $\nu\rightarrow 1$. That means that $\bar{C}_2$ and $\bar{C}_4$ are the the lowest components of the remaining chirals. Assuming that this holds also at $\nu\ne 1$, this identification assigns flavor symmetry charges $-\mathbbm{i}\nu^{\pm1/2}/2$, with multiplicity $2$, to the dynamical chiral fields. Note that there is no adjoint valued chiral multiplet in this calculation, as
it has been integrated out to produce the ABJM superpotential \cite{Ohta:2012ev}.

Let 
\[
Q=\nu^{1/2}+\nu^{-1/2}.
\]
The one loop determinant coming from the index theorem in \cite{Drukker:2012sr},
for the collection of fields with the quantum numbers discussed above,
is\footnote{$s_{b}$ is the double sine function defined in e.g. \cite{Hama:2011ea}.}
\begin{align*}
Z_{\text{scalar}}\left(\sigma\right) & =\prod_{\omega\in\left(\boxempty,\bar{\boxempty}\right)}s_{\sqrt{\nu}}\left(\mathbbm{i}\frac{Q}{2}\pm\omega\left(\sigma^{\left(1,2\right)}\right)-\mathbbm{i}\frac{\nu^{\pm1/2}}{2}\right)\\
 & =\prod_{\omega\in\left(\boxempty,\bar{\boxempty}\right)}\frac{1}{2\cosh\left(\pi\sqrt{\nu}\omega\left(\sigma^{\left(1,2\right)}\right)\right)2\cosh\left(\frac{\pi}{\sqrt{\nu}}\omega\left(\sigma^{\left(1,2\right)}\right)\right)}.
\end{align*}
where the notation is meant to imply that we multiply the $s_{b}$
functions for all values of $\pm$ corresponding to the different
scalars. In the second line, we have used the following special function
identities
\[
s_{b}\left(-x\right)=s_{b}\left(x\right)^{-1},\quad\frac{s_{b}\left(\frac{\mathbbm{i}b}{2}+x\right)}{s_{b}\left(x-\frac{\mathbbm{i}b}{2}\right)}=\frac{1}{2\cosh\left(\pi bx\right)}.
\]
The latter identity has recently played a role in the IR formula for
correlation functions of Higgs and Coulomb branch operators in $\mathcal{N}=4$
SCFTs proposed in \cite{Gaiotto:2019mmf}.\footnote{A derivation of the IR formula using holomorphic factorisation was provided in \cite{Bullimore:2020jdq}. We thank the authors for bringing this work to our attention.} We have already seen a
connection between the latitude loop and the SQM used in \cite{Dedushenko:2016jxl,Dedushenko:2017avn}
to compute the same correlators. However, that connection was valid
only in the limit $\nu\rightarrow0$.

We can easily generalize the calculation to the Gaiotto-Witten type theories discussed in Section \ref{subsec:Gaiotto-Witten-Theories}. To every gauge group factor in such a theory we associate a moduli space given by an adjoint valued scalar $\sigma$, and a one loop determinant given by Eq \ref{eq:one_loop_vector_GW}. A hypermultiplet in a representation $\mathfrak{R}$ contributes 
\[Z_{\text{hyper}}\left(\sigma\right)  =\prod_{\omega\in\mathfrak{R}}\frac{1}{2\cosh\left(\pi\sqrt{\nu}\omega\left(\sigma\right)\right)}\, ,\]
while a twisted hypermultiplet contributes
\[Z_{\text{twisted hyper}}\left(\sigma\right)  =\prod_{\omega\in\mathfrak{R}}\frac{1}{2\cosh\left(\frac{\pi}{\sqrt{\nu}}\omega\left(\sigma\right)\right)}\, .\]
This result coincides with the form for the matrix models for some Chern-Simons-matter theories of this type conjectured in \cite{Drukker:2020dvr}.

The unnormalized expectation value of the bosonic latitude loop corresponds simply to an insertion of 
$$\frac{1}{N_1}\sum_{\rho\in\mathfrak{R}}\exp\left(2\pi\sqrt{\nu}\rho\left(\sigma\right)\right)$$
in the matrix model. The various factors of $2,\pi,\sqrt{\nu}$ can be deduced by comparing $\Phi_{1,2}$ at $\theta=0$ with the original bilinear expressions appearing in the bosonic loop \ref{eq:bosonic}. The complete computation of the expectation value of the bosonic latitude loop in ABJM therefore coincides with Eq \ref{eq:matrixlat} and the conjecture put forth in \cite{Bianchi:2018bke}. We emphasize that this result holds assuming that the off-shell closure procedure goes through.

\section{\label{sec:Latitude-loop-in-standard-N4-theories}Latitude loops in standard $\mathcal{N}=4$ theories}
In this section we explore latitude type loops in \textit{standard} $\mathcal{N}=4$ theories: those without Chern-Simons terms. In Section \ref{subsec:N4_standard}, we define latitude supercharges in standard theories. We show
that the latitude supercharges are related, in this context, to the supercharges used to define the topological quantum
mechanics investigated in \cite{Dedushenko:2016jxl}. We then classify
generic loop operators preserving the latitude supercharge in this class of theories. In Section \ref{subsec:SQM}, we exhibit the latitude superalgebra as it acts on the supersymmetric quantum mechanics on a loop operator worldvolume. 

In Section \ref{subsec:Localization-in-a-dual-theory}, we consider an $\mathcal{N}=4$
gauge theory in the same universality class as ABJM. We
use the form of the latitude supersymmetry algebra to conjecturally
identify the loop operator representing the bosonic latitude
in this IR dual theory. We then perform localization in the dual model and derive a matrix model expression for the expectation value of the dual loop operator. The matrix models for
the dual theories looks somewhat different from the one derived in ABJM, but yields the same result
for the expectation value of the loop.  This constitutes our primary
evidence for the validity of the identification of the dual loop operator.
\subsection{\label{subsec:N4_standard}The bosonic latitude supercharges in standard $\mathcal{N}=4$ theories}

The latitude supercharges lie in an $\mathcal{N}=4$ subalgebra of
the $\mathcal{N}=6$ supersymmetry algebra of ABJM. In this section,
we study some properties of the bosonic latitude supercharges when these are realized on the three sphere in standard $\mathcal{N}=4$ theories. This is to be contrasted with the realization of these supercharges in the GW type theories of section \ref{subsec:Gaiotto-Witten-Theories} and in ABJM. We will use the $\mathcal{N}=4$
supersymmetry conventions of \cite{Dedushenko:2016jxl}. These are summarized in Appendix \ref{subsec:3d_N4_theories_on_S3}.

We begin by defining an $\mathcal{N}=4$ latitude
spinor for the latitude supercharge on $\mathbb{S}^{3}$, and the analogous spinors for
$Q_{2,3}$\footnote{Note the difference in normalization of the spinor, by a factor of $\nu^{1/4}$, from the ABJM rescaled spinor in Eq \ref{eq:ABJM_rescaled_spinor}}
\[
\left(\xi_{\nu}^{L}\right)_{\alpha a\dot{a}}\equiv\frac{\mathbbm{i}}{8}\bar{\Theta}_{i}^{IJ}\left(\nu\right)\Gamma_{IJ}^{p}{\left(\bar{\sigma}_{p}\right)_{a\dot{a}}}\epsilon_{\alpha}^{\left(i\right)}\,,\label{eq:latitude_embedding}
\]
where 
\[
\bar{\sigma}_{i}\equiv\mathbbm{i}\tau_{i}\,,\quad i\in\left\{ 1,2,3\right\}\,,\qquad\bar{\sigma}_{4}\equiv\mathbbm{1}_{2}\,.
\]
Recall that $\epsilon_{\alpha}^{\left(i\right)}$ is a basis for the conformal Killing spinors on $\mathbb{S}^3$, and $\Gamma_{IJ}^{p}$ are the 6d Euclidean gamma matrices. Exchanging the indices $a$ and $\dot{a}$ would produce a mirror latitude supercharge.

The spinor $\xi_{\nu}^{L}$ sits inside a linear space of spinors
whose action on the fields generates a Poincar\'{e} subalgebra of the $\mathfrak{osp}\left(4|4\right)$
superconformal algebra on $\mathbb{S}^{3}$, of the type discussed
in \cite{Dedushenko:2016jxl}. This Poincar\'{e}  subalgebra is characterized
by the absence of Weyl transformations, and of diffeomorphisms associated with conformal Killing
vectors, which generically appear in the anticommutator of conformal supercharges. As shown in \cite{Dedushenko:2016jxl}, a Poincar\'{e}  subalgebra of this type can be obtained by demanding that a Killing spinor $\xi$ satisfy
\[
\nabla_{\mu}\xi_{a\dot{a}}=\gamma_{\mu}\xi_{a\dot{a}}^{'}\,,\qquad\xi_{a\dot{a}}^{'}=\frac{\mathbbm{i}}{2r}{h_{a}}^{b}\xi_{b\dot{b}}{\bar{h}^{\dot{b}}}_{\dot{a}}\,,
\]
where 
\[
{h_{a}}^{b}\in\mathfrak{su}\left(2\right)_{H}\,,\quad{h_{\dot a}}^{\dot b}\in\mathfrak{su}\left(2\right)_{C}\,.
\]
For the latitude spinors, both $h$ and $\bar{h}$ can be taken to
be $\tau_{3}$. Note that the Killing spinors associated with $Q_{2,3}$ are separately inside the Poincar\'{e}  subalgebra
defined by $h,\bar{h}$, while those of $Q_{1,4}$ are not. 

The bosonic symmetries which appear in the square of the transformation using 
$\delta_{\xi_{\nu}^{L}}$ are
\begin{enumerate}
\item A diffeomorphism with Killing vector $v=-\nu\,\partial_{\phi}+\partial_{\tau}$\, ;
\item R-symmetry transformations $-\nu\,R_{C}$ and $R_{H}$, where $R_{C,H}$
act on doublets of $SU\left(2\right)_{C}$ and $SU\left(2\right)_{H}$
as matrices
\begin{equation}
{{R_{C}}_{\dot{a}}}^{\dot{b}}=\frac{\mathbbm{i}}{2}\begin{pmatrix}1 & 0\\
0 & -1
\end{pmatrix},\qquad{{R_{H}}_{a}}^{b}=\frac{\mathbbm{i}}{2}\begin{pmatrix}1 & 0\\
0 & -1
\end{pmatrix}\, ;\label{eq:latitude-R-symmetries}
\end{equation}
\item and a gauge transformation with parameter $\Lambda=\frac{1}{2}{{\left(\xi_{\nu}^{L}\right)}^{c\dot{a}}}{{\left(\xi_{\nu}^{L}\right)}_{c}}^{\dot{b}}\Phi_{\dot{a}\dot{b}}-v^{\mu}A_{\mu}$
with 
\[
{{\left(\xi_{\nu}^{L}\right)}^{c\dot{a}}}{{\left(\xi_{\nu}^{L}\right)}_{c}}^{\dot{b}}=\frac{1}{2}\begin{pmatrix}-e^{\mathbbm{i}\varphi}\sqrt{1-\nu^{2}}\sin\theta & -\mathbbm{i}\\
-\mathbbm{i} & e^{-\mathbbm{i}\varphi}\sqrt{1-\nu^{2}}\sin\theta
\end{pmatrix}\,.
\]
\end{enumerate}

\subsubsection{Relationship to topological quantum mechanics}

Superconformal theories with 16 supercharges, in $3$ and $4$ dimensions,
admit special local operator algebras whose correlators enjoy enhanced
spacetime symmetry \cite{Beem:2013sza,Beem:2016cbd}. In \cite{Dedushenko:2016jxl}, the authors define such a set of ``Higgs branch operators'' in
any standard 3d $\mathcal{N}=4$ gauge theory. The Higgs
branch operators are non-trivial elements of the cohomology of a supercharge
$\mathcal{Q}_{\beta}^{H}$, which is itself a combination of Poincar\'{e}
 and conformal supercharges. When placed along a line, correlators
of Higgs branch operators are position independent, though they may
still depend on operator ordering.

The authors of \cite{Dedushenko:2016jxl} go on to define an analogous
cohomology on $\mathbb{S}^{3}$, where $\mathcal{Q}_{\beta}^{H}$
is part of a specific Poincar\'{e}  subalgebra of the full $\mathcal{N}=4$
superconformal algebra. This fact allows them to deform the
$\mathcal{N}=4$ gauge theory by an appropriate $\mathcal{Q}_{\beta}^{H}$-exact
Yang-Mills term, and to perform a localization computation which captures
the expectation values of the operators. In order to preserve $\mathcal{Q}_{\beta}^{H}$,
Higgs branch operators must be placed along a great circle in $\mathbb{S}^{3}$.
The result of the localization computation can be interpreted as a
one dimensional topological field theory living on the circle, i.e.
a topological quantum mechanics. 

In the notation of \cite{Dedushenko:2016jxl}, the supercharge
\[
\mathcal{Q}_{\beta}^{H}\equiv\mathcal{Q}_{1}^{H}+\beta\mathcal{Q}_{2}^{H}\,,
\]
on $\mathbb{S}^{3}$ has the following properties
\[
\left(\mathcal{Q}_{1}^{H}\right)^{2}=\left(\mathcal{Q}_{2}^{H}\right)^{2}=0\,,\qquad\left(\mathcal{Q}_{\beta}^{H}\right)^{2}=4\mathbbm{i}\beta\left(P_{\tau}+R_{C}\right)\,,
\]
where $P_{\tau}$ is a translation along the $\tau$ direction and
$R_{C}$ is an $R$ symmetry transformation inside $SU\left(2\right)_{C}\subset SO\left(4\right)_{R}$.

$\mathcal{Q}_{\beta}^{H}$ is represented on $\mathbb{S}^{3}$ by
a spinor $\xi_{\beta a\dot{a}}^{H}$ which, after re-scaling, can
be written in our notation as
\begin{gather*}
\xi_{\beta11}^{H}=-\frac{1}{\sqrt{2}}{R_{i}}^{4}e^{\Omega/2}\epsilon_{\mathbb{R}^{3}}^{\left(i\right)}\,,\\
\xi_{\beta12}^{H}=\frac{\beta}{\sqrt{2}}{R_{i}}^{3}e^{\Omega/2}\epsilon_{\mathbb{R}^{3}}^{\left(i\right)}\,,\\
\xi_{\beta21}^{H}=-\frac{1}{\sqrt{2}}{R_{i}}^{2}e^{\Omega/2}\epsilon_{\mathbb{R}^{3}}^{\left(i\right)}\,,\\
\xi_{\beta22}^{H}=-\frac{\beta}{\sqrt{2}}{R_{i}}^{1}e^{\Omega/2}\epsilon_{\mathbb{R}^{3}}^{\left(i\right)}\,.
\end{gather*}

The mirror spinor $\xi_{\beta a\dot{a}}^{C}\equiv\xi_{\beta\dot{a}a}^{H}$
generates the mirror supercharge $\mathcal{Q}^{C}$ used to define
the Coulomb branch version of the cohomology in \cite{Dedushenko:2017avn,Dedushenko:2018icp}.
The square of the transformation $\delta_{\xi_{\beta}^{C}}$ gives
\begin{enumerate}
\item A translation with Killing vector $v_{C}=\partial_{\tau}$\,;
\item An R-symmetry transformation acting on doublets of $SU\left(2\right)_{H}$
as the matrix
\[
{{R_{H}}_{a}}^{b}=\frac{\mathbbm{i}\beta}{2}\begin{pmatrix}-1 & 0\\
0 & 1
\end{pmatrix}\,,
\]
with $R_{C}$ vanishing;
\item and a gauge transformation with parameter $\Lambda_{C}=\frac{1}{2}{{\left(\xi_{\beta}^{C}\right)}^{c\dot{a}}}{{\left(\xi_{\beta}^{C}\right)}_{c}}^{\dot{b}}\Phi_{\dot{a}\dot{b}}-v_{C}^{\mu}A_{\mu}$\,.
\end{enumerate}
In fact, we can show that $\mathcal{Q}^{C}$ is related to the latitude
supercharge in the limit $\nu\rightarrow0$ and $\beta\rightarrow1$.\footnote{We chose an embedding of the latitude in the UV $\mathcal{N}=4$ algebra
which relates the latitude supercharge to the mirror supercharge $\mathcal{Q}^{C}$.
Taking the mirror embedding would yield a supercharge related to $\mathcal{Q}^{H}$.} We do this by exhibiting a \emph{global} $SO\left(4\right)_{R}$
rotation, defined by matrices
\[
{\left(W_{1}\right)_{a}}^{b}\in SU\left(2\right)_{l}\,,\quad{\left(W_{2}\right)_{\dot{a}}}^{\dot{b}}\in SU\left(2\right)_{r}\,,
\]
such that 
\[
\left(\xi_{\nu}^{L}\right)_{\alpha a\dot{a}}={\left(W_{1}\right)_{a}}^{b}{\left(W_{2}\right)_{\dot{a}}}^{\dot{b}}\left(\xi_{\beta}^{C}\right)_{\alpha b\dot{b}}\,.
\]
The explicit matrices are given by 
\[
W_{1}=\begin{pmatrix}\mathbbm{i} & 0\\
0 & -\mathbbm{i}
\end{pmatrix}\,,\quad W_{2}=\frac{1}{\sqrt{2}}\begin{pmatrix}1 & -\mathbbm{i}\\
-\mathbbm{i} & 1
\end{pmatrix}\,.
\]
Hence, the latitude supercharge at generic $\nu$, or its mirror dual,
\textit{interpolate} between the ordinary $\mathcal{N}=2$ supercharges used to compute
the partition functions in \cite{Kapustin:2009kz} and the supercharges
used to define the Higgs and Coulomb branch cohomologies in \cite{Dedushenko:2017avn,Dedushenko:2018icp,Dedushenko:2016jxl}. 

\subsubsection{\label{subsec:Latitude-loops-in-generic-N4-theories}Latitude loops
in standard $\mathcal{N}=4$ theories}

In this section, we exhibit supersymmetric loops of bosonic latitude
type in standard $\mathcal{N}=4$ gauge theories. 3d $\mathcal{N}=4$
gauge theories of this kind admit more than one type of supersymmetric
loop observable. Some of these observables are of Wilson loop type, while
others are defect operators, which are sometimes called vortex loops \cite{Kapustin:2012iw,Drukker:2012sr}.\footnote{Vortex loops appeared early on in the context of bosonic Chern-Simons
theory \cite{Witten:1988hf,Moore:1989yh}. Supersymmetric surface
operators, which are the four dimensional analogue of vortex loops,
were considered in \cite{Gukov:2006jk,Gukov:2008sn} (see \cite{Gukov:2014gja}
for a review).} Both types of loop observable can be tuned to preserve the bosonic latitude
supercharges. We will define prototypical operators of both types, and tabulate
the supercharges they preserve.

The worldvolume of a loop operator on $\mathbb{S}^{3}$, invariant
under the latitude supercharge, must itself be invariant under the
translation present in the square of the supercharge. Let $\ell^{\mu}\left(\tau_{0}\right)$
be coordinates on the loop with affine parameter $\tau_{0}$. $\ell^{\mu}$
is the orbit under the action of the vector field $\dot{\ell}$
which must coincide with $-\nu\,\partial_{\phi}+\partial_{\tau}$
restricted to some $\theta$. Whether or not $\ell$ is a closed curve
depends on the value of $\theta$ where the loop is placed, as well as 
$\nu$. The types of orbits are detailed in Table \ref{table:loop_worldvolume}.
\begin{table}
\begin{centering}
\label{table:loop_worldvolume}%
\begin{tabular}{|c|c|c|c|}
\hline 
orbit $\dot{\ell}$ & $\theta=0$ & $\theta=\pi/2$ & $\theta$ generic\tabularnewline
\hline 
\hline 
$\nu=\pm1$ & closed + maximal & closed + maximal & closed + maximal\tabularnewline
\hline 
$\nu=0$ & closed + maximal & point & closed\tabularnewline
\hline 
$\nu$ generic & closed + maximal & closed + maximal & non-compact\tabularnewline
\hline 
\end{tabular}
\par\end{centering}
\caption{The type of curve represented by $\ell$ as a function of the fixed
angle $\theta$. A ``maximal'' closed curve is a great circle on
$\mathbb{S}^{3}$.}
\end{table}
In addition, there may be closed orbits at special values of $\nu$
and $\cos\theta$.

\paragraph{Wilson loops}

In a standard $\mathcal{N}=4$ gauge theory, a supersymmetric
Wilson loop can be constructed using only vector multiplet fields.
For an ordinary $\mathcal{N}=4$ vector multiplet, the relevant fields
are the connection $A_{\mu}$ and the $SU\left(2\right)_{C}$ triplet
of scalars $\Phi_{\dot{a}\dot{b}}$. We take the following ansatz
for a supersymmetric Wilson loop along a contour with coordinates
$\ell^{\mu}$
\[
\mathcal{W}=\frac{1}{\text{dim}\left(\mathfrak{R}\right)}\text{tr}_{\mathfrak{R}}\mathcal{P}\exp\ointctrclockwise_{\ell}\left(\dot{\ell}^{\mu}A_{\mu}+\left|\dot{\ell}\right|M^{\dot{a}\dot{b}}\Phi_{\dot{a}\dot{b}}\right)\,,
\]
where $M^{\dot{a}\dot{b}}$ is a coordinate dependent, but field independent,
symmetric matrix. One can show that the most general such loop preserving
the latitude supercharge runs along $\dot{\ell}^{\mu}\partial_{\mu}=-\nu\partial_{\varphi}+\partial_{\tau}$ with fixed $\theta$, and has
\begin{equation}
M^{\dot{a}\dot{b}}=M_{\text{Wilson}}^{\dot{a}\dot{b}}\equiv\left|\dot{\ell}\right|^{-1}\begin{pmatrix}\frac{1}{2}e^{\mathbbm{i}\varphi}\sqrt{1-\nu^{2}}\sin\theta & \frac{\mathbbm{i}}{2}\\
\frac{\mathbbm{i}}{2} & -\frac{1}{2}e^{-\mathbbm{i}\varphi}\sqrt{1-\nu^{2}}\sin\theta
\end{pmatrix}\,,\label{eq:Wilson_loop_matrix}
\end{equation}
which is a familiar expression. Note that at $\theta=0$ the Wilson loop is $\nu$-independent. 

The total number of supercharges preserved by $\mathcal{W}$ are listed in Table \ref{table:Wilson_loop_supercharges}.
A subset of these are within the Poincar\'{e}  subalgebra containing the latitude
supercharge. Their number and type depend on the value of $\theta$ where the loop is placed
and on $\nu$. 
\begin{center}
\begin{table}
\begin{centering}
\begin{tabular}{|c|c|c|c|c|c|c|}
\hline 
 & \multicolumn{2}{c|}{$\theta=0$} & \multicolumn{2}{c|}{$\theta=\pi/2$} & \multicolumn{2}{c|}{$\theta$ generic}\tabularnewline
\hline 
supercharges preserved: & total & Poincar\'{e} & total & Poincar\'{e} & total & Poincar\'{e}\tabularnewline
\hline 
\hline 
$\nu=1$ & $8$ & $Q_{2,3}+2$ & $8$ & $Q_{2,3}+2$ & $4$ & $Q_{2,3}$\tabularnewline
\hline 
$\nu=0$ & $8$ & $Q_{2,3}+2$ & N/A & N/A & $4$ & $2$\tabularnewline
\hline 
$\nu$ generic & $8$ & $Q_{2,3}+2$ & $4$ & $2$ & N/A & N/A\tabularnewline
\hline 
\end{tabular}
\par\end{centering}
\caption{\label{table:Wilson_loop_supercharges}Supercharges preserved by a
generic $\mathcal{N}=4$ latitude Wilson loop: total number preserved
and number preserved inside the Poincar\'{e} subalgebra defined by
the latitude supercharge, with $Q_{2,3}$ singled out.}
\end{table}
\par\end{center}

\paragraph{Vortex loops}

The supersymmetric vortex loops we consider are defect operators associated
to singular classical BPS configurations embedded in a vector multiplet,
of the type described in e.g. \cite{Kapustin:2012iw,Drukker:2012sr}.
In these configurations, the field strength $F$ is taken to be proportional to a delta function
supported on the loop worldvolume, and an imaginary auxiliary field
is turned on to preserve supersymmetry. The effects of the connection associated with $F$ can be felt by charged local operators away from the loop worldvolume, while the profile for the auxiliary field cannot. Nevertheless, this profile should be considered part of the configuration describing the defect for some applications \cite{Kapustin:2012iw}.  

A BPS defect embedded in a vector multiplet is a singular fixed point of the gaugino transformations. A typical $1/2$ BPS abelian vortex loop defect of charge $q$ in an $\mathcal{N}=2$
gauge theory solves the BPS equation for a vector multiplet in the
following way \cite{Kapustin:2012iw}
\[
F=2\pi q\delta_{\ell},\quad D=-2\pi q\mathbbm{i}\star\left(\delta_{\ell}\wedge d\ell\right)\,,
\]
where $d\ell$ is the one form dual to $\dot{\ell}$, and $\delta_{\ell}$
is the Poincar\'{e}  dual to the loop worldvolume. To get non-abelian
vortex loops, one replaces the number $q$ with an element of the
Lie algebra $\mathfrak{g}$.\footnote{The number $q$, or the Lie algebra element generalizing it, are subject
to discrete identifications arising from large gauge transformations
\cite{Gukov:2006jk}.} Supersymmetry requires that the loop worldvolume be the integral
curve of the vector field one gets from squaring the supercharge.
In the case of $\mathcal{N}=2$ supersymmetry, this worldvolume is
always a maximal circle on $\mathbb{S}^{3}$.

The $\mathcal{N}=4$ version of the supersymmetric vortex loop is
entirely analogous. We set $F$ as above, and take the triplet of
auxiliary fields in the $\mathcal{N}=4$ vector multiplet to be 
\begin{gather}
D^{ab}=-2\pi q\star\left(\delta_{\ell}\wedge d\ell\right)M_{\text{vortex}}^{ab}\,,\label{eq:vortex_loop_matrix}\\
M_{\text{vortex}}^{ab}\equiv\begin{pmatrix}-\frac{\mathbbm{i}}{2}e^{\mathbbm{i}\tau}\sqrt{1-\nu^{2}}\cos\theta & \frac{\nu}{2}\\
\frac{\nu}{2} & -\frac{\mathbbm{i}}{2}e^{-\mathbbm{i}\tau}\sqrt{1-\nu^{2}}\cos\theta
\end{pmatrix}\,. \label{eq:M_vortex}
\end{gather}
The resulting singular background solves the BPS equation from \cite{Dedushenko:2016jxl},
i.e. the variation of the gaugino, specialized to the latitude supercharge
\[
\delta_{\xi_{\nu}^{L}}\lambda_{a\dot{b}}=-\frac{\mathbbm{i}}{2}\varepsilon^{\mu\mu'\rho}\gamma_{\rho}\xi_{\nu a\dot{b}}^{L}F_{\mu\mu'}-{D_{a}}^{c}\xi_{\nu c\dot{b}}^{L}=0\,,\quad\forall a,\dot{b}\,.
\]
All of the vortex loops defined by the above configuration preserve
$Q_{2,3}$ separately, as well as the latitude supercharge. Additional
supercharges are given in Table \ref{table:vortex_loop_supercharges}. 

In a twisted $\mathcal{N}=4$ vector multiplet, the roles of the dotted
and undotted indices, and the matrices $M_{\text{Wilson}}$ and $M_{\text{vortex}}$,
are exchanged. The relationship between the Wilson and vortex loops
is basically mirror symmetry \cite{Kapustin:2012iw,Assel:2015oxa}.

\begin{center}
\begin{table}
\begin{centering}
\begin{tabular}{|c|c|c|c|c|c|c|}
\hline 
 & \multicolumn{2}{c|}{$\theta=0$} & \multicolumn{2}{c|}{$\theta=\pi/2$} & \multicolumn{2}{c|}{$\theta$ generic}\tabularnewline
\hline 
supercharges preserved: & total & Poincar\'{e}  & total & Poincar\'{e}  & total & Poincar\'{e} \tabularnewline
\hline 
\hline 
$\nu=1$ & $8$ & $Q_{2,3}+2$ & $8$ & $Q_{2,3}+2$ & $4$ & $Q_{2,3}$\tabularnewline
\hline 
$\nu=0$ & $4$ & $Q_{2,3}$ & N/A & N/A & $4$ & $Q_{2,3}$\tabularnewline
\hline 
$\nu$ generic & $4$ & $Q_{2,3}$ & $8$ & $Q_{2,3}+2$ & N/A & N/A\tabularnewline
\hline 
\end{tabular}
\par\end{centering}
\caption{\label{table:vortex_loop_supercharges}Supercharges preserved by a
generic $\mathcal{N}=4$ latitude vortex loop: total number preserved
and number preserved inside the Poincar\'{e} subalgebra defined by
the latitude supercharge, with $Q_{2,3}$ singled out.}
\end{table}
\par\end{center}
\subsection{\label{subsec:SQM}Supersymmetric Quantum Mechanics}

An alternative definition of BPS line operators is provided by coupling a 1d supersymmetric quantum mechanics (SQM) supported on the defect to the 3d bulk theory. The coupling can be realized by gauging the flavor symmetries of the 1d system with the 3d vector multiplet. 
For instance, in \cite{Assel:2015oxa}, a large class of $1/2$-BPS loop operators has been described in this way. However, in that case, the 1d theory always preserves 4 supercharges, whereas our SQM leaves unbroken an $\mathcal{N}=2$ SUSY algebra.

To be concrete, we introduce the action of the latitude supercharges:
\begin{equation}
\delta=\epsilon\, Q_2 + \bar{\epsilon}\,Q_3\,.
\end{equation}
The latitude superalgebra in the $\mathcal{N}=4$ language reads as:
\begin{align}
\acomm{Q_2}{Q_3}=\nu F+\tilde{H}\,,
\end{align}
where $F\equiv R_C-P_\phi$ and $\tilde{H}\equiv P_\tau-R_H$ \footnote{Up to a $\nu$ rescaling, the generators $\tilde{H}$ and $F$ can be identified respectively with the central charge $Z$ and the combination $\mathcal{T}+2L_z$ as $\theta_0\to 0$.}. The only non-vanishing commutators are:
\begin{equation}
\comm*{R_C}{Q_2}=-Q_2\,,\qquad \comm*{R_C}{Q_3}=Q_3\,.
\end{equation}
We want to interpret this algebra as a centrally extended $\mathcal{N}=2$ 1d Supersymmetry algebra. Since $\tilde{H}$ generates the translations along the loop, it plays the role of the Hamiltonian of the SQM. Furthermore, as $P_\phi$ acts only on the normal bundle in $S^3$ with respect to the loop worldvolume, $\nu F$ behaves as a central extension of the algebra and it generates an $U(1)$ flavor symmetry for the SQM. The reader can find a complete description of the $\mathcal{N}=2$ and $\mathcal{N}=4$ SQM in \cite{Hori:2014tda}. For now, we limit ourselves to point out that our generator $\nu F$ should be identified with their generator $J_-$, which is an R-symmetry from the $\mathcal{N}=4$ point of view, but it is a flavor symmetry from the $\mathcal{N}=2$ perspective.

While the outlined procedure is straightforward for a line operator, we might need to turn on background fields on a curved manifold. This usually occurs when we place supersymmetric theories on curved spaces.
Moreover, the determination of the backgrounds fields becomes relevant when we will define the refined Witten index of the SQM. The action of $Q_2$ and $Q_3$ allows us to organize the 3d $\mathcal{N}=4$ vector multiplet into 1d $\mathcal{N}=2$ multiplets. Thus, we can read the action of the flavor symmetries looking at the action of $\delta^2$ on the reduced multiplets. 

On general grounds, the degrees of freedom of the 3d $\mathcal{N}=4$ vector multiplet are recast into one $\mathcal{N}=2$ 1d vector multiplet, two $\mathcal{N}=2$ 1d chiral multiplets, and two $\mathcal{N}=2$ 1d Fermi multiplets. The explicit decomposition is spelled out in Appendix \ref{app_sqm}.

Let us discuss how $\nu$ affects the 1d algebra. On a hand, a chiral multiplet has charge 1 under $\nu F$. On the other hand, $\nu$ never acts on the 1d vector multiplet. It follows that $\nu F$ is a flavor symmetry of the $\mathcal{N}=2$ algebra. Also the loops described in \cite{Assel:2015oxa} exhibit an analogous flavor symmetry, generated by $J_-$. Thus, we identify $\nu F$ with $J_-$. However, for them, $\nu$ can take only integer values. In that case, $\nu$ can always be reabsorbed into a redefinition of the 1d Killing spinors by a factor $e^{\mathbbm{i}\nu\tau}$. Thus, the $\mathcal{N}=2$ 1d vector and $\mathcal{N}=2$ chiral constitute an $\mathcal{N}=4$ vector multiplet and the full $\mathcal{N}=4$ 1d SUSY is restored. For latitude loops, the non integer value of $\nu$ prevents this kind of enhancement and breaks down the Supersymmetry to $\mathcal{N}=2$. 

Then, we claim that if we want to describe the latitude as 3d/1d defect system, we need to gauge the SQM with a vector given by the embedding described in \ref{app_sqm}. In addition, a background flavor symmetry for the generator $\nu F$ must be turned on. We will provide later the explicit action for the SQM.

Our considerations hold until $\nu\neq1$. The case $\nu=1$ deserves a specific investigation. In fact, as we have seen in \ref{sub_lat_alg_nu_1}, when $\nu \to 1$ the loop exhibits a supersymmetry enhancement. In particular, two extra supercharges $Q_5$ and $Q_6$ annihilates the loop. The subalgebra spanned by $Q_2$, $Q_3$, $Q_5$, and $Q_6$ do not sit in any three-dimensional Poincar\'e subalgebra. The resulting 1d superalgebra is indeed a 1d superconformal algebra, rather than an $\mathcal{N}=4$ Poincar\'e algebra as in \cite{Assel:2015oxa}. 
Therefore, also the Supersymmetric Quantum Mechanics becomes a Superconformal Quantum Mechanics at $\nu=1$.
We present this repeating the same steps as $\nu$ generic, but considering the two additional supercharges. Details on the decomposition are discussed in $\ref{sqm1}$.

Thus, let us define the action of the supercharges:
\begin{equation}\label{d1}
\delta=\epsilon\, Q_2+ \bar{\epsilon}\, Q_3+\rho\, Q_5+\bar{\rho}\, Q_6\,.
\end{equation}
It is convenient to recast the four parameters $\epsilon$, $\bar\epsilon$, $\rho$, $\bar\rho$ into two ``superconformal'' parameters:
\begin{equation}\label{antiper_spin}
\zeta=e^{-\frac{\mathbbm{i}}{2}\tau} \rho +e^{\frac{\mathbbm{i}}{2} \tau} \epsilon\,,\qquad \bar\zeta=e^{-\frac{\mathbbm{i}}{2}\tau} \bar\rho +e^{\frac{\mathbbm{i}}{2} \tau} \bar\epsilon
\end{equation}
These 1d spinors are indeed anti-periodic. Even though this choice is somehow unconventional, it reproduces the expected results\footnote{Anti-periodic Killing spinors have made an appearance in the context of the hyperbolic index, aka the supersymmetric Renyi entropy \cite{Hosseini:2019and}, and the 4d superconformal index \cite{Cabo-Bizet:2018ehj}}. 

A first hint of the realization of an underlying superconformal symmetry comes from the space-time symmetries of $\delta^2$. In fact, the most general diffeomorphism contained in $\delta^2$ generates the 1d conformal group $SL(2,\mathbb{R})$. Besides, $\delta^2$ contains a non zero dilatation, as well as some R-symmetries and a gauge transformation. 
Finally, in the appendix, we provide some concrete examples of the action of the supercharges on the SQM multiplets. They turn out to agree with the algebra found for the latitude Wilson loop at $\nu=1$ of section \ref{sub_lat_alg_nu_1}. 
Then, we conclude that our dual operator reproduces the expected symmetry enhancement, namely the $\mathcal{N}=2$ $\nu$-dependent Poincar\'e SUSY algebra ``flows'' to a conformal $\mathcal{N}=2$ superconformal algebra as $\nu\to 1$.

\subsection{\label{subsec:Localization-in-a-dual-theory}Localization in a dual
theory}

ABJM at Chern-Simons level $1$ with gauge group $U\left(N\right)\times U\left(N\right)$
is a dual infra-red description of $U\left(N\right)$ $\mathcal{N}=8$
super-Yang-Mills \cite{Aharony:2008ug}. There exists an additional
$\mathcal{N}=4$ gauge theory with no Chern-Simons terms within the
same universality class. This theory has the following $\mathcal{N}=4$
content: a $U\left(N\right)$ vector multiplet, one adjoint hypermultiplet,
and one fundamental hypermultiplet \cite{Aharony:2008ug}. We will call this theory ``the
UV theory''. An alternative to localization in ABJM, in the presence
of the latitude Wilson loop, is localization in the UV theory with
an insertion of the dual loop operator.\footnote{The UV theory is a better candidate for localization than $\mathcal{N}=8$
SYM for reasons explained in e.g. \cite{Kapustin:2009kz,Kapustin2010}.} In this section, we carry out this localization and the identification of the dual
loop operator.

\subsubsection{The localizing term}

The Yang-Mills action $S_{\text{YM}}$ as defined in \cite{Dedushenko:2016jxl}, Eq \ref{eq:Yang_Mills_action}, with
appropriate matrices $h,\bar{h}$, is closed under $Q_{2,3}$ and
under the latitude supercharge. We can show that $S_{\text{YM}}$
is exact, i.e. it is the variation under the latitude supercharge
of an appropriate fermionic functional. It can therefore be used as
a localizing term.

$S_{\text{YM}}$ was shown in \cite{Dedushenko:2016jxl} to be
exact using the auxiliary spinor 
\[
\xi_{-\beta a\dot{a}}^{H}\equiv\xi_{\beta a\dot{b}}^{H}{\left(\tau_{3}\right)^{\dot{b}}}_{b}\,.
\]
We define the analogous spinor for the latitude 
\[
\left(\tilde{\xi}_{\nu}^{L}\right)_{a\dot{a}\alpha}\equiv\left(\xi_{\nu}^{L}\right)_{a\dot{b}\alpha}{{\tau_{3}}^{\dot{b}}}_{\dot{a}}\,.
\]
We can now try to form the following localizing term
\begin{equation}
S_{\text{localizing}}\equiv\delta_{\xi_{\nu}^{L}}\delta_{\tilde{\xi}_{\nu}^{L}}\left(\frac{1}{2g_{YM}}\tau_{3}^{ab}\tau_{3}^{\dot{a}\dot{b}}\int d^{3}x\sqrt{g}\text{Tr}\left(\lambda_{a\dot{a}}\lambda_{b\dot{b}}-2D_{ab}\Phi_{\dot{a}\dot{b}}\right)\right)\,.\label{eq:latitude_UV_localizing_term}
\end{equation}
Somewhat surprisingly, $S_{\text{localizing}}$ is indeed proportional
to $S_{\text{YM}}$. 

The localization locus for $S_{\text{YM}}$ was worked out in \cite{Dedushenko:2016jxl}. It consists of the usual spacetime independent vev for an adjoint
valued scalar field, which in this case is $\Phi_{\dot{1}\dot{2}}$.\footnote{The auxiliary field $D_{12}$ also has a vev.}
The one loop determinant, which gives the effective action for this modulus, is identical to the one derived in the $\mathcal{N}=2$
formalism in \cite{Kapustin:2009kz}
\[
\sigma\equiv\left\langle \Phi_{\dot{1}\dot{2}}\right\rangle ,\quad Z_{\text{vector}}\left(\sigma\right)=\prod_{\alpha>0}4\sinh^{2}\left(\pi\alpha\left(\sigma\right)\right)=\prod_{i<j}^{N}4\sinh^2\left(\pi\left(\sigma_{i}-\sigma_{j}\right)\right)\,.
\]

At this point, it may seem strange that the effective action for $\sigma$
coming from the vector multiplet fields does not depend on $\nu$.
For instance, we could have used the index theorem to derive
the one loop determinant in the cohomological formalism, as we did
in \ref{subsec:Localization-in-the-ABJM-model} for ABJM, the results
of which surely depend on $\nu$. The resolution is that, from the
$\mathcal{N}=2$ perspective, the $\mathcal{N}=4$ vector multiplet
contains an additional dynamical adjoint chiral multiplet, whose lowest
component is given by $\Phi_{\dot{1}\dot{1}}$, which was not present in \ref{subsec:Localization-in-the-ABJM-model}.\footnote{More accurately, the adjoint chiral multiplet in the ABJM model had
the right quantum numbers to form an invariant $\delta$-exact mass
term and could therefore be integrated out, or ignored in the localization calculation.
In the $\mathcal{N}=2$ formalism, this is the situation for a chiral
multiplet with R-charge $1$.} As can be seen from the action of the square of the latitude supercharge
on $\Phi_{\dot{1}\dot{1}}$, this additional multiplet has the correct
quantum numbers to correct the total one loop determinant to the original
$\mathcal{N}=2$ expression via the special function identity
\begin{align*}
\prod_{\alpha>0}4\sinh^{2}\left(\pi\alpha\left(\tilde{\sigma}\right)\sqrt{\nu}\right) & =\prod_{\alpha>0}s_{\sqrt{\nu}}\left(\mathbbm{i}\frac{Q}{2}-\alpha\left(\tilde{\sigma}\right)-\mathbbm{i}\sqrt{\nu}\right)s_{\sqrt{\nu}}\left(\mathbbm{i}\frac{Q}{2}+\alpha\left(\tilde{\sigma}\right)-\mathbbm{i}\sqrt{\nu}\right)\times\\
 & \qquad\times\sinh\left(\pi\sqrt{\nu}\alpha\left(\tilde{\sigma}\right)\right)\sinh\left(\pi\frac{\alpha\left(\tilde{\sigma}\right)}{\sqrt{\nu}}\right)\,,
\end{align*}
which again appears in \cite{Gaiotto:2019mmf}.\footnote{In this expression, $\tilde{\sigma}\equiv\left\langle \Phi_{\dot{1}\dot{2}}\right\rangle /\sqrt{\nu}$, so that the argument of the $\text{sinh}$ function coincides with the
one derived above. This has been done in order to match the normalization of the supercharge used in the cohomological localization of ABJM in Section \ref{subsec:Localization-in-the-ABJM-model}. }

As argued in \cite{Dedushenko:2016jxl}, the hypermultiplets in a
theory of this type do not require localization since their action
is quadratic when evaluated in the background of the localized vector
multiplet. The fields can therefore be integrated out exactly at one
loop without adding any deformation term. The resulting one loop determinants
coincide with the ones derived in \cite{Kapustin:2009kz} using the
$\mathcal{N}=2$ formalism and are, in particular, $\nu$-independent\footnote{From the point of view of the equivariant index theorem of Section \ref{subsec:Localization-in-the-ABJM-model}, the $\nu$ independence here is a simple consequence of having only hypermultiplets and no twisted-hypermultiplets, and of the rescaling of the Killing spinor used in this model.}
\[
Z_\text{adjoint hyper}\left(\sigma\right)=\prod_{i,j}^{N}\frac{1}{2\cosh\left(\pi\left(\sigma_{i}-\sigma_{j}\right)\right)},\quad Z_\text{fund. hyper}\left(\sigma\right)=\prod_{i}^{N}\frac{1}{2\cosh\left(\pi\sigma_{i}\right)}\, .
\]
In fact, the entire matrix model is the same as the one derived for the UV theory in \cite{Kapustin:2010xq}. This is an analogue of the result obtained with the Higgs branch supercharge, equivalently at $\nu\rightarrow 0$, in \cite{Dedushenko:2016jxl}. We have shown that it holds for arbitrary values of $\nu$.

\subsubsection{A UV avatar for the latitude}

At first blush, it seems that the Wilson loop described in \ref{subsec:Latitude-loops-in-generic-N4-theories},
placed at $\theta=0$, could serve as a UV avatar for the bosonic latitude
loop in ABJM, i.e. the two operators would be identified at the IR
fixed point. However, after localization the expectation value of
this loop, in the fundamental representation, is simply
\begin{equation}
\left\langle \sum_{i}e^{2\pi\sigma_{i}}\right\rangle _{\text{KWY}}\,,\label{eq:KWY_loop_expectation}
\end{equation}
where $\text{KWY}$ indicates the original matrix model for the UV
theory described in \cite{Kapustin:2010xq}. In particular, the expectation
value is $\nu$-independent. Moreover,
this Wilson loop preserves far more supersymmetry than does the ABJM bosonic latitude
loop.

Another option is to identify the latitude loop with a vortex loop
preserving the latitude supercharge at $\theta=0$. In fact, insertion
of a vortex loop alters the one loop determinants used in the matrix
model in a way which is qualitatively similar to the expressions Eq \ref{latitudenu}, c.f. \cite{Kapustin:2012iw}. However,
the identification of the latitude loop with a vortex loop would imply,
for instance, that the additional supercharges preserved by the loop
at $\nu\rightarrow1$, denoted in \ref{subsec:Embedding-in-N4} as
$Q_{5,6}$, sit in the same Poincar\'{e}  subalgebra as $Q_{2,3}$,
which turns out not to be correct.

In order to identify a UV avatar for the latitude loop we must therefore search for a different BPS loop operator. A clue comes from the comparison of the partition function of the UV theory to ABJM, carried out in \cite{Kapustin2010}. It implies that the relationship
between the theories involves a specific $SL\left(2,\mathbb{Z}\right)$
duality transformation. This transformation is also visible in the original type
IIB brane construction in \cite{Aharony:2008ug}. The transformation is not merely
mirror symmetry, i.e. a transformation using the $S$ element of the
type IIB S-duality group, but rather an action which involves the $T$ generator
as well, of the type studied in e.g. \cite{Jensen:2009xh}. In particular,
FI and mass terms are not exchanged by the duality, but rather mixed. 

Under mirror symmetry, Wilson loops are mapped to vortex loops \cite{Kapustin:2012iw, Assel:2015oxa},
and it seems reasonable to expect that the full action of $SL\left(2,\mathbb{Z}\right)$
maps Wilson loops to combined vortex-Wilson loop operators. Such mixed operators can presumably be defined in a variety of ways,
but we are not aware of any previous attempts to do so. We will argue that this possibility is realized for the bosonic latitude
loop of ABJM and its avatar in the UV theory. As preliminary evidence, we note that the supercharges preserved
by a combination of Wilson and vortex loops of the UV theory exactly
match the supercharges preserved by the latitude loop, even when taking
into account the additional supercharges present in the limit $\nu\rightarrow1$. In this statement, the supercharges preserved by the combined line operator are assumed
to be the ones in the overlap of those preserved by the constituent
lines. Note that the fermionic latitude loop in ABJM \textit{does} preserve the right
amount of supersymmetry, at $\nu\rightarrow1$, to be mapped to a
pure Wilson loop. The fermionic
latitude at generic $\nu$ presumably maps to an as-yet-unidentified
loop operator of vortex-Wilson type, in the same cohomology class as
the bosonic latitude, but preserving more of the supersymmetry.

We will characterize the specific BPS operator dual to the bosonic latitude in the next section. We will make frequent use of the results for $1/2$ BPS Wilson and vortex operators produced in \cite{Assel:2015oxa}, whose derivation was aided by the type IIB String Theory construction of the relevant quiver gauge theories \cite{Hanany:1996ie}. It should be noted, however, that the relevance of supersymmetric Wilson and vortex operators, and presumably of any mixed versions, goes beyond the class of theories which can be engineered in type IIB \cite{Dey:2021jbf}. This conclusion follows from the applicability of mirror symmetry, with both the $S$ and $T$ generators, in more general classes of quiver gauge theories \cite{Dey:2020hfe}. More generally, mixed operators should be relevant even to non-supersymmetric theories in the context of the $SL\left(2,\mathbb{Z}\right)$ action on conformal field theories with abelian symmetry introduced by Witten in \cite{Witten:2003ya}.

\subsubsection{Latitude loops from SQM }

According to the authors of \cite{Assel:2015oxa}, we can think of Wilson and vortex
loop insertions in 3d $\mathcal{N}=4$ theories in two useful ways 
\begin{enumerate}
\item as type IIB $1$-branes ending on a Hanany-Witten type setup of D3,
NS5, and D5 branes engineering the theory \cite{Hanany:1996ie}, and on other branes away from the main setup;
\item or as the coupling of the 3d fields to a supersymmetric quantum mechanics
(SQM) living on the loop worldvolume.
\end{enumerate}
In order to recover the effect of the loop operator from the brane description, one starts by
reading off the worldvolume SQM. This SQM is the effective theory living on the $1$-branes, which have one compact direction. The SQM couples to the bulk theory using gauge and superpotential
terms. One then integrates out the quantum mechanical degrees of freedom
to obtain a deformation of the bulk theory localized on the loop.

It is often useful to preform localization before integrating out the SQM. This
involves a computation of the supersymmetric index (Witten index)
of the supersymmetric quantum mechanics with certain deformations.
The complete picture is useful, for example, because the action of
3d mirror symmetry can be identified with type IIB S-duality, whose
action on all of the branes in the setup is known. We refer the reader
to very interesting analysis in reference \cite{Assel:2015oxa} for more details. 

In order to take advantage of the brane description, we need a Hanany-Witten
type setup for the UV thoery, and a 1-brane which describes the
operator. The former was described in \cite{Aharony:2008ug}.
The authors of \cite{Aharony:2008ug} described two different setups,
one which engineers the ABJM model and another which engineers the
UV theory. The two setups are related by S-duality of type IIB string
theory. Unfortunately, we do not have a description of the latitude
loop as a 1-brane which attaches to these brane setups. We will therefore
make an educated guess about the SQM governing the bosonic latitude
loop, based on the symmetry algebra and the corresponding theories
for Wilson and vortex loops presented in \cite{Assel:2015oxa}. We
then show that integrating out the SQM degrees of freedom reproduces
the correct matrix model for the expectation value of the bosonic
latitude loop. While we have good reason to expect that the SQM we
present is the correct one, we would like to stress that we have no
evidence for the realization of this theory on the worldvolume of
a compact type IIB 1-brane. We nevertheless make some comments below.

\paragraph{Brane setup}

The type IIB setup engineering the ABJM model at level $1$ includes
\cite{Aharony:2008ug}
\begin{itemize}
\item A stack of $N$ coincident D3-branes along the directions $0126$
with the $6$ direction compactified to a circle. 
\item One NS5-brane spanning $012345$ and situated at some point in
the $6$ direction.
\item One $\left(1,1\right)$ brane spanning $012\left[3,7\right]_{\theta}\left[4,8\right]_{\theta}\left[5,9\right]_{\theta}$
and situated at a different point in the $6$ direction. A $\left(1,1\right)$
brane is a bound state of an NS5-brane and a D5-brane. The subscript
$\theta$ indicates that the brane is rotated by an angle $\theta$
in the relevant plane. For level $1$ we have $\theta=\pi/4$.
\end{itemize}
Performing an S-duality transformation, and shifting the type IIB
axion, brings the ABJM setup to the following one engineering the UV theory \cite{Aharony:2008ug}
\begin{itemize}
\item A stack of $N$ coincident D3-branes along the directions $0126$
with the $6$ direction compactified to a circle. 
\item One D5-brane spanning $012345$ and situated at some point in the
$6$ direction.
\item One NS5-brane spanning $012\left[3,7\right]_{\theta}\left[4,8\right]_{\theta}\left[5,9\right]_{\theta}$
and situated at a different point in the $6$ direction. 
\item A constant value for the type IIB axion $\chi=2\pi$.
\end{itemize}

A Wilson loop in a gauge theory living on D3-branes can
be engineered by adding fundamental strings that end on the D3-branes.
For a Wilson loop in the fundamental representation, a single string
suffices. A vortex loop can be engineered by including D1-branes instead
of fundamental strings. Both 1-branes have one compact direction and
must end on a 5-brane which is situated away from the main setup \cite{Assel:2015oxa}. It seems reasonable to expect that the latitude loop can be engineered
in a similar fashion. A 1-brane realizing the latitude loop in the
UV theory should be a combination of those realizing the Wilson and
vortex loops, for instance it could be a bound state. Such bound states
indeed exist in type IIB string theory \cite{Witten:1995im}. However,
in order to derive the worldvolume SQM theory on such a compact brane one must
analyze the boundary conditions imposed by the branes in the compact
direction, which is beyond the scope of this paper. We will content
ourselves with finding a supersymmetric quantum mechanics with the
desired properties. 

The latitude supercharges at generic $\nu$ can be shown to be linear
combinations of the supercharges preserved by the $1/2$ BPS Wilson
loop treated in \cite{Assel:2015oxa}, while the mirror supercharges
are linear combinations of the those preserved by the $1/2$ BPS vortex.
Our starting point for constructing a SQM for the bosonic latitude
will be the worldvolume theory for the vortex loop, engineered by
a single D1-brane, as described in \cite{Assel:2015oxa}. We will
deform this theory as necessary to accommodate the properties of the latitude.

\paragraph{The latitude worldvolume theory}

The setup engineering $1/2$ BPS vortex loops in 3d $\mathcal{N}=4$
theories described in \cite{Assel:2015oxa} has a D1-brane ending on a
D3 brane stack on the one end and on an NS5'-brane on the other. A
single D1-brane carries a $U\left(1\right)$ gauge field on its
worldvolume. In the absence of any other branes, the worldvolume theory
has 2d $\mathcal{N}=\left(8,8\right)$ supersymmetry. The compact
direction causes the worldvolume theory on the D1 to be, effectively,
an $\mathcal{N}=4$ gauged supersymmetric quantum mechanics. The $S$
generator of Type IIB S-duality exchanges this D1-brane with a fundamental
string ending on the D3 brane stack, which is the standard description
for a supersymmetric Wilson loop. 

According to the discussion in section 3 of \cite{Assel:2015oxa},
the worldvolume theory of a D1-brane ending on a D3-brane stack has the following
matter content, in terms of 1d $\mathcal{N}=4$ multiplets
\begin{itemize}
\item a $U\left(1\right)$ vector multiplet;
\item $N$ charge $1$ chiral multiplets. Their $U\left(N\right)$ flavor
symmetry is gauged by the bulk $U\left(N\right)$ gauge symmetry living
on the D3 branes where they sit in the fundamental representation; 
\item $N$ charge $-1$ chiral multiplets. Their $U\left(N\right)$ flavor
symmetry is gauged by the bulk $U\left(N\right)$ gauge symmetry living
on the D3 branes where they sit in the anti-fundamental representation. 
\end{itemize}
The setup for the latitude loop at generic $\nu$ must, by definition,
preserve only the supercharges $Q_{2,3}$. Hence, we are looking for
a $\nu$-dependent deformation of the D1-brane worldvolume theory giving
rise to a 1d $\mathcal{N}=2$ $U\left(1\right)$ gauge theory with
matter. Moreover, as argued in Section \ref{subsec:SQM},
the theory must be conformal in the limit $\nu\rightarrow1$. We have
already seen that $\nu$ appears in the latitude superalgebra as the
coefficient of a central term associated to a flavor symmetry. Such
a term indeed breaks conformal invariance. The only continuous $\mathcal{N}=2$ preserving global
symmetry in the D1 theory is an $\mathcal{N}=4$ R-symmetry which commutes
with an $\mathcal{N}=2$ subalgebra. The generator of this R-symmetry
was called $J_{-}$ in \cite{Assel:2015oxa,Hori:2014tda}, and was
indeed identified with a combination of bulk charges of the type that
make up $F$. Weakly gauging $J_{-}$ in the D1-brane worldvolume theory, with a background gauge field with holonomy $\exp\left(2\pi\mathbbm{i}\nu\right)$,
breaks $\mathcal{N}=4$ to $\mathcal{N}=2$. The full $\mathcal{N}=4$
supersymmetry is recovered in the limit $\nu\rightarrow1$.\footnote{Actually, as explained in \cite{Assel:2015oxa}, $\mathcal{N}=4$
supersymmetry is recovered for any integer value of $\nu$ by using
a different Killing spinor in the SQM.} 

An additional part of the worldvolume action for the bosonic latitude must break half of the
supersymmetry, but not conformal invariance, even at $\nu\rightarrow1$.
A minimal guess is that the worldvolume theory has, in addition to
the usual minimally coupled action, a level $1$ 1d Chern-Simons term
for the $U\left(1\right)$ gauge field. We will see below that postulating
this action leads to the correct result for the expectation value
of the latitude loop.\footnote{It is tempting to ascribe the Chern-Simons term in a D1-brane worldvolume theory
to the presence of a non-vanishing type IIB axion in the quiver engineering
the UV theory, but we were not able to verify this intuition.}

Including a topological term for the worldvolume gauge field in order
to produce the ``electric'' part of a defect operator is familiar
from the construction of generic surface operators in 4d $\mathcal{N}=4$
gauge theories described in \cite{Gukov:2006jk}. The relevant electric
parameter of the surface operator is $\eta$, which is defined as
the coefficient of a term measuring the first Chern class of an abelian
bulk gauge field restricted to the surface operator worldvolume \cite{Gukov:2014gja}.
When the surface operator is realized by coupling a 4d theory, with
gauge group $SU\left(2\right)$ broken near the defect to $U\left(1\right)$,
to a 2d GLSM with $SU\left(2\right)$ flavor symmetry, $\eta$ is
realized as the theta parameter of the dynamical $U\left(1\right)$
gauge field in the worldvolume theory. Indeed, the construction of
the same surface operator in type IIA string theory using additional
D2-branes, described in \cite{Alday:2009fs}, may be related by
T duality to our sought-after type IIB setup. 

\paragraph{The index}

We now evaluate the 1d supersymmetric index for the latitude worldvolume theory, following \cite{Hori:2014tda} and \cite{Assel:2015oxa}. The index can be
deduced from the computations of the example in Appendix B.1 of \cite{Assel:2015oxa}.
We must make a few changes to this example. In the notation of Appendix B.1, the changes are as follows.
\begin{enumerate}
\item We set $k\rightarrow1$ so that we describe a single D1-brane.
\item We set $M=N$ and $\vec{m}=\vec{\sigma}$, since we have only one D3-brane segment whose bulk gauge field gauges both chirals. Due to the symmetry in this setup, we set $r_{+}=r_{-}=1$ and $q_{+}=q_{-}=1/4$.
\item We add a localized level $1$ Chern-Simons term. This is simply an
insertion of $\exp\left(-2\pi\mathbbm{i}u\right)$ into the matrix
model.
\item We set $z$, the $J_{-}$ flavor fugacity, to $\nu$ in order to account
for the latitude superalgebra. 
\end{enumerate}
The resulting index is, up to an overall factor \cite{Assel:2015oxa}
\[
\mathcal{I}=\sum_{c=1}^{N}\left(e^{\mathbbm{i}\pi\nu}e^{2\pi\sigma_{c}}\prod_{i<j}^{N}\frac{\sinh\left(\pi\left(\sigma_{i}-\sigma_{j}+\mathbbm{i}\nu\left({\delta^{c}}_{i}-{\delta^{c}}_{j}\right)\right)\right)}{\sinh\left(\pi\left(\sigma_{i}-\sigma_{j}\right)\right)}\prod_{i,j}^{N}\frac{\cosh\left(\pi\left(\sigma_{i}-\sigma_{j}\right)\right)}{\cosh\left(\pi\left(\sigma_{i}-\sigma_{j}+\mathbbm{i}\nu{\delta^{c}}_{i}\right)\right)}\right)\,.
\]
Note that $\mathcal{I}$ is not equal to the product of factors one would get from inserting the $1/2$-BPS Wilson and vortex loops separately. Instead, $\mathcal{I}$ represents a mixed loop.

Although not incorporated in \cite{Assel:2015oxa}, a natural way of normalizing $\mathcal{I}$ is to divide it by the value of the Witten index for the decoupled worldvolume theory. By this we mean the theory with both $\nu$ and the mass parameters $\sigma_i$ set to $0$.\footnote{It is not entirely clear to us why this is the correct limit.} For our $\mathcal{I}$, the result is simply $N$. We adopt this normalization below. There is also an option of multiplying the answer by a term $\mathcal{W}^{fl}$ corresponding to a finite counterterm: a ``flavor'' Wilson loop \cite{Assel:2015oxa}.

\paragraph{The latitude loop expectation value}

We are now in a position to compute the expectation value for the
UV avatar to the bosonic latitude loop in the UV theory. It is given
by a coupled 3d-1d calculation, equation 5.48 of reference \cite{Assel:2015oxa}
\[
\left\langle \text{UV avatar loop}\right\rangle =\frac{1}{\left|\mathcal{W}\right|}\int\prod_{i=1}^{N}d\sigma_{i}\,Z_{\text{vector}}\,Z_{\text{adjoint hyper}}\,Z_{\text{fund. hyper}}\,\mathcal{I}\,.
\]
The form of $\mathcal{I}$ means that it almost completely cancels
the bulk term $Z_{\text{vector}}\,Z_{\text{adjoint hyper}}$, and
replaces it with shifted terms. The resulting integral expression is
\begin{align*}
\left\langle \text{UV avatar loop}\right\rangle  & =\frac{1}{N\left(N!\right)}\int\prod_{i=1}^{N}d\sigma_{i}\,\sum_{c=1}^{N}\bigg(e^{\mathbbm{i}\pi\nu}e^{2\pi\sigma_{c}}\prod_{i<j}^{N}2\sinh\left(\pi\left(\sigma_{i}-\sigma_{j}+\mathbbm{i}\nu\left({\delta^{c}}_{i}-{\delta^{c}}_{j}\right)\right)\right)\\
 & \times\prod_{i<j}^{N}2\sinh\left(\pi\left(\sigma_{i}-\sigma_{j}\right)\right)\prod_{i,j}^{N}\frac{1}{2\cosh\left(\pi\left(\sigma_{i}-\sigma_{j}+\mathbbm{i}\nu{\delta^{c}}_{i}\right)\right)}\prod_{i}^{N}\frac{1}{2\cosh\left(\pi\sigma_{i}\right)}\Bigg)\,.
\end{align*}

We can now attempt to identify the UV avatar loop with the bosonic latitude. We denote the expectation value of the avatar loop, normalized by the partition function of the UV theory, as $W_\text{UV}(\nu)$. Using the Cauchy determinant formula \ref{eq:cauchy} with $\lambda_i=\sigma_i+i\nu \delta_i^c$ and $\mu_j=\sigma_j$, we can write the integral expression for $W_\text{UV}(\nu)$ as
\begin{align}  
 \langle W_{\text{UV}}(\nu) \rangle= \frac{1}{Z \,N(N!)}\sum_{\rho\in S_N}(-1)^\rho\int d\sigma^N
 \sum_{c=1}^N\,\,\,\frac{e^{\mathbbm{i}\pi\nu} e^{2\pi\sigma_c} }{\prod_{j=1}^N 2\cosh\pi\sigma_j\prod_{k=1}^N 2\cosh\pi(\sigma_k-\sigma_{\rho(k)}+\mathbbm{i}\nu\delta_k^c)}\,.
 \end{align}
Taking into account the fact that the partition functions of ABJM and the UV theory coincide at any $\nu$, the expression above is precisely what we found for $W_{B}(\nu)$ in  Eq \ref{latitudenu}.

We have shown that the expectation value of the UV avatar loop in the UV theory matches that of the bosonic latitude in ABJM. Together with the matching of the supersymmetry algebra at all values
of $\nu$, this constitutes the evidence we have for the correct identification
of the UV avatar for the bosonic latitude loop and for the use of
the particular SQM used to define it. 
\section{\label{sec:Conclusion}Conclusion}

We have examined several aspects of latitude Wilson loops in the ABJM model: a family of BPS loop operators parameterized by a real number $\nu$. We have exhibited the superalgebra preserved by these operators in flat space, and when conformally mapped to the three sphere. We have also investigated the limit $\nu\rightarrow 1$, in which the latitude loops degenerate to the standard Gaiotto-Yin BPS Wilson loop of 3d gauge theories with $\mathcal{N}=2$ supersymmetry \cite{Gaiotto:2007qi}. 

We have shown that the supercharges preserved by the latitude loop, at generic $\nu$, fit inside an $\mathcal{N}=4$ supersymmetry algebra, but not inside $\mathcal{N}\le 3$. Consequently, latitude type BPS Wilson loops can be defined in more general Chern-Simons-matter theories of Gaiotto-Witten (GW) type. However, using localization to derive their exact expectation values requires a procedure for closing a generic supercharge off-shell in this class of theories, for which we did not find any previous reference. We have argued that this closure can be achieved using a cohomological formulation of Chern-Simons theories introduced by K\"{a}ll\'{e}n in \cite{Kallen:2011ny} and extended to include matter in \cite{Ohta:2012ev}. However, we did not exhibit an explicit mapping between this formulation and the fields in GW theories. Nevertheless, we performed localization assuming that off-shell closure could be achieved in this way, and recovered the matrix model for the latitude expectation value conjectured in \cite{Bianchi:2018bke}.

We have further shown that BPS loop operators preserving the latitude supercharges can be defined in ``standard'' $\mathcal{N}=4$ gauge theories. Standard theories are gauge theories without Chern-Simons terms, which can be constructed using unconstrained off-shell vector multiplets. The latitude-like loops in this class of theories come in two primary types: Wilson and vortex. We have argued that there should also exist latitude-like loops which are a mixture of the two primary types. Specifically, we examined a standard theory which is known to be IR dual to the ABJM model: the theory with one adjoint and one fundamental hypermultiplet \cite{Aharony:2008ug}. Using symmetry arguments, we showed that this theory does not possess a loop operator which could serve as a dual to the latitude Wilson loop in ABJM, and which is of one of the primary types. Instead, we argued that the dual operator must be of mixed Wilson-vortex type. We subsequently defined mixed type BPS latitude operators using a specific supersymmetric quantum mechanics (SQM), which includes a novel worldvolume Chern-Simons term, and which is closely related to the SQM models introduced for BPS vortex loops in \cite{Assel:2015oxa}. Applying localization to the combined system, consisting of the standard theory and the SQM, we showed that the matrix model of \cite{Bianchi:2018bke} is once again reproduced.

A number of issues have come up during this work into which a more general investigation would be desirable. One of these is the need for a procedure for closing a single generic supercharge off-shell in GW type theories, or indeed in any supersymmetric theory. Such off-shell closure makes a number of arguments based on supersymmetry more transparent, and is specifically a prerequisite for localization. We are not aware of any general work in this direction. Another issue is the need for a general definition of mixed Wilson-vortex type loop operators, including BPS versions of such, in gauge theories in three dimensions. Such a definition would include a study of the moduli space of such operators, of the type carried out for the analogous loop operators in four dimensions by Kapustin \cite{Kapustin:2005py}. It would also be interesting to identify the extended objects in string/M-theory which are responsible for mixed loops, and which generalize the F-strings and D-strings used to elucidate the properties of BPS Wilson and vortex loops in \cite{Assel:2015oxa}. Finally, we think that an examination of the duality properties of mixed loops, under both mirror symmetry \cite{Intriligator:1996ex} and Aharony duality \cite{Aharony:1997gp}, could prove very illuminating. 

\section*{Acknowledgments}

We would like to thank Ofer Aharony, Sara Pasquetti, Silvia Penati, Jian Qiu and Domenico Seminara for many interesting discussions. This work has been supported in part by Italian Ministero dell'Istruzione, Universit\'a e Ricerca (MIUR), and Istituto Nazionale di Fisica Nucleare (INFN) through the ``Gauge and String Theory'' (GAST). The work of IY was financially supported by the European Union's Horizon 2020 research and innovation programme under the Marie Sklodowska-Curie grant agreement No. 754496 - FELLINI.  
 
\appendix
\section{\label{sec:3d_supersymmetric_theories}3d supersymmetric theories}

In this appendix we review the relevant supersymmetric theories for this paper, providing the actions and the related supersymmetry algebras. We summarize the relevant conventions.

\subsection{ABJM Supersymmetry conventions}\label{app1}

We start with the $\mathcal{N}=6$ superconformal algebra $\mathfrak{osp}(6|4)$. The bosonic part of the algebra is given by $\mathfrak{so}(3,2)\oplus\mathfrak{so}(6)$. $\mathfrak{so}(3,2)$ is the conformal algebra in three dimensions and $\mathfrak{so}(6)$ is the R-symmetry algebra. The non-trivial commutators are given by
\begin{align}
\comm{M^{\mu\nu}}{M^{\rho\sigma}}&=\delta^{\sigma[\mu}M^{\nu]\rho}+\delta^{\rho[\nu}M^{\mu]\sigma}\,,\\
 \comm{P^\mu}{M^{\nu\rho}}&=\delta^{\mu[\nu}P^{\rho]}\,, \quad &\comm{K^\mu}{M^{\nu\rho}}&=\delta^{\mu[\nu}K^{\rho]} \,,\\
\comm{P^\mu}{K^\nu}&=2\delta^{\mu\nu}D+2M^{\mu\nu} \,,&\comm{D}{P^\mu}&=P^\mu, \quad &\comm{D}{K^\mu}&=-K^\mu\, .
\end{align}
Exploiting the isomorphism $\mathfrak{so}(6)\simeq\mathfrak{su}(4)$, we represent the R-symmetry generators as matrices $J\indices{_I^J}$ transforming in the adjoint of $\mathfrak{su}(4)$  
\begin{equation}
\comm{J\indices{_I^J}}{J\indices{_K^L}}=\delta_I^L J\indices{_K^J}-\delta_K^J J\indices{_I^L}\,.
\end{equation} 
We take the odd generators $\bar{Q}_{IJ,\alpha}$ and $\bar{S}_{IJ,\alpha}$ as spacetime spinors transforming in the antisymmetric representation of $\mathfrak{su}(4)$. The odd-odd commutation relations are
\begin{align}
\acomm*{\bar{Q}_{IJ,\alpha}}{\bar{Q}_{KL}^\beta}&=2\epsilon_{IJKL}(\gamma^\mu)\indices{_\alpha^\beta}P_\mu\,,\\
\acomm*{\bar{S}_{IJ,\alpha}}{\bar{S}_{IJ^\beta}}&=2\epsilon_{IJKL}(\gamma^\mu)\indices{_\alpha^\beta}K_\mu\,,  \\
\acomm*{\bar{Q}_{IJ,\alpha}}{\bar{S}_{KL,\beta}}&=\epsilon_{IJKL}\left((\gamma^{\mu\nu})\indices{_\alpha^\beta}M_{\mu\nu}+2\delta_{\alpha}^\beta D\right)+2\delta_{\alpha}^\beta\epsilon_{IJMN}\left(\delta_K^N J\indices{_L^M}-\delta_L^N J\indices{_K^M} \right) \,.
\end{align}
The mixed commutators are given by
\begin{align}
\comm*{D}{\bar{Q}_{IJ,\alpha}}&=\frac{1}{2}\bar{Q}_{IJ,\alpha}\,, \quad &\comm*{D}{\bar{S}_{IJ,\alpha}}&=-\frac{1}{2}\bar{S}_{IJ,\alpha} \,,\\
\comm*{M^{\mu\nu}}{\bar{Q}_{IJ,\alpha}}&=-\frac{1}{2}(\gamma^{\mu\nu})\indices{_\alpha^\beta}\bar{Q}_{IJ,\beta}\,, \quad &\comm*{M^{\mu\nu}}{\bar{S}_{IJ,\alpha}}&=-\frac{1}{2}(\gamma^{\mu\nu})\indices{_\alpha^\beta}\bar{S}_{IJ,\beta}\,,  \\
\comm*{K^\mu}{\bar{Q}_{IJ,\alpha}}&=(\gamma^\mu)\indices{_\alpha^\beta}\bar{S}_{IJ,\beta}\,, \quad  &\comm*{P^\mu}{\bar{S}_{IJ,\alpha}}&=(\gamma^\mu)\indices{_\alpha^\beta}\bar{Q}_{IJ,\beta}\,, 
\end{align}
and
\begin{align}
\comm*{J\indices{_I^J}}{\bar{Q}_{KL,\alpha}}&=\frac{1}{2}\delta^J_I \bar{Q}_{KL,\alpha}-\delta^J_K \bar{Q}_{IL,\alpha}-\delta^J_L \bar{Q}_{KI,\alpha}\,,\\
\comm*{J\indices{_I^J}}{\bar{S}_{KL,\alpha}}&=\frac{1}{2}\delta^J_I \bar{S}_{KL,\alpha}-\delta^J_K \bar{S}_{IL,\alpha}-\delta^J_L \bar{S}_{KI,\alpha}\,.
\end{align}
Finally, we also write explicitly the action of $J\indices{_I^J}$on the (anti-)fundamental representation
\begin{equation}
\comm*{J\indices{_I^J}}{O_K}=\frac{1}{4}\delta_I^JO_K-\delta_K^JO_I  \,,\qquad \comm*{J\indices{_I^J}}{O^K}=\delta_I^KO_J-\frac{1}{4}\delta_I^JO^K\,.
\end{equation}

The ABJM theories form a class of Lagrangian $\mathcal{N}=6$ superconformal theories.  They are Chern-Simons matter theories with gauge group $U(N_1)_k\cross U(N_2)_{-k}$.
In the paper we follow the conventions of \cite{Drukker:2009hy}. We denote the two gauge fields with $A_\mu$ and $\hat{A}_\mu$, respectively for $U(N_1)$ and $U(N_2)$. The matter scalar fields $C_I$, $\bar{C}^I$ $I=1,\,\dots,\,4$, transforming respectively in the bifundamental and antibifundamental of the gauge group. The lower index $I$ defines the fundamental of the $SU(4)$ R-symmetry group. Finally, the matter spinor fields $\bar{\psi}^I$ and $\psi_I$ transforms respectively in the bifundamental and antibifundamental of the gauge group.

The flat space action is given by
\begin{equation}
S=S_{\mathrm{CS}}+S_{\mathrm{mat}} + S_{\mathrm{int}}\,,
\end{equation}
where
\begin{equation}
\begin{aligned}
S_{\mathrm{CS}}&=-\mathbbm{i}\frac{k}{4\pi}\int d^3x \ \varepsilon^{\mu\nu\rho}\left[\Tr\left(A_\mu\partial_\nu A_\rho+\frac{2\mathbbm{i}}{3}A_\mu A_\nu A_\rho\right)-\Tr\left(\hat{A}_\mu\partial_\nu\hat{A}_\rho+\frac{2\mathbbm{i}}{3}\hat{A}_\mu\hat{A}_\nu\hat{A}_\rho\right)\right] \,,\\
S_{\mathrm{mat}}&=\int d^3x \ \Tr\left[D_\mu C_I D^\mu\bar{C}^I+\mathbbm{i}\bar{\psi}^I\gamma^\mu D_\mu\psi_I\right]\,. \\
\end{aligned}
\end{equation}
The covariant derivatives are defined as
\begin{equation}
\begin{aligned}
D_\mu C_I =\partial_\mu C_I +\mathbbm{i} A_\mu C_I-\mathbbm{i}C_I \hat{A}_\mu \,, \qquad D_\mu \bar{C}^I = \partial_\mu \bar{C}^I +\mathbbm{i} \hat{A}_\mu \bar{C}^I-\mathbbm{i}\bar{C}^I A_\mu\,. \\
\end{aligned}
\label{covd}
\end{equation}
\(S_{\mathrm{int}}\) contains the superpotential terms.

The ABJM action is invariant under the following SUSY transformations
\begin{subequations}\label{abjmsusy}
\begin{align}
\delta A_\mu&=\frac{4\pi \mathbbm{i}}{k}\bar{\Theta}^{IJ, \alpha}{(\gamma_\mu)_\alpha}^\beta\bigg(C_I\psi_{J\beta} +\frac{1}{2}\varepsilon_{IJKL}\bar\psi^K_\beta\bar C^L\bigg)\,,\\
\delta\hat A_\mu&=\frac{4\pi \mathbbm{i}}{k}\bar{\Theta}^{IJ, \alpha}{(\gamma_\mu)_\alpha}^\beta\bigg(\psi_{J\beta}C_I +\frac{1}{2}\varepsilon_{IJKL}\bar C^L\bar\psi^K_\beta\bigg)\,,\\
\delta C_K&=\bar{\Theta}^{IJ,\alpha}\ \varepsilon_{IJKL}\ \bar \psi^{L}_\alpha \,,\\
\delta\bar C^K&=2\bar{\Theta}^{KL,\alpha}\ \psi_{L,\alpha}\,,\\
\delta\bar\psi^{K,\beta}&=-2\mathbbm{i}\bar{\Theta}^{KL,\alpha}{(\gamma^\mu)_\alpha}^\beta D_\mu C_L-\frac{4\pi \mathbbm{i}}{k}\bar{\Theta}^{KL,\beta}(C_L\bar C^M C_M-C_M\bar C^MC_L) +\\
&- \frac{8\pi \mathbbm{i}}{k}\bar{\Theta}^{IJ,\beta}C_I\bar C^KC_J
-2i\bar\epsilon^{KL,\beta} C_L \,,\notag \\
\delta\psi^\beta_K&=-\mathbbm{i}\bar{\Theta}^{IJ,\alpha}\varepsilon_{IJKL}{(\gamma^\mu)_\alpha}^\beta D_\mu \bar C_L+\frac{2\pi \mathbbm{i}}{k}\bar{\Theta}^{IJ,\beta}\varepsilon_{IJKL}(\bar C^LC_M\bar C^M-\bar C^MC_M\bar C^L)\\
&\quad +\frac{4\pi  \mathbbm{i}}{k}\bar{\Theta}^{IJ,\beta}\varepsilon_{IJML}\bar C^M C_K\bar C^L-\mathbbm{i}\bar\epsilon^{IJ,\beta}\varepsilon_{IJKL}\bar C^L\, .
\end{align}
\end{subequations}
The flat space Killing spinors are taken to be
\begin{equation}
\bar\Theta^{IJ}=\bar\theta^{IJ}-x_\mu\gamma^\mu\bar\epsilon^{IJ}\,.
\end{equation}
With these conventions, Eq \ref{abjmsusy} closes the $\mathfrak{osp}(6|4)$ superconformal algebra on-shell.

\subsection{\label{subsec:3d_N4_theories_on_S3}3d $\mathcal{N}=4$ theories on $S^3$}

We review some elements of the 3d $\mathcal{N}=4$ theories on $S^3$, following \cite{Dedushenko:2018icp}.
We provide the relevant multiplets, their supersymmetry transformations and their actions. Fields are labeled by Lorentz spin and gauge group $G$ and the R-symmetry $\mathfrak{su}(2)_C\oplus\mathfrak{su}(2)_H$ representations.\footnote{We denote with $\alpha=1,2$ Lorentz spinor indices, with $a,\dot{a}=1,2$ respectively $\mathfrak{su}(2)_C$, $\mathfrak{su}(2)_H$ indices. } 
We are interested in Lagrangian theories involving vector multiplets and hypermultiplets. The vector multiplet component fields are
\begin{equation}
\mathcal{V}=\left(A_\mu, \, \lambda_{\alpha,a\dot{a}}, \, \Phi_{\dot{a}\dot{b}}, \, D_{ab}\right)\,.
\end{equation} 
They transform in the adjoint representations of the gauge group $G$ and in the following representations of the R-symmetry group
\begin{itemize}
\item $A_\mu$ is the vector field  transforming in the $(\mathbf{1},\mathbf{1})$ of $\mathfrak{su}(2)_C\oplus\mathfrak{su}(2)_H$;
\item $\lambda_{\alpha,a\dot{a}}$ (the gaugino) is a spinor transforming in the $(\mathbf{2},\mathbf{2})$ of $\mathfrak{su}(2)_C\oplus\mathfrak{su}(2)_H$;
\item $\Phi_{\dot{a}\dot{b}}$ is a scalar field transforming in the $(\mathbf{3},\mathbf{1})$ of $\mathfrak{su}(2)_C\oplus\mathfrak{su}(2)_H$;
\item $D_{ab}$ is a scalar field transforming in the $(\mathbf{1},\mathbf{3})$ of $\mathfrak{su}(2)_C\oplus\mathfrak{su}(2)_H$.
\end{itemize}   
The hypermultiplet $\mathcal{H}$ transforms in a unitary representation $\mathcal{R}$ of the gauge group $G$ and its field components are
\begin{equation}
\mathcal{H}=\left( q_a, \tilde{q}^a, \psi_{\dot{a}},\tilde{\psi}_{\dot{a}} \right)\,,
\end{equation}
where
\begin{itemize}
\item $q_a$ are scalar field  transforming in the $(\mathbf{1},\mathbf{2})$ of $\mathfrak{su}(2)_C\oplus\mathfrak{su}(2)_H$ and in the $\mathcal{R}$ of $G$;
\item $\tilde{q}^a$  are scalar fields transforming $(\mathbf{1},\bar{\mathbf{2}})$ of $\mathfrak{su}(2)_C\oplus\mathfrak{su}(2)_H$ and in the $\bar{\mathcal{R}}$ of $G$;
\item $\psi_{\dot{a}}$ are spinor fields transforming in the $(\mathbf{2},\mathbf{1})$ of $\mathfrak{su}(2)_C\oplus\mathfrak{su}(2)_H$ and in the $\mathcal{R}$ of $G$;
\item $\tilde{\psi}^{\dot{a}}$ are spinor fields transforming in the $(\bar{\mathbf{2}},\mathbf{1})$ of $\mathfrak{su}(2)_C\oplus\mathfrak{su}(2)_H$ and in the $\bar{\mathcal{R}}$ of $G$.
\end{itemize}

The supersymmetry transformations for the vector multiplet are given by
\begin{subequations}\label{susyN4}
\begin{align}
\delta_\xi A_\mu&=\frac{\mathbbm{i}}{2}\xi^{a\dot{b}}\gamma_\mu\lambda_{a\dot{b}}\,, \\
\delta_\xi\lambda_{a\dot{b}}&=-\frac{\mathbbm{i}}{2}\epsilon^{\mu\nu\rho}\gamma_{\rho}\xi_{a\dot{b}}F_{\mu\nu}-{D_{a}}^c\xi_{c\dot{b}}-\mathbbm{i}\gamma^\mu{\xi_a}^{\dot{c}}\mathcal{D} _\mu\Phi_{\dot{c}\dot{b}}+2\mathbbm{i}{\Phi_{\dot{b}}}^{\dot{c}}\xi'_{a\dot{c}}+ \\
&+\frac{\mathbbm{i}}{2}\xi_{a\dot{d}}\comm*{{\Phi_{\dot{b}}}^{\dot{c}}}{{\Phi_{\dot{c}}}^{\dot{d}}} \,,    \notag   \\ 
\delta_\xi\Phi_{\dot{a}\dot{b}}&={\xi^c}_{(\dot{a}}\lambda_{|c|\dot{b})} \,, \\
\delta_\xi D_{ab}&=-\mathbbm{i}\mathcal{D}_\mu ({\xi_{(a}}^{\dot{c}}\gamma^\mu\lambda_{b)\dot{c}})-2\mathbbm{i}{\xi'_{(a}}^{\dot{c}}\lambda_{b)\dot{c}}+\mathbbm{i}\comm*{{\xi_{(a}}^{\dot{c}}\lambda_{b)}^{\dot{d}}}{\Phi_{\dot{c}\dot{d}}}  \,.
\end{align}
\end{subequations}
For the hypermultiplet, we have
\begin{subequations}
\begin{align}
\delta_\xi q^a&=\xi^{a\dot{b}}\psi_{\dot{b}}\,,  &\delta_\xi\psi_{\dot{a}}&=\mathbbm{i}\gamma_\mu\xi_{a\dot{a}}\mathcal{D} _\mu q^a+\mathbbm{i}\xi'_{a\dot{a}}q^a-\mathbbm{i}\xi_{a\dot{c}}\Phi\indices{^{\dot{c}}_{\dot{a}}}q^a \,, \\
\delta_\xi\tilde{q}^a&=\xi^{a\dot{b}}\tilde{\psi}_{\dot{b}}\,,         &\delta_\xi \tilde{\psi}_{\dot{a}}&=\mathbbm{i}\gamma_\mu\xi_{a\dot{a}}\mathcal{D} _\mu \tilde{q}^a+\mathbbm{i}\xi'_{a\dot{a}}\tilde{q}^a-\mathbbm{i}\xi_{a\dot{c}}\Phi\indices{^{\dot{c}}_{\dot{a}}}\tilde{q}^a  \,.
\end{align}
\end{subequations}
When the SUSY parameter $\xi_{\alpha,a \dot{a}}$ satisfies the $S^3$ conformal Killing spinor equation, these transformations realize the whole superconformal algebra $\mathfrak{osp}(4|4)$. 

The following is an invariant Lagrangian for the hypermultiplet coupled to the vector multiplet, which is derived from the flat space expression by covariantizing derivatives and by adding specific conformal masses
\begin{align}
S_{\mathrm{hyper}}&=\int d^3x\sqrt{g}\Bigg[ \mathcal{D}^\mu\tilde{q}^a\mathcal{D} _\mu q_a - \mathbbm{i}\tilde{\psi}^{\dot{a}}\gamma^\mu\mathcal{D}_\mu\psi_{\dot{a}}+\frac{3}{4r^2}\tilde{q}^a q_a+\mathbbm{i}\tilde{q}^a{D_a}^b q_b-
\frac{1}{2}\tilde{q}^a\Phi^{\dot{a}\dot{b}}\Phi_{\dot{a}\dot{b}} q_a+\\
&-\mathbbm{i}\tilde{\psi}^{\dot{a}}{\Phi_{\dot{a}}}^{\dot{b}}\psi_{\dot{b}}+\mathbbm{i}\left(\tilde{q}^a{\lambda_a}^{\dot{b}}\psi_{\dot{b}}+\tilde{\psi}^{\dot{a}}{\lambda^{b}}_{\dot{a}} q_b\right)\Bigg]\notag \,.
\end{align}

To the best of our knowledge, there is no vector multiplet action invariant under the off-shell $\mathcal{N}=4$ superconformal symmetry. However, if the conformal Killing spinors are further restricted to obey the condition \cite{Dedushenko:2018icp}
\begin{equation}\label{susycond}
\xi'_{a\dot{a}}\equiv \frac{1}{3}\gamma^\mu\nabla_\mu\xi_{a\dot{a}}=\frac{\mathbbm{i}}{2r}h\indices{_a^b}\xi_{b\dot{b}}\bar{h}\indices{^{\dot{b}}_{\dot{a}}}\,,
\end{equation}
then the following action turns out to be closed under the transformations \ref{susyN4}
\begin{equation}
\begin{aligned}
\label{eq:Yang_Mills_action}S_{\mathrm{YM}}=&\frac{1}{g^2_{\mathrm{YM}}}\int d^3x\sqrt{g}\Tr\bigg[F^{\mu\nu}F_{\mu\nu}-\mathcal{D}^\mu\Phi^{\dot{a}\dot{b}}\mathcal{D}_\mu\Phi_{\dot{a}\dot{b}}+\mathbbm{i}\lambda^{a\dot{a}}\gamma^\mu\mathcal{D}_\mu \lambda_{a\dot{a}}-D^{ab}D_{ab}+\\
&-\mathbbm{i}\lambda^{a\dot{a}}\comm*{{\lambda_a}^{\dot{b}}}{\Phi_{\dot{a}\dot{b}}}-\frac{1}{4}\comm*{{\Phi^{\dot{a}}}_{\dot{b}}}{{\Phi^{\dot{c}}}_{\dot{d}}}\comm*{{\Phi^{\dot{b}}}_{\dot{a}}}{{\Phi^{\dot{d}}}_{\dot{c}}}-\frac{1}{2r}h^{ab}\bar{h}^{\dot{a}\dot{b}}\lambda_{a\dot{a}}\lambda_{b\dot{b}}+\\
&+\frac{1}{r}\left({h_a}^b{D_b}^a\right)\Bigl({\bar{h}^{\dot{a}}}_{\dot{b}}{\Phi^{\dot{b}}}_{\dot{a}}\Bigr)-\frac{1}{r^2}\Phi^{\dot{a}\dot{b}}\Phi_{\dot{a}\dot{b}}\bigg] \,.
\end{aligned}
\end{equation}
where $h\indices{_a^b}$ and $\bar{h}\indices{^{\dot{b}}_{\dot{a}}}$ are respectively $\mathfrak{su}(2)_C$ and $\mathfrak{su}(2)_H$ matrices, normalized such that $h\indices{_a^c}h\indices{_c^b}=\delta_a^b$ and 
$\bar{h}\indices{^{\dot{b}}_{\dot{c}}}\bar{h}\indices{^{\dot{c}}_{\dot{a}}}=\delta_{\dot{a}}^{\dot{b}}$.
The condition \ref{susycond} selects the half of the conformal Killing spinors generating the Poincar\'e subalgebra $\mathfrak{su}(2|1)_\ell\oplus\mathfrak{su}(2|1)_r$. Indeed, one can check that these supercharges generate the isometry group of $S^3$  $\mathfrak{su}(2)_\ell\oplus\mathfrak{su}(2)_r$, as well as a $\mathfrak{u}(1)_\ell\oplus\mathfrak{u}(1)_r$ R-symmetry, specified by the choice of $h\indices{_a^b}$ and $\bar{h}\indices{^{\dot{b}}_{\dot{a}}}$.

\subsubsection{Closure of the Supersymmetry algebra}\label{clos_N_4}

Here we describe explicitly the closure of the 3d $\mathcal{N}=4$ SUSY algebra following  \cite{Dedushenko:2018icp}. This requires evaluating the action of the bosonic generator $\acomm*{\delta_{\xi}}{\delta_{\tilde{\xi}}}$ on the supermultiplets, denoted generically by $\mathcal{B}$. The action is given by
\begin{equation}\label{eq4.6}
\acomm*{\delta_{\xi}}{\delta_{\tilde{\xi}}} \mathcal{B}=\left(\hat{\mathcal{K}}_{\xi,\tilde{\xi}}+ \mathcal{G}_\Lambda+\mathrm{e.o.m.} \right) \mathcal{B}\,.
\end{equation}
$\mathcal{G}_\Lambda$ is a gauge transformation with parameter $\Lambda$ defined as
\begin{equation}
\Lambda=\mathbbm{i}\left(\tilde{\xi}\indices{^c_{\dot{a}}} \xi_{c\dot{b}}\right)\Phi^{\dot{a}\dot{b}}-\mathbbm{i}(\tilde{\xi}^{a\dot{a}}\gamma^\mu\xi_{a\dot{a}})A_\mu,
\end{equation}
and $\hat{\mathcal{K}}_{\xi,\tilde{\xi}}$ are the representation of bosonic symmetries on the field space. Their explicit form is given by
\begin{equation}
\hat{\mathcal{K}}_{\xi,\tilde{\xi}}=\hat{\mathcal{L}}_v+\hat{R}_C+\hat{R}_H+\hat{\rho}\Delta\,,
\end{equation}
where
\begin{itemize} 
\item $\hat{\mathcal{L}}_v$ is the Lie derivative along the vector $v^\mu=\mathbbm{i}\tilde{\xi}^{a\dot{a}}\gamma^\mu\xi_{a\dot{a}}$; 
\item $\hat{R}_{C/H}$ is an $\mathfrak{su}(2)_{C/H}$ transformation, acting as
\begin{align}
\bar{R}_{\dot{a}\dot{b}}=&\mathbbm{i}\left(\tilde{\xi}\indices{^c_{(\dot{a}}} \xi'_{|a|\dot{b})}+{\xi^c}_{(\dot{a}}\tilde{\xi}'_{|c|\dot{b})}  \right)\\
R_{ab}=&\mathbbm{i}\left(\tilde{\xi}\indices{_{(a}^{\dot{c}}} \xi'_{b)\dot{c}}+\xi\indices{_{(a}^{\dot{c}}} \tilde{\xi}'_{b)\dot{c}} \right)\,,
\end{align}
according to the rule $(\hat{R}_H q)_a=R_{ab}\,q^b$ (the same for $\hat{R}_C$);
\item $\hat\rho$ is the dilatation parameter 
\begin{equation}\label{diloper}
\hat{\rho}=\mathbbm{i}\left(\tilde{\xi}^{a\dot{b}}\xi'_{a\dot{b}}+\xi^{a\dot{b}}\tilde{\xi}'_{a\dot{b}} \right)\,, 
\end{equation}
and $\Delta$ represents the dimension of the fields, and takes the values $\Delta[\mathcal{V}]=(0,3/2,1,2)$ and $\Delta[\mathcal{H}]=(1/2,1)$.
\end{itemize}
The term denoted by e.o.m. stands for equation of motion. We include it for those multiplets whose closure is realized only on-shell. This is indeed the case for the hypermultiplet:
\begin{equation}
\begin{aligned}
\acomm*{\delta_{\xi}}{\delta_{\tilde{\xi}}}\psi_{\dot{a}} &=\left(\hat{\mathcal{B}}_{\xi,\tilde{\xi}}+ \mathcal{G}_\Lambda\right)\psi_{\dot{a}}+\tilde{\xi}^{a\dot{b}}[\xi_{a\dot{a}}(\mathrm{e.o.m.}(\psi))_{\dot{b}}]+\xi^{a\dot{b}}[\tilde{\xi}_{a\dot{a}}(\mathrm{e.o.m.}(\psi))_{\dot{b}}] \\
\acomm*{\delta_{\xi}}{\delta_{\tilde{\xi}}}\tilde{\psi}_a &=\left(\hat{\mathcal{B}}_{\xi,\tilde{\xi}}+ \mathcal{G}_\Lambda\right)\tilde{\psi}_{\dot{a}}-\tilde{\xi}^{a\dot{b}}[\xi_{a\dot{a}}(\mathrm{e.o.m.}(\tilde{\psi}))_{\dot{b}}]-\xi^{a\dot{b}}[\tilde{\xi}_{a\dot{a}}(\mathrm{e.o.m.}(\tilde{\psi}))_{\dot{b}}] 
\end{aligned}
\end{equation} 
where
\begin{equation}
\begin{aligned}
(\mathrm{e.o.m.}(\psi))_{\dot{b}}=&-\mathbbm{i}\left[\gamma^\mu \mathcal{D}_\mu\psi_{\dot{a}}+{\Phi_{\dot{a}}}^{\dot{b}}\psi_{\dot{b}}+\lambda_{a\dot{a}}q^a   \right] \\
 (\mathrm{e.o.m.}(\tilde{\psi}))_{\dot{b}}=&\mathbbm{i}\left[\gamma^\mu \mathcal{D}_\mu\tilde{\psi}_{\dot{a}}-{\Phi_{\dot{a}}}^{\dot{b}}\tilde{\psi}_{\dot{b}}-\tilde{q}^a\lambda_{a\dot{a}}   \right]
\end{aligned}
\end{equation}

\subsubsection{Off-shell closure for hypermultiplets}

The 3d $\mathcal{N}=4$ supersymmetry algebra admits two inequivalent off-shell
vector multiplets: ordinary and twisted. The two multiplets are related
to each other by the outer automorphism of the R-symmetry group $SU\left(2\right)_{l}\times SU\left(2\right)_{r}$
which exchanges the two $SU\left(2\right)$ factors. There are also
two types of hypermultiplets, ordinary and twisted. These do not sit
in any off-shell multiplet with a finite number of fields. Nevertheless,
a single supersymmetry in an $\mathcal{N}=4$ theory incorporating
regular vector multiplets and hypermultiplets can sometimes be closed
off-shell by adding appropriate auxiliary fields \cite{Dedushenko:2016jxl}.
The same is true for the twisted multiplets.

Let $\xi_{\alpha a\dot{a}}$ be the $\mathcal{N}=4$
conformal Killing spinor associated with a supersymmetry transformation
$\delta$. Off-shell closure of $\delta$ on a hypermultiplet can
be achieved by finding another spinor $\chi_{\alpha a\dot{a}}$ satisfying
\cite{Dedushenko:2016jxl}
\begin{equation}
{\xi^{\alpha c}}_{\dot{a}}\xi_{\beta c\dot{b}}={\chi^{\alpha c}}_{\dot{b}}\chi_{\beta c\dot{a}}\,,\quad{\xi_{a}}^{\dot{c}}\chi_{b\dot{c}}=0\,,\quad{\xi_{(a}}^{\dot{c}}\gamma^{\mu}\nabla_{\mu}\xi_{b)\dot{c}}=\frac{3\mathbbm{i}}{2}{\chi_{(a}}^{\dot{c}}\gamma^{\mu}\nabla_{\mu}\chi_{b)\dot{c}}\,.\label{eq:hypermultiplet_closure_spinor}
\end{equation}
In order to close off-shell the hypermultiplet transformations, one
should add auxiliary fields $G_{a},\tilde{G}_{a}$. One must then
make the following modification to the supersymmetry transformations
\begin{gather*}
\delta\psi_{\dot{a}}\rightarrow\delta\psi_{\dot{a}}+\mathbbm{i}{\chi^{a}}_{\dot{a}}G_{a}\,,\quad\delta G^{a}=\mathbbm{i}\chi^{a\dot{a}}\Psi_{\dot{a}}^{\text{eom }}\,,\\
\delta\tilde{\psi}_{\dot{a}}\rightarrow\delta\tilde{\psi}_{\dot{a}}+\mathbbm{i}{\chi^{a}}_{\dot{a}}\tilde{G}_{a}\,,\quad\delta\tilde{G}^{a}=-\mathbbm{i}\chi^{a\dot{a}}\tilde{\Psi}_{\dot{a}}^{\text{eom}}\,,
\end{gather*}
where $\Psi_{\dot{a}}^{\text{eom }},\tilde{\Psi}_{\dot{a}}^{\text{eom }}$
are proportional to the fermion equations of motion \cite{Dedushenko:2016jxl}.

We would like to comment on some of
the differences in performing localization using the the latitude
supercharge versus the supercharge employed in \cite{Dedushenko:2016jxl},
the latter being equivalent to the limit $\nu\rightarrow0$. An interesting
difference between $\xi_{\nu}^{L}$ and $\xi_{\beta}^{C,H}$ is the
existence, or lack thereof, of good solutions to equation \ref{eq:hypermultiplet_closure_spinor}.
Specifically, the existence, or lack thereof, of a \textit{Killing spinor} satisfying \ref{eq:hypermultiplet_closure_spinor}. Equivalently, that the co-kernel of the supersymmetry transformation
by $\xi$ on a hypermultiplet scalar, which yields a subset of the
hypermultiplet fermions, is spanned by contraction with some other
Killing spinors $\chi_{\alpha a\dot{a}}$. $\chi_{\alpha a\dot{a}}$
can then be used to define a canonical orthogonal subspace for the
fermions: a subspace whose fields transform into an auxiliary field.
For $\xi_{\beta}^{H}$, a solution is given by 
\[
\chi=\xi_{-\beta}^{H}\,.
\]
This solution is used in \cite{Dedushenko:2016jxl} to close the hypermultiplet
transformations off-shell. For $\xi_{\nu}^{L}$, however, a solution
does not exist unless $\nu$ is $0$ or $1$. This presumably makes
closing the algebra a more complicated, or even impossible, task.
We therefore conclude that $\xi_{\beta}^{C,H}$ are less generic elements
of the Poincar\'{e}  subalgebra than $\xi_{\nu\ne0,1}^{L}$.

\section{Geometry conventions}

\subsection{\label{subsec:Euclidean-spinor-conventions}Euclidean spinor conventions}

The Pauli matrices are 
\[
\tau_{1}=\begin{pmatrix}0 & 1\\
1 & 0
\end{pmatrix}\,,\quad\tau_{2}=\begin{pmatrix}0 & -\mathbbm i\\
\mathbbm i & 0
\end{pmatrix}\,,\quad\tau_{3}=\begin{pmatrix}1 & 0\\
0 & -1
\end{pmatrix}.
\]
We take the 3d Clifford algebra to be generated by
\[
\gamma_{a}=\tau_{a}\,,
\]
with index structure 
\[
{\left(\gamma_{a}\right)_{\beta}}^{\alpha}\,.
\]
Tangent space indices are raised/lowered by the metric $\delta_{ab}$.
The gamma matrices satisfy 
\[
\gamma_{a}\gamma_{b}=\delta_{ab}+\mathbbm{i}\varepsilon_{abc}\gamma^{c}\,.
\]
Spinors are sections of the spin bundle with index structure $\epsilon_{\alpha}$.
Raising and lowering of spinor indices is done with 
\[
\varepsilon^{\alpha\beta}=\begin{pmatrix}0 & 1\\
-1 & 0
\end{pmatrix}\,,\quad\varepsilon_{\alpha\beta}=\begin{pmatrix}0 & -1\\
1 & 0
\end{pmatrix}\,.
\]

\subsection{\label{sec:Flat-space-spinor}Flat space spinor algebra}

We identify the flat space coordinates with those of the tangent space:
$x_{a}$. The following is a basis for the conformal Killing spinors
on $\mathbb{R}^{3}$
\[
\epsilon_{\mathbb{R}^{3}}^{\left(1\right)}=\begin{pmatrix}1\\
0
\end{pmatrix}\,,\quad\epsilon_{\mathbb{R}^{3}}^{\left(2\right)}=\begin{pmatrix}0\\
1
\end{pmatrix}\,,\epsilon_{\mathbb{R}^{3}}^{\left(3\right)}=-x_{a}\gamma^{a}\begin{pmatrix}1\\
0
\end{pmatrix}\,,\epsilon_{\mathbb{R}^{3}}^{\left(4\right)}=-x_{a}\gamma^{a}\begin{pmatrix}0\\
1
\end{pmatrix}\,.
\]
These satisfy 
\[
\partial_{a}\epsilon_{\mathbb{R}^{3}}^{\left(i\right)}=\frac{1}{3}\gamma_{a}\gamma^{b}\partial_{b}\epsilon_{\mathbb{R}^{3}}^{\left(i\right)}\,.
\]
The conformal group is generated by the vectors
\[
w^{\mu,ij}\equiv\epsilon_{\mathbb{R}^{3}}^{\left(i\right)}\gamma^{\mu}\epsilon_{\mathbb{R}^{3}}^{\left(j\right)}\,,
\]
acting as infinitesimal diffeomorphisms. Suppressing the vector index,
the subsets 
\[
P^{m}\equiv{\tau^{m}}_{ij}w^{ij},\quad M^{mn}\equiv\varepsilon^{mnp}{\tau_{p}}_{ij}w^{i,j+2}\,,
\]
generate the isometry algebra of translations and rotations, which
is the algebra of the 3d Euclidean Poincar\'{e}  group. The remaining
combinations yield conformal Killing transformations.

\subsection{\label{sec:Geometry-of-S3}Geometry of $\mathbb{S}^{3}$}

We will use toroidal coordinates and the following metric on the round
unit radius three sphere
\begin{gather*}
ds^{2}=d\theta^{2}+\sin^{2}\theta d\varphi^{2}+\cos^{2}\theta d\tau^{2}\,,\\
\theta\in\left[0,\pi/2\right),\quad\varphi\in\left[0,2\pi\right),\quad\tau\in\left[0,2\pi\right)\,.
\end{gather*}
Two maximal circles are located at $\theta=0,\pi/2$.

We choose the following vielbein
\begin{gather*}
e^{1}=\sin\left(\varphi+\tau\right)d\theta+\cos\theta\sin\theta\cos\left(\varphi+\tau\right)d\varphi-\cos\theta\sin\theta\cos\left(\varphi+\tau\right)d\tau\,,\\
e^{2}=-\cos\left(\varphi+\tau\right)d\theta+\cos\theta\sin\theta\sin\left(\varphi+\tau\right)d\varphi-\cos\theta\sin\theta\sin\left(\varphi+\tau\right)d\tau\,,\\
e^{3}=\sin^{2}\theta d\varphi+\cos^{2}\theta d\tau\,.
\end{gather*}
In this frame, the spin connection is given by 
\[
\omega_{abc}=\varepsilon_{abc}\,.
\]
The spin covariant derivative is 
\begin{align*}
\nabla_{\mu}\epsilon & \equiv\partial_{\mu}\epsilon+\frac{1}{8}{\omega_{\mu}}^{ab}\left[\gamma_{a},\gamma_{b}\right]\epsilon\\
 & =\partial_{\mu}\epsilon+\frac{\mathbbm{i}}{2}\gamma_{\mu}\epsilon\,.
\end{align*}

A basis for the conformal Killing spinors on $\mathbb{S}^{3}$ is
given by
\[
\epsilon^{\left(1\right)}=\begin{pmatrix}1\\
0
\end{pmatrix},\quad\epsilon^{\left(2\right)}=\begin{pmatrix}0\\
1
\end{pmatrix}\,,\epsilon^{\left(3\right)}=\begin{pmatrix}e^{-\mathbbm{i}\tau}\cos\theta\\
-e^{\mathbbm{i}\varphi}\sin\theta
\end{pmatrix}\,,\epsilon^{\left(4\right)}=\begin{pmatrix}e^{-\mathbbm{i}\varphi}\sin\theta\\
e^{\mathbbm{i}\tau}\cos\theta
\end{pmatrix}\,.
\]
These satisfy 
\[
\nabla_{\mu}\epsilon^{\left(i\right)}=\frac{1}{3}\gamma_{\mu}\gamma^{\nu}\nabla_{\nu}\epsilon^{\left(i\right)}\,.
\]
Defining 
\begin{equation}
\eta^{\left(i\right)}\equiv\frac{1}{3}\gamma^{\mu}\nabla_{\mu}\epsilon^{\left(i\right)}\,,\label{eq:eta}
\end{equation}
this becomes the conformal Killing spinor equation
\[
\nabla_{\mu}\epsilon^{\left(i\right)}=\gamma_{\mu}\eta^{\left(i\right)}\,.
\]

The Lie algebra of the conformal group is generated by the action
of the vectors
\[
v^{\mu,ij}=\epsilon^{\left(i\right)}\gamma^{\mu}\epsilon^{\left(j\right)}\,,
\]
acting as infinitesimal diffeomorphisms. The subsets 
\begin{equation}
J_{l}^{\mu a}\equiv{\tau^{a}}_{ij}v^{ij}\,,\quad J_{r}^{\mu a}\equiv{\tau^{a}}_{ij}v^{i+2,j+2}\,,\label{eq:left_right_diffeo}
\end{equation}
generate the isometry algebra, which is isomorphic to $so\left(4\right)\simeq su_{l}\left(2\right)\oplus su_{r}\left(2\right)$.
The remaining combinations yield conformal Killing transformations.
Our choice of vielbein was motivated by
\[
{e_{\mu}}^{a}=-\frac{1}{2}J_{l,\mu}^{a}\,.
\]

\subsection{\label{sec:Change-of-coordinates}Change of coordinates from flat
space}

Define the function 
\[
\exp\Omega\equiv1+\sin\theta\cos\varphi\,.
\]
The round $\mathbb{S}^{3}$, in toroidal coordinates, is related to
$\mathbb{R}^{3}$ by the following change of coordinates 
\begin{gather*}
x_{1}\rightarrow e^{-\Omega}\cos\theta\cos\tau\,,\\
x_{2}\rightarrow e^{-\Omega}\cos\theta\sin\tau\,,\\
x_{3}\rightarrow e^{-\Omega}\sin\theta\sin\varphi\,,
\end{gather*}
followed by a Weyl transformation with parameter $\Omega$.\footnote{In order to compare our conventions to those of \cite{Dedushenko:2016jxl},
one should take $r=1/2$ in \cite{Dedushenko:2016jxl}, and also rescale
$\exp\Omega_{\text{here}}=2\exp\Omega_{\text{there}}$. }

Let $B$ be the change of variables matrix. The flat metric and the
flat vielbein ${e_{\mu}}^{a}={\delta_{\mu}}^{a}$ transform as 
\[
g\rightarrow e^{2\Omega}B^{T}g_{\mathbb{R}^{3}}B\,,\quad e\rightarrow e^{'}\equiv e^{\Omega}B^{T}e_{\mathbb{R}^{3}}\,.
\]
The frame rotation matrix $F\in SO\left(3\right)$ is defined as
\[
{F_{a}}^{b}\equiv{e_{\mu}^{\mathbb{S}^{3}}}^{b}g_{\mathbb{S}^{3}}^{\mu\nu}{e_{\nu}^{'}}_{a}\,.
\]

Define the spinor bilinears 
\[
A_{\mathbb{S}^{3}}^{ij}\equiv\epsilon^{\left(i\right)}\epsilon^{\left(j\right)}\,,\quad A_{\mathbb{R}^{3}}^{ij}\equiv\epsilon_{\mathbb{R}^{3}}^{\left(i\right)}\epsilon_{\mathbb{R}^{3}}^{\left(j\right)}\, .
\]
There exists a numerical matrix ${R^{i}}_{j}$, unique up to sign,
which relates the spinors on $\mathbb{R}^{3}$ and on $\mathbb{S}^{3}$
with their chosen coordinate systems, such that 
\begin{gather*}
A_{\mathbb{S}^{3}}^{ij}={R^{i}}_{k}{R^{j}}_{l}A_{\mathbb{R}^{3}}^{kl}\,,\\
v_{a}^{ij}={F_{a}}^{b}{R^{i}}_{k}{R^{j}}_{l}w_{b}^{kl}\,.
\end{gather*}
Given $R$, we can associate the $\mathbb{S}^{3}$ spinor $\epsilon^{\left(i\right)}$
with 
\[
e^{\Omega/2}{R^{i}}_{j}\epsilon_{\mathbb{R}^{3}}^{\left(j\right)}\,.
\]

One may check that with the current choice of basis for the spinors,
\[
R=\begin{pmatrix}-\frac{1}{2}-\frac{\mathbbm{i}}{2} & 0 & -\frac{1}{2}+\frac{\mathbbm{i}}{2} & 0\\
0 & -\frac{1}{2}+\frac{\mathbbm{i}}{2} & 0 & \frac{1}{2}+\frac{\mathbbm{i}}{2}\\
0 & \frac{1}{2}-\frac{\mathbbm{i}}{2} & 0 & \frac{1}{2}+\frac{\mathbbm{i}}{2}\\
-\frac{1}{2}-\frac{\mathbbm{i}}{2} & 0 & \frac{1}{2}-\frac{\mathbbm{i}}{2} & 0
\end{pmatrix}\,.
\]
The matrix $R$ satisfies
\[
R^{\dagger}R=\mathbbm{1}_{4}\,.
\]
One could derive $R$ by lifting the $SO\left(3\right)$ frame rotation
$F$ to $SU\left(2\right)$, and acting on the spinor indices.
\section{Details on the SQM}\label{app_sqm}

Below we describe how the SQM is gauged by the bulk vector multiplet.
Our strategy goes as follows. First, we determine the 3d submultiplet of the vector multiplet generated by the action of $Q_2$ and $Q_3$. This can be dimensionally reduced on the curve supporting the defect. Finally, we compare our result $\mathcal{N}=2$ SQM, and we read which symmetries are turned on. 
Since the authors of \cite{Hori:2014tda} define the SQM with a Lorentzian time, while our coordinate $\tau$ is a Euclidean time, we introduce a \emph{real time} $t=-\mathbbm{i}\tau $. Therefore, we also need to set\footnote{In our convention $D_\tau=\partial_\tau-\mathbbm{i}A_\tau$ and $D_t=\partial_t-\mathbbm{i}A_t$. }
\begin{equation}
A_\tau=-\mathbbm{i}A_t\,, \qquad D_\tau=-\mathbbm{i}D_t\,.
\end{equation}
Moreover, to avoid coordinate singularities as $\theta\to 0$, we use frame indices, which are well defined on the whole $S^3$.

On general grounds, the 3d $\mathcal{N}=4$ vector multiplet is decomposed as follows:
\begin{itemize}
\item 1 $\mathcal{N}=2$ vector multiplet 
\begin{equation}
\mathbb{V}=(v_t,\,\sigma,\, \lambda,\, \bar\lambda,\, D )\,.
\end{equation}
\item 2 $\mathcal{N}=2$ chiral multiplets \footnote{$ i=1,2$ labels the multiplet. The corresponding anti-chirals are denoted by $\tilde{\Phi}_i=(\tilde\phi_i, \tilde\psi_i )$.}
\begin{equation}
\Phi_i=(\phi_i, \psi_i )\,.
\end{equation}
\item 2 $\mathcal{N}=2$ Fermi multiplets
\begin{equation}
\mathbb{F}=(\eta, F )\,, \qquad \tilde{\mathbb{F}}=(\tilde\eta, \tilde F )\,.
\end{equation}
\end{itemize}
In fact, if we write 3d $\mathcal{N}=4$ vector multiplet as a $\mathcal{N}=2$ vector multiplet plus a $\mathcal{N}=2$ chiral, we can reduce it to a 2d $\mathcal{N}=(2,2)$ vector plus an $\mathcal{N}=(2,2)$ chiral. From an  $\mathcal{N}=(0,2)$ point of view, the vector multiplet contains a vector one plus a chiral one. Similarly, the $\mathcal{N}=(2,2)$ chiral multiplet is decomposed into a $\mathcal{N}=(0,2)$ chiral one plus a Fermi multiplet \cite{Witten:1993yc}.  Since the $\mathcal{N}=2$ SQM is the dimensional reduction of the 2d $\mathcal{N}=(0,2)$, a last dimensional reduction leads to the above decomposition.

Let us provide the explicit SQM structure.
Since on the circle $A_\tau=A_3$, it is natural to build the 1d vector multiplet $\mathbb{V}$ acting on it with $\delta=\epsilon\, Q_2 + \bar{\epsilon}\,Q_3$. The multiplet turns out to be given by
\begin{subequations}
\begin{align}
v_\tau&=\mathbbm{i}A_3\,,\\
\sigma&=-\Phi_{\dot{1}\dot{2}}\,,\\
\lambda&=\left(\frac{1+\mathbbm{i}}{2}\right) \left(\sqrt{1-\nu }  \lambda _{2,1\dot{1}}+\mathbbm{i}\e^{-\mathbbm{i} \tau } \sqrt{1+\nu} \lambda _{2,2\dot{1}}\right) \,,\\
\bar{\lambda}&=\left(\frac{1+\mathbbm{i}}{2}\right) \left(\sqrt{1-\nu }\, \lambda _{1,2\dot{2}}+\mathbbm{i} \e^{\mathbbm{i} \tau } \sqrt{1+\nu}  \lambda _{1,1\dot{2}}\right)\,,\\
D&=--\mathbbm{i}M^{ab}_{\text{vortex}}D_{ab}-\mathbbm{i}\nu\Phi_{\dot{1}\dot{2}}+F_{12}+\frac{1}{2}\comm*{\Phi\indices{_{\dot{2}}^{\dot{c}}}}{\Phi_{\dot{c}\dot{1}}}\, .
\end{align}
\end{subequations}
Notice that the auxiliary fields in $D$ are related to the vortex loop defined in Eq \ref{eq:M_vortex}. Similarly, $\sigma$ coincides the scalar part of the Wilson loop connection in Eq \ref{eq:Wilson_loop_matrix}, up to an $\mathbbm{i}$ factor.
The corresponding supersymmetry transformations are:
\begin{subequations}\label{eq3.2.1}
\begin{align}
\delta v_\tau&=\frac{\mathbbm{i}}{2}\epsilon\bar{\lambda}+\frac{\mathbbm{i}}{2}\bar{\epsilon}\lambda\,,  \\
\delta \sigma&= -\delta v_\tau \,, \\
\delta \lambda&=  \epsilon\left(D_t\sigma+\mathbbm{i}D\right)\,,\\
\delta\bar{\lambda}&=\bar{\epsilon}\left(D_t\sigma-\mathbbm{i}D\right)\,,\\
\delta D&=\frac{1}{2}\epsilon D_t^+\bar\lambda-\frac{1}{2}\bar\epsilon D_t^+\lambda\,,
\end{align}
\end{subequations}
where $D_t^+=D_t-\mathbbm{i}\comm{\sigma}{\,\,\,\,}$. The crucial point of this decomposition is that $\delta^2\mathbb{V}$ \emph{does not contain $\nu$}. For instance:
\begin{equation}
\delta^2\sigma=\mathbbm{i}\epsilon \bar\epsilon D_\tau\sigma\,.
\end{equation}
The operator $\delta^2$ contains a gauge transformation generated by
\begin{equation}
\Lambda=\mathbbm{i}(\sigma+v_t)\, .
\end{equation}

The other two scalar fields are recast as lowest components of a chiral and an anti-chiral multiplet $\Phi_1$ and $\tilde{\Phi}_1$. Let us first consider the chiral one
\begin{subequations}
\begin{align}
\phi_1&=\Phi_{\dot{1}\dot{1}}\,,\\
\psi_1&=\left(\frac{1+\mathbbm{i}}{2}\right) \left(\e^{\mathbbm{i} \tau }\sqrt{1+\nu}  \lambda _{1,1\dot{1}}-\mathbbm{i} \sqrt{1-\nu } \lambda _{1,2\dot{1}}\right) \,.
\end{align}
\end{subequations}
The supersymmetry transformations are
\begin{subequations}
\begin{align}
\delta \phi_1 &=- \epsilon \psi_1\,,\\
\delta \psi_1&=\bar{\epsilon} \left( \mathbbm{i}D_t^+ \phi_1+\mathbbm{i}\nu\phi_1 \right)\,.\label{eq3.2.2}
\end{align}
\end{subequations}
It is easy to deduce from the action of $\delta^2$ the presence of a background symmetry proportional to $\nu$
\begin{equation}
\delta^2\phi_1=\epsilon \bar\epsilon (D_\tau^+\phi_1-\mathbbm{i}\nu\phi_1)\,, \qquad  \delta^2\psi_1=-\epsilon \bar\epsilon (D_\tau^+\,\psi_1-\mathbbm{i}\nu\,\psi_1)\,.
\end{equation}
The result is understood giving  charge $-1$ to $\Phi_1$ under the $U(1)$ flavor symmetry generated by $\nu F$.

Similarly, we obtain the antichiral multiplet:
\begin{subequations}
\begin{align}
\tilde{\phi}_1&=\Phi_{\dot{2}\dot{2}}\,,\\
\tilde{\psi}_1&=\left(\frac{1+\mathbbm{i}}{2}\right)  \left( -\mathbbm{i}\sqrt{1-\nu }  \lambda _{2,1\dot{2}}+\e^{-\mathbbm{i} \tau } \sqrt{1+\nu} \lambda _{2,2\dot{2}}\right)\,.
\end{align}
\end{subequations}
The supersymmetry variations are given by
\begin{subequations}
\begin{align}
\delta \tilde{\phi}_1 &=- \bar{\epsilon} \tilde{\psi}_1\,,\\
\delta \tilde{\psi}_1&=\bar{\epsilon} \left( \mathbbm{i}D_t^+ \tilde{\phi}_1-\mathbbm{i}\nu\tilde{\phi}_1 \right)\,.
\end{align}
\end{subequations}
We can interpret these variations assigning charge 1 to $\tilde{\Phi}_1$ under to $\nu F$. Since $\nu$ is not integer, the $\nu$-background cannot be reabsorbed in any redefinition of the Killing spinors. This fact prevents $\mathbb{V}$ and $\Phi_1$ to constitute an $\mathcal{N}=4$ vector multiplet.

The other chiral multiplets come from the components of the orthogonal components of the gauge field. In particular, the first one is given by
\begin{subequations}
\begin{align}
\phi_2&=A_1+\mathbbm{i} A_2\,,\\
\psi_2&=-\left(\frac{1+\mathbbm{i}}{2}\right)  \left(\sqrt{1-\nu } \lambda _{2,2\dot{2}}+\mathbbm{i}\e^{\mathbbm{i} \tau } \sqrt{1+\nu}  \lambda _{2,1\dot{2}}\right)\,.
\end{align}
\end{subequations}
The variations are given by
\begin{subequations}
\begin{align}
\delta \phi_2 &=- \epsilon \psi_2\,,\\
\delta \psi_2&=-i\bar{\epsilon}(\mathbbm{i} D_t^+\phi_2+ 2\mathbbm{i}\phi_2)+ \epsilon\, \mathbbm{i}\,\e^{\mathbbm{i}\tau}\sqrt{1-\nu^2}\,\Phi_{\dot{2}\dot{2}}\,.\label{nonchir1}
\end{align}
\end{subequations}

The remaining component of the gauge field becomes the lowest component of an anti-chiral field
\begin{subequations}
\begin{align}
\tilde{\phi}_2&=A_1-\mathbbm{i} A_2\,,\\
\tilde\psi_2&=\left(\frac{1+\mathbbm{i}}{2}\right) \left(\sqrt{1-\nu }  \lambda _{1,1\dot{1}}+i \e^{-\mathbbm{i} \tau }\sqrt{\nu +1} \lambda _{1,2\dot{1}}\right)\,.
\end{align}
\end{subequations}
The variations are given by\footnote{The non-chiral parts of the transformations in Eq \ref{nonchir1} and in Eq \ref{nonchir2} do not appear in the 1d algebra of \cite{Hori:2014tda}. While these terms are due to a gauge transformations for $Q_2^2$ and $Q_3^2$ for the bulk theory, we do not have a clear interpretation from the worldvolume perspective. However, we claim that they do not spoil our arguments for the computation of the index.}
\begin{subequations}
\begin{align}
\delta \tilde{\phi}_2 &=- \bar{\epsilon} \tilde{\psi}_2\,,\\
\delta \tilde{\psi}_2&=\epsilon (\mathbbm{i} D_t^+\tilde\phi_2+2\mathbbm{i}\tilde\phi_2 )+ \bar\epsilon\, \mathbbm{i}\,\e^{-\mathbbm{i}\tau}\sqrt{1-\nu^2}\,\Phi_{\dot{1}\dot{1}}\,.\label{nonchir2}
\end{align}
\end{subequations}

The remaining fermionic degrees of freedom are recast in two Fermi fields. At this stage, we can choose them to be any non singular linear combinations of $\lambda_{a\dot{a}}$.
In other words, the only constraint we impose is that the change of variables from the 3d degrees of freedom to the 1d ones is invertible. Our choice is given by
\begin{subequations}
\begin{align}
\eta&=e^{-\mathbbm{i}\tau}\,\lambda_{2,2\dot{1}}\,,\\ 
F&=\frac{1+\mathbbm{i}}{2}\left(\mathbbm{i}e^{-\mathbbm{i}\tau}\sqrt{1-\nu}D_{22}+\sqrt{1+\nu}\left( \mathbbm{i}D_\tau \Phi_{\dot{1}\dot{2}}-\Phi_{\dot{1}\dot{2}} -D_{12}-\mathbbm{i}F_{12}+\frac{\mathbbm{i}}{2}\comm*{\Phi_{\dot{1}\dot{1}}}{\Phi_{\dot{2}\dot{2}}}  \right)\right)\,,
\end{align}
\end{subequations}
and
\begin{subequations}
\begin{align}
\tilde{\eta}&=e^{\mathbbm{i}\tau}\,\lambda_{1,1\dot{2}}\,, \\
\tilde{F}&=\frac{1-\mathbbm{i}}{2}\left(e^{\mathbbm{i}\tau}\sqrt{1-\nu}D_{11}+\sqrt{1+\nu}\left(\mathbbm{i}D_{12} - F_{12}-D_\tau \Phi_{\dot{1}\dot{2}}+\mathbbm{i}\Phi_{\dot{1}\dot{2}} +\frac{1}{2}\comm*{\Phi_{\dot{1}\dot{1}}}{\Phi_{\dot{2}\dot{2}}}  \right)\right)\,.
\end{align}
\end{subequations}
The variations are \footnote{These 1d transformations are chosen to reproduce the expected $\delta^2$ on $\eta$ and $\tilde\eta$. We eliminated parts related to orthogonal derivatives, which cancel among themselves as we take the double variations.}
\begin{subequations}
\begin{align}
\delta\eta&=\epsilon F-\bar\epsilon \frac{1-\mathbbm{i}}{2}\sqrt{1-\nu}\, e^{-\mathbbm{i}\tau}\, \comm*{\phi_2}{\phi_1}\,,\\
\delta\tilde{\eta}&=\bar\epsilon\tilde F+\epsilon\frac{1-\mathbbm{i}}{2}\sqrt{1-\nu} \,e^{\mathbbm{i}\tau} \, \comm*{\tilde{\phi}_2}{\tilde{\phi}_1}\,,\\
\delta F&= \bar\epsilon\left( -\mathbbm{i}D_t^+\eta - \frac{1-\mathbbm{i}}{2}\sqrt{1-\nu}\, e^{-\mathbbm{i}\tau}\, \left( \comm*{\psi_2}{\phi_1}+ \comm*{\phi_2}{\psi_1} \right)\right).\\
\delta \tilde F&=\epsilon\left( -\mathbbm{i}D_t^+\tilde\eta+
 \frac{1-\mathbbm{i}}{2}\sqrt{1-\nu}\, e^{\mathbbm{i}\tau}\, \left( \comm*{\tilde\psi_2}{\tilde\phi_1}+ \comm*{\tilde\phi_2}{\tilde\psi_1} \right) \right)\,.
\end{align}
\end{subequations}

\subsection{Embedding at $\nu=1$}\label{sqm1}

We study in detail the properties of the algebra generated by $Q_2$, $Q_3$, $Q_5$ and $Q_6$. We define an operator acting on the fields as
\begin{equation}
\delta=\epsilon\, Q_2+ \bar{\epsilon}\, Q_3+\rho\, Q_5+\bar{\rho}\, Q_6\,.
\end{equation}
In particular, we discuss the emergence of the conformal symmetry at $\nu=1$.  We compute the action of $\delta^2$ using the variations in Eq \ref{clos_N_4}, restricted to the circle $\theta=0$. We find that $\delta$ squares to:
\begin{itemize}
\item A diffeomorphism generated by
\begin{equation}
v^a_{\nu=1}=\zeta\bar{\zeta}(0,0,1)\,,
\end{equation}
where $\zeta$ and $\bar{\zeta}$ are anti-periodic 1d spinors defined in Eq \ref{antiper_spin}.
The vector $v^a_{\nu=1}$ can be expanded on the subgroup of $\mathrm{Diff}\left(S^1\right)$ generated by $L_{-1},L_0,L_1$, where $L_m=\e^{\mathbbm{i}m\tau}\partial_\tau$. The generators $L_m$ close the Witt algebra
\begin{equation}\label{L_algebra}
\comm{L_m}{L_n}=\mathbbm{i}(m-n)L_{m+n}\,.
\end{equation}
Thus, $\mathbbm{i}L_0, \,L_{-1},\,L_1$ constitute a representation of the generators $M$, $P$ and $K$ introduced in Section \ref{sub_lat_alg_nu_1}. 

\item A dilatation, given by
\begin{equation}
\hat{\rho}= \mathbbm{i}   \left(e^{ \mathbbm{i} \tau } \epsilon  \bar{\rho }-e^{-\mathbbm{i} \tau }\rho  \bar{\epsilon }\right)\,.
\end{equation}
This corresponds to a Weyl transformation contained in the commutators $\acomm*{Q_2}{Q_6}$ and $\acomm*{Q_3}{Q_5}$, which, from the point of view of the bulk theory, corresponds to the Weyl transformation stored in $K_1\pm\mathbbm{i}K_2$. 

\item A gauge transformation
\begin{equation}
\Lambda=-\bar\zeta\zeta(A_3+\mathbbm{i}\Phi_{\dot{1}\dot{2}})\,,
\end{equation}

\item The following R-symmetry transformations
\begin{align}
R_H=\frac{\mathbbm{i}}{2}\left(
\begin{array}{cc}
 \bar\rho\rho- \bar\epsilon\epsilon  & 0 \\
 0 & \bar\epsilon\epsilon-\bar\rho\rho\\
\end{array}
\right)  \,,\qquad 
R_C=\frac{\mathbbm{i}}{2}\left(
\begin{array}{cc}
\bar\epsilon\epsilon -\bar\rho\rho & 0 \\
 0 & \bar\rho \rho-\bar\epsilon \epsilon  \\
\end{array}
\right)\,.
\end{align}
\end{itemize}
In order to explore the superconformal algebra, we represent the $\delta$-action on the vector multiplet and on the chiral one. We will denote the 1d component fields in the same way as $\nu$ generic, even tough the embedding is slightly different from the limit $\nu\to 1$ of the one described in the previous section\footnote{For instance, the embedding proposed at $\nu-$generic becomes singular for the Fermi fields as $\nu\to 1$.}. Moreover, all the fermions are now taken anti-periodic.

At $\nu=1$, the embedding for the 1d vector multiplet turns out to be
\begin{align}
v_\tau&=\mathbbm{i}A_3=A_t\,,\\
\sigma&=-\Phi_{\dot{1}\dot{2}}\,,\\
\lambda&=\mathbbm{i} \frac{1+\mathbbm{i}}{\sqrt{2}}  \,e^{-\frac{\mathbbm{i}}{2}\tau}\,\lambda _{2,2\dot{1}} \,,\\
\bar{\lambda}&=\mathbbm{i}\frac{1+\mathbbm{i}}{\sqrt{2}}\, e^{\frac{\mathbbm{i}}{2}\tau}\, \lambda _{1,1\dot{2}}\,,\\
D&=F_{12}-\mathbbm{i}D_{12}+\frac{1}{2}\comm*{\Phi\indices{_{\dot{2}}^{\dot{c}}}}{\Phi_{\dot{c}\dot{1}}}\,.
\end{align}
The supersymmetry transformations are given by
\begin{align}
\delta v_\tau&= \frac{\mathbbm{i}}{2}\zeta\bar{\lambda}+\frac{\mathbbm{i}}{2}\bar{\zeta}\lambda\,,\\
\delta \sigma&= -\delta v_\tau \,, \\
\delta \lambda&=\zeta\left(D_t\sigma+\mathbbm{i}D\right)+2\mathbbm{i} \sigma\partial_\tau\zeta\,,\\
\delta\bar{\lambda}&=\bar{\zeta}\left(D_t\sigma-\mathbbm{i}D\right)+2\mathbbm{i} \sigma\partial_\tau\bar\zeta\,,\\
\delta D&=\frac{1}{2}\left(D_t^+\left(\zeta\bar\lambda\right)+2\mathbbm{i}\bar\lambda\partial_\tau\zeta\right)-\frac{1}{2}\left(D_t^+\left(\bar\zeta\lambda\right)+2\mathbbm{i} \lambda\partial_\tau\bar\zeta\right)\,.
\end{align}
We can use these variations to compute the action of $\delta^2$ on the field components:
\begin{equation}\label{alg1}
\delta^2\sigma=D_\tau\left(\bar\zeta\zeta\sigma\right)\,,
\end{equation}
We see that it reproduces the algebra in Eq \ref{1d_conf_ferm} \footnote{We are omitting the gauge transformations, which are not relevant for the discussion.}
\begin{subequations}
\begin{align}
\acomm{Q_2}{Q_3}\sigma&= L_0\sigma\,, &\acomm{Q_5}{Q_6}\sigma&=L_0\sigma\,,\\
\acomm{Q_2}{Q_6}\sigma&=\left( L_1+\mathbbm{i}\e^{\mathbbm{i}\tau}\right)\sigma\,, &\acomm{Q_3}{Q_5}\sigma&=\left( L_{-1}-\mathbbm{i}\e^{-\mathbbm{i}\tau}\right)\sigma\,.
\end{align}
\end{subequations}
For example, $\sigma$ is consistently uncharged under the R-symmetry $J$ and that $\sigma$ has Weyl weight 1.
For the gauginos we obtain
\begin{align}\label{alg2}
\delta^2\lambda=D_\tau^+\left(\bar\zeta\zeta\lambda\right)+ \bar\zeta\lambda\partial_\tau\zeta\,, \qquad \delta^2\bar\lambda=D_\tau^+\left(\bar\zeta\zeta\bar\lambda\right)+\zeta\bar\lambda\partial_\tau\bar\zeta\,,
\end{align}
which yields
\begin{subequations}
\begin{align}
\acomm{Q_2}{Q_3}\lambda&=\left(L_0+\frac{\mathbbm{i}}{2}\right) \lambda \,, &\acomm{Q_5}{Q_6} \lambda&=\left( L_0-\frac{\mathbbm{i}}{2}\right) \lambda\,,\\
\acomm{Q_2}{Q_6} \lambda&=\left( L_1+\frac{3\mathbbm{i}}{2}\e^{\mathbbm{i}\tau}\right) \lambda\,, &\acomm{Q_3}{Q_5} \lambda&=\left( L_{-1}-\frac{3\mathbbm{i}}{2}\e^{-\mathbbm{i}\tau}\right) \lambda\,.
\end{align}
\end{subequations}
Thus, $\lambda$ has Weyl weight $3/2$ and has charge $-1/2$ under $J$. For $\bar\lambda$ we get the same result but with opposite $J$-charge.

We also provide the embedding for the chiral field
\begin{align}
\phi_1&=\Phi_{\dot{1}\dot{1}}\,,\\
\psi_1&=e^{-\frac{\mathbbm{i}}{2}\tau}\frac{\left(1+\mathbbm{i}\right)}{\sqrt{2}}   \lambda _{1,1\dot{1}}\,,
\end{align}
with the supersymmetry transformations
\begin{align}
\delta \phi_1 &=- \zeta \psi_1\,,\\
\delta \psi_1&=\left( iD_t^+ \left(\bar{\zeta} \phi_1\right) -\phi_1\partial_\tau\bar\zeta \right)\,.
\end{align}
The action of $\delta^2$ on $\phi_1$ is
\begin{subequations}
\begin{align}
\acomm{Q_2}{Q_3}\phi_1&=\left( L_0-\mathbbm{i}\right)\phi_1\,, &\acomm{Q_5}{Q_6}\phi_1&=\left( L_0+\mathbbm{i}\right)\phi_1\,,\\
\acomm{Q_2}{Q_6}\phi_1&=\left( L_1+\mathbbm{i}\e^{\mathbbm{i}\tau}\right)\phi_1\,, &\acomm{Q_3}{Q_5}\phi_1&=\left( L_{-1}-\mathbbm{i}\e^{-\mathbbm{i}\tau}\right)\phi_1\,.
\end{align}
\end{subequations}
We see that $\phi_1$ is charged under $J$ with charge 1 and that $\phi_1$ has Weyl weight 1.
For $\psi_1$, we get
\begin{subequations}
\begin{align}
\acomm{Q_2}{Q_3}\psi_1&=\left( L_0 -\frac{\mathbbm{i}}{2}\right)\psi_1\,, &\acomm{Q_5}{Q_6}\psi_1&=\left(L_0+\frac{\mathbbm{i}}{2}\right)\psi_1\,,\\
\acomm{Q_2}{Q_6}\psi_1&=\left( L_1+\frac{3}{2}\mathbbm{i}\e^{\mathbbm{i}\tau}\right)\psi_1\,, &\acomm{Q_3}{Q_5}\psi_1&=\left( L_{-1}-\frac{3}{2}\mathbbm{i}\e^{-\mathbbm{i}\tau}\right)\psi_1\,.
\end{align}
\end{subequations}
Thus, we see that we can assign a Weyl weight $3/2$ to $\psi_1$ and a $J$ charge $1/2$. In conclusion, we have shown concretely that our $\delta$ reproduces the algebra of the latitude Wilson loop at $\nu=1$.

\bibliographystyle{JHEP}
\bibliography{General_bibliography}
\end{document}